\documentclass[referee]{raa}           
\usepackage[pdftex]{graphicx}
\usepackage{times}
\usepackage{natbib}
\usepackage{amssymb,amsmath}
\usepackage{multirow}
\usepackage{bm}
\usepackage{booktabs}
\usepackage{threeparttable}

\bibpunct{(}{)}{;}{a}{}{,}
\usepackage[a4paper]{geometry}
\usepackage[a4paper=true,pagebackref=true]{hyperref}
\hypersetup{pdftitle = The title of my PDF, pdfauthor = My name, pdfsubject= The subject, pdfkeywords = keyword1 keyword2 keyword3} 
\hypersetup{colorlinks = true, linkcolor = green, anchorcolor = red, citecolor = blue, filecolor = red, pagecolor = red, urlcolor = red}

\begin{document}

   \title{A new correction method for quasi-Keplerian orbits}

 \volnopage{ {\bf 20XX} Vol.\ {\bf X} No. {\bf XX}, 000--000}
   \setcounter{page}{1}

  \author{Yue Chen\inst{1,2}, Da-Zhu Ma\inst{3}, Fang Xia\inst{1}
   }

   \institute{Purple Mountain Observatory, Chinese Academy of Sciences, Nanjing 210033, China\\
        \and
              School of Astronomy and Space Science, University of Science and Technology of China, Hefei, Anhui 230026, China\\
	\and
School of Information and Engineering, Hubei Minzu University, Enshi 445000, China; {\it mdzhbmy@126.com}\\
\vs \no
   {\small Received 20XX Month Day; accepted 20XX Month Day}
}

\abstract{A pure two-body problem has seven integrals including the Kepler energy, the Laplace vector, and the angular momentum vector. However, only five of them are independent. When the five independent integrals are preserved, the two other dependent integrals are naturally preserved from a theoretical viewpoint; but they may not be either from a numerical computational viewpoint. Because of this, we use seven scale factors to adjust the integrated positions and velocities so that the adjusted solutions strictly satisfy the seven constraints. Noticing the existence of the two dependent integrals, we adopt the Newton iterative method combined with the singular value decomposition to calculate these factors. This correction scheme can be applied to perturbed two-body and $N$-body problems in the solar system. In this case, the seven quantities of each planet slowly vary with time. More accurate values can be given to the seven slowly-varying quantities by integrating the integral invariant relations of these quantities and the equations of motion. They should be satisfied with the adjusted solutions. Numerical tests show that the new method can significantly reduce the rapid growth of numerical errors of all orbital elements.
}

   \authorrunning{Y. Chen et al. }            
   \maketitle

%
\section{Introduction}           
\label{sec:introduction}
Numerical integration methods are convenient tools to study complex nonlinear dynamics problems. \citep[e.g.][]{2003PhLA..313...77W,PhysRevD.74.083001,2006ApJ...652.1466W,2007PhRvD..76l4004W,2008PhRvD..77j3012W,2014PhRvD..89l4034H,2015MNRAS.452.3167W,2015PhRvD..91b4042W,2019AnP...53100136L}. Above all, geometric integration algorithms can keep some physical or geometric properties. Therefore, they have been widely used in celestial mechanics, general relativity, cosmology, and post-Newtonian spinning compact binaries. Here, we list some classes of geometric algorithms. As one of the geometric integration                                                                                                                     algorithms, symplectic integrators \citep{1983ITNS...30.2669R,1990PhyD...43..105F,1991AJ....102.1528W,2010PhRvD..82l4040Z,2010PhRvD..81h4045W,2013EPJC...73.2413M,2013MNRAS.435.2246M} can maintain the symplectic structure of the system. Extended phase-space methods \citep{2015CeMDA.121..211P,2016MNRAS.459.1968L,2017ApJ...834...64L,2017MNRAS.469.3031L} as explicitly symplectic-like or symmetric schemes are mainly used for inseparable Hamiltonian systems. Energy-preserving algorithms \citep{2018ApJS..237....6B,2019ApJS..240...40B,2019ApJ...887..191H} are generally implicit and nonsymplectic. Furthermore, they can exactly conserve the energy integral of a conservative Hamiltonian. The manifold correction schemes \citep{Nacozy1971,2007CoPhC.177..500H} adopt invariant manifolds to correct errors in the numerical solutions. In this paper, we only focus on the manifold correction schemes.

Nacozy's manifold correction scheme \citep{Nacozy1971} uses the least-squares method and pulls the solution back to the original integral hypersurface along the shortest path. In this way, the error of an integral given by the correction method doubles that given by the uncorrected method.
This is why Nacozy's manifold correction scheme can improve the precision of numerical integration.
However, \cite{Hairer1999} found that this algorithm was not effective in the simulation of a five-body problem of the sun and four outer planets. \cite{2006CoPhC.175...15W,2007AJ....133.2643W} pointed out that this algorithm doesn't work very well if only the total energy integral is preserved,
but can exhibit good performance if all the individual quasi-integrals are corrected.
The quasi-integrals are slowly-varying quantities of each body that moves in a Kepler orbit affected by a small perturbation. The slowly-varying quantities obtained from their integral-invariant relations \citep{huang1983,Mikkola2002} are regarded as reference values to correct the numerical solution. This is because they are more accurate than those that are directly determined by the integrated positions and velocities. Here are some details of the related manifold correction methods as follows.
For the pure Keplerian problem, there are seven conserved quantities including five dependent integrals in relative coordinates, the Kepler energy $K$, the momentum vector $\bm{L}$, and the Laplace vector $\bm{P}$.
The seven conserved quantities are closely related to the orbital elements.
The Kepler energy directly determines the semimajor axis and the mean anomaly.
The eccentricity is calculated from the magnitude of $\bm{P}$,
and the argument of perihelion is determined by the $z$ component of $\bm{P}$.
The orbital inclination and longitude of ascending node are given by the magnitude and three components of $\bm{L}$.
It means that the precision of orbital elements of each body can be improved effectively if the Kepler energy $K$, the Laplace vector $\bm{P}$, and the angular momentum vector $\bm{L}$ are conserved simultaneously at each integration step.
For an $N$-body problem, these integrals are no longer invariant quantities.
However, with the help of the integral invariant relations, the varying quantities can also be used as the correction reference values. Nacozy's manifold correction method is still effective.
Based on this, numerous extended Nacozy's manifold correction methods have been developed. Some examples are the velocity-position scaling method \citep{Fukushima2003a,Liu1994,Fukushima2003b,Fukushima2003c,Fukushima2004,Ma2008c} and velocity scaling method \citep{Ma2008a,Ma2008b,2007AJ....133.2643W}.
These methods have greatly improved the accuracy of numerical integration.
The manifold correction scheme of \cite{Ma2008c} has been applied to the elliptic restricted three-body problems  \citep{2016MNRAS.463.1352W} and the dissipative circular restricted three-body problems \citep{2018AJ....155...67W}.
As a point to note, there are two correction methods, the linear transformation with single-axis rotation method of \cite{Fukushima2004} and the extended approximate manifold correction method of \cite{Ma2008c}. It has been reported that they can improve the accuracy of all the orbital elements of each body.
The first method \citep{Fukushima2004} is a rigorous method that requires two steps to keep the Kepler energy $K$, the momentum vector $\bm{L}$,
and the Laplace vector $\bm{P}$.
In the first step, the rotation matrix $\mathbf{R}$:
$(\bm{r},\bm{v})\rightarrow(\bm{r}^{\prime},\bm{v}^{\prime})$ is introduced to maintain the consistency of the orbital angular momentum vector,
so as to adjust the direction of position $\bm{r}$ and velocity $\bm{v}$.
In the second step, the rotated position $\bm{r}^{\prime}$ and velocity $\bm{v}^{\prime}$ are linearly transformed $(\bm{r}^{\prime},\bm{v}^{\prime})\to(s_x\bm{r}^{\prime},s_v(\bm{v}^{\prime}-\alpha{\bm{r}}))$,
so that the corrected position and velocity strictly satisfy the three equations related to $K$, $\bm{P}$, and $\bm{L}$.
Unlike the method of \cite{Fukushima2004}, 
the method of \cite{Ma2008c} is a one-step correction method.
Five independently integrals, $K$, the
three components of $\bm{L}$ $(L_x$, $L_y$, $L_z)$ and the
z-component of $\bm{P}$ ($P_z$) are approximately and simultaneously satisfied in the method of \cite{Ma2008c}. The consistency of these integrals
means the improvement of errors of all the orbital elements of each
body.   

Unlike the methods of \cite{Fukushima2004} and \cite{Ma2008c}, a new manifold correction method will be given in this paper.
The correction vector of the new method is obtained directly by solving a set of nonlinear equations.
Although the nonlinear equations are underdetermined, the Newton iterative method with the singular value decomposition (SVD) is helpful to solve them.

\section{A new manifold correction scheme to pure Keplerian systems}

In this section, we construct a  new correction scheme for a pure Keplerian problem and evaluate the effectiveness of the new scheme.

\subsection{A pure Keplerian system}
\label{sec:applied}

A pure Keplerian problem is a two-body problem without perturbation. In the relative coordinate system, the Kepler energy is
\begin{equation}\label{eq:111}
K=\frac{\bm{v}^2}{2}-\frac{\mu}{r}.
\end{equation}
$K$ is an integral constant. The equation of the relative motion  is
\begin{equation}\label{eq:5}
\frac{d\bm{v}}{dt} = -\left(\frac{\mu}{r^3}\right)\bm{r},
\end{equation}
where $\bm{r}=(x,y,z)^{\mathrm{T}}, \bm{v}=(\dot{x},\dot{y},\dot{z})^{\mathrm{T}}, \mu$ = $G(M+m)$, and $r=\vert\bm{r}\vert$
represent position vector, velocity vector, the gravitational parameter, and the radius, respectively.

Clearly, the angular momentum vector and the Laplace vector are also integral constants in the pure Keplerian problem. They are written as
\begin{equation}\label{eq:6}
\bm{L}\equiv\bm{r}\times\bm{v},\quad \bm{P}\equiv\bm{v}\times\bm{L}-\left(\frac{\mu}{r}\right)\bm{r},\quad
\end{equation}
In fact, only five of these conserved quantities are completely independent because
\begin{equation}\label{eq:611}
\bm{P}\cdot\bm{L}=0,\quad P-2KL=\mu^2.
\end{equation}
Note that $K$, $\bm{L}$, and $\bm{P}$ can directly determine the orbital elements $a$, $e$, $I$, $\omega$, and $\Omega$:
\begin{equation}\label{eq:7}
\begin{split}
a=&-\frac{\mu}{2K},\quad e=\frac{P}{\mu},\\
I=&\arccos\left (\frac{L_z}{L} \right),\quad \omega= \arcsin\left(\frac{P_z}{e\sin I}\right),\\ \Omega=&\arctan\left(\frac{L_x}{-L_y}\right).
\end{split}
\end{equation}
Here, $I$, $\Omega$, and $\omega$ are in the ranges of $0<I<\pi$, $0<\Omega<2\pi$, and $0<\omega<2\pi$, respectively. The location of $\Omega$ on the orbital plane is decided by the signs of $L_x$ and $L_y$, and the location of $\omega$ is based on the signs of $P_z$ and $P_x cos \Omega + P_y sin \Omega$.
For the sixth orbital element, the mean anomaly $M$ is related to the mean motion specified by the Keplerian energy.
\subsection{The construction of the algorithm}
\label{subsec:construction}
For the pure Keplerian system, $\bm{\phi}=(K,L_x,L_y,L_z,P_x,P_y,P_z)^{\mathrm{T}}$ are conserved quantities and can be expressed as $(\phi_1(\bm{x}),\phi_2(\bm{x}),...,\phi_7(\bm{x}))^{\mathrm{T}}$. They are
\begin{equation}\label{eq:1}
\Delta\bm{\phi}(\bm{x}) = \bm{\phi}(\bm{x})-\bm{c} = \bm{0},
\end{equation}
where $\bm{\phi}(\bm{x}) = (\phi_1(\bm{x}),\phi_2(\bm{x}),...,\phi_7(\bm{x}))^{\mathrm{T}}$, $\bm{x} = (x,y,z,\dot{x},\dot{y},\dot{z})^{\mathrm{T}}$, and $\bm{c}=(c_1,c_1,...,c_7)^{\mathrm{T}}$ represent the seven conserved quantities, the state vector, and the integral constant vector, respectively. However, usually $\Delta\bm{\phi}(\bm{x}_1)\not = \bm{0}$ because of the errors in the numerical calculation.
In order to pull the solution back to the hypersurface,
the seven parameters $\bm{s}=(s_1,s_2,s_3,s_4,s_5,s_6,s_7)^{\mathrm{T}}$ are introduced to construct a correction vector $\bm{\varepsilon}(\bm{s})=(s_1x_1,s_2y_1,s_3z_1,s_4\dot{x_1}+s_7x_1,s_5\dot{y_1}+s_7y_1,s_6\dot{z_1}+s_7z_1)^{\mathrm{T}}$. $\bm{\varepsilon}(\bm{s})$ are used to adjust the numerical solution $\bm{x}_1$ with the form of
\begin{equation}\label{eq:9}
\bm{x}^{\ast}=
\left(\begin{array}{cc}
\bm{r}^{\ast}\\
\bm{v}^{\ast}
\end{array}
\right )=
\left(\begin{array}{cc}
\bm{r}_1\\
\bm{v}_1
\end{array}
\right )+\bm{\varepsilon}(\bm{s}),
\end{equation}
which satisfies Eq.~(\ref{eq:1}):
\begin{equation}\label{eq:3}
\Delta\bm{\phi}(\bm{x}^{\ast}(\bm{s}))= \bm{0}.
\end{equation}
Eq.~(\ref{eq:3}) can also be written as
\begin{equation}\label{eq:10}
\begin{split}
K-\frac{\bm{v}^{\ast 2}}{2}-\frac{\mu}{r^{\ast}}=0,\quad \bm{L}-\bm{r}^{\ast}\times\bm{v}^{\ast}=\bm{0},\quad \bm{P}-&\bm{v}^{\ast}\times\bm{L}-\left(\frac{\mu}{r^{\ast}}\right)\bm{r}^{\ast}=\bm{0}.
\end{split}
\end{equation}

Obviously, Eq.~(\ref{eq:3}) can be expanded to a set of nonlinear equations about $\bm{s}$
\begin{equation}\label{eq:12}
\Delta\bm{\phi}(\bm{s})=\begin{cases}
\frac{1}{2}\left((s_4\dot{x_1}+s_7x_1)^2+(s_5\dot{x_1}+s_7x_1)^2(s_6\dot{x_1}+s_7x_1)^2 \right)\\
-\frac{\mu}{\sqrt{(s_1x_1)^2+(s_2y_1)^2+(s_3z_1)^2}}-K &=0\\
s_2y_1(s_6\dot{x_1}+s_7x_1)-s_3z_1(s_5\dot{x_1}+s_7x_1)&=0\\
s_3z_1(s_4\dot{x_1}+s_7x_1)-s_1x_1(s_6\dot{x_1}+s_7x_1)&=0\\
s_1x_1(s_5\dot{x_1}+s_7x_1)-s_2y_1(s_4\dot{x_1}+s_7x_1)&=0\\
L_z(s_5\dot{x_1}+s_7x_1)-L_y(s_6\dot{x_1}+s_7x_1)\\
-\frac{\mu s_1 x_1}{\sqrt{(s_1x_1)^2+(s_2y_1)^2+(s_3z_1)^2}}-P_x &=0\\
L_x(s_6\dot{x_1}+s_7x_1)-L_z(s_4\dot{x_1}+s_7x_1)\\
-\frac{\mu s_2 y_1}{\sqrt{(s_1x_1)^2+(s_2y_1)^2+(s_3z_1)^2}}-P_y &=0\\
L_y(s_4\dot{x_1}+s_7x_1)-L_x(s_5\dot{x_1}+s_7x_1)\\
-\frac{\mu s_3 z_1}{\sqrt{(s_1x_1)^2+(s_2y_1)^2+(s_3z_1)^2}}-P_z &=0.
\end{cases}
\end{equation}

 Because only five of the seven equations are completely independent, Eq.~(\ref{eq:12}) is underdetermined. A method to solve such a problem is given in the Appendix~\ref{sec:iteration}. When $\bm{s}$ is obtained, the adjusted vector $\bm{x^{\ast}(\bm{s})}$ will be used as the initial solution for the next step integration. For simplicity, we call the new method as ``M1''.  Note that the form of the correction vector is not arbitrary, and the reason for such an operation will be elaborated in section~\ref{sec:parameter}.
 
For comparison, the existed methods of \cite{Fukushima2004} and \cite{Ma2008c} will be called ``M2'' and ``M3'' in the next section, respectively.

\subsection{Numerical tests}
\label{sec:experimental}

In order to evaluate the numerical performance of M1, we take the simplest two-body problem~(\ref{eq:111}) with $\mu$=1 as a test model. The initial orbital elements are $a=2$, $e=0.1$, $I=23^{\circ}$, $\Omega=50^{\circ}$, $\omega=30^{\circ}$, and $M=40^{\circ}$.
A fifth-order Runge-Kutta integrator (RK5) with a fixed time step of $1/100$ of the period $T$ is selected as a basic numerical integrator.
The analytical solution is taken as the reference value in the pure Keplerian problem.

\begin{figure}
\centering
    \includegraphics[width=0.45\textwidth]{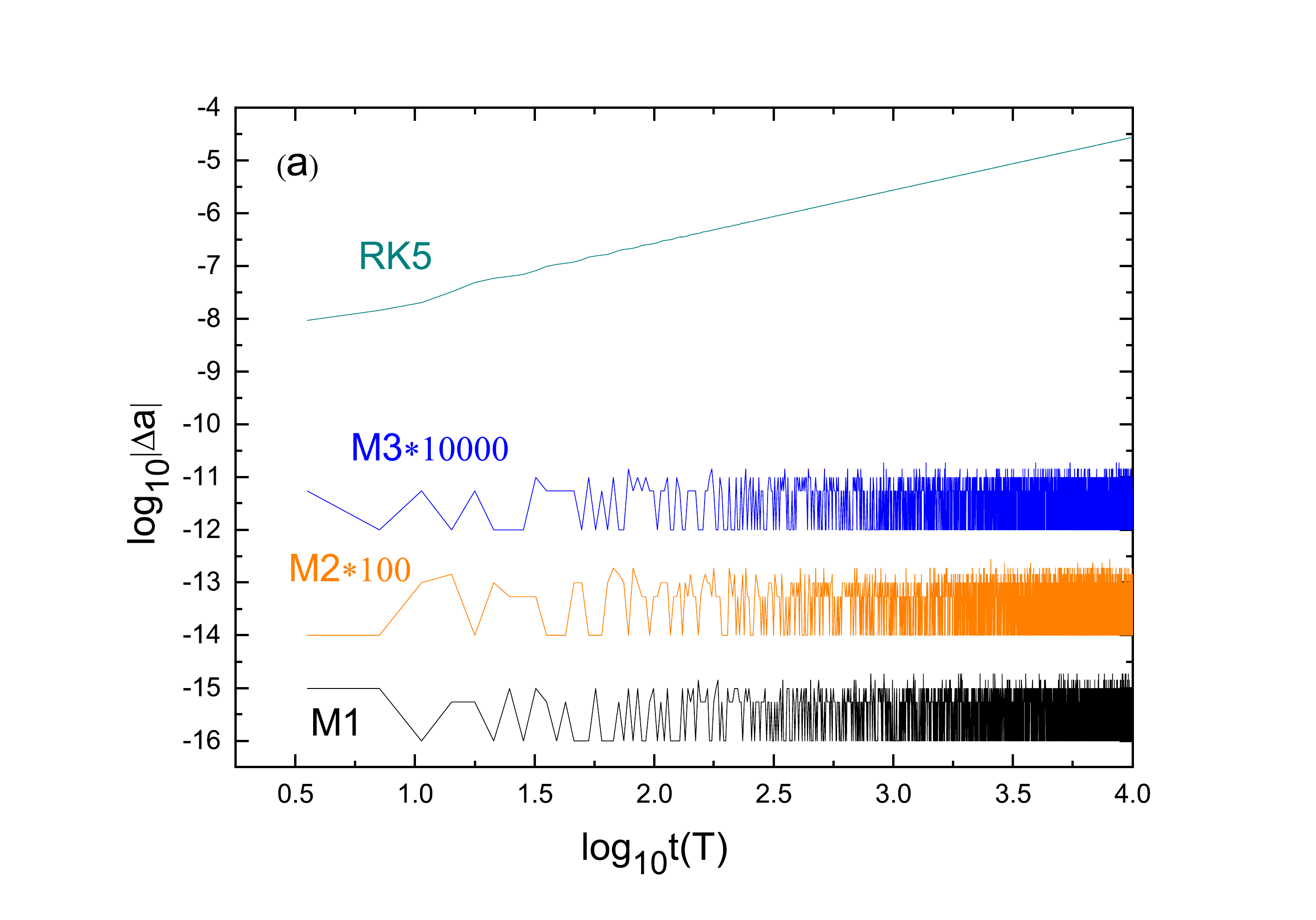}
    \includegraphics[width=0.45\textwidth]{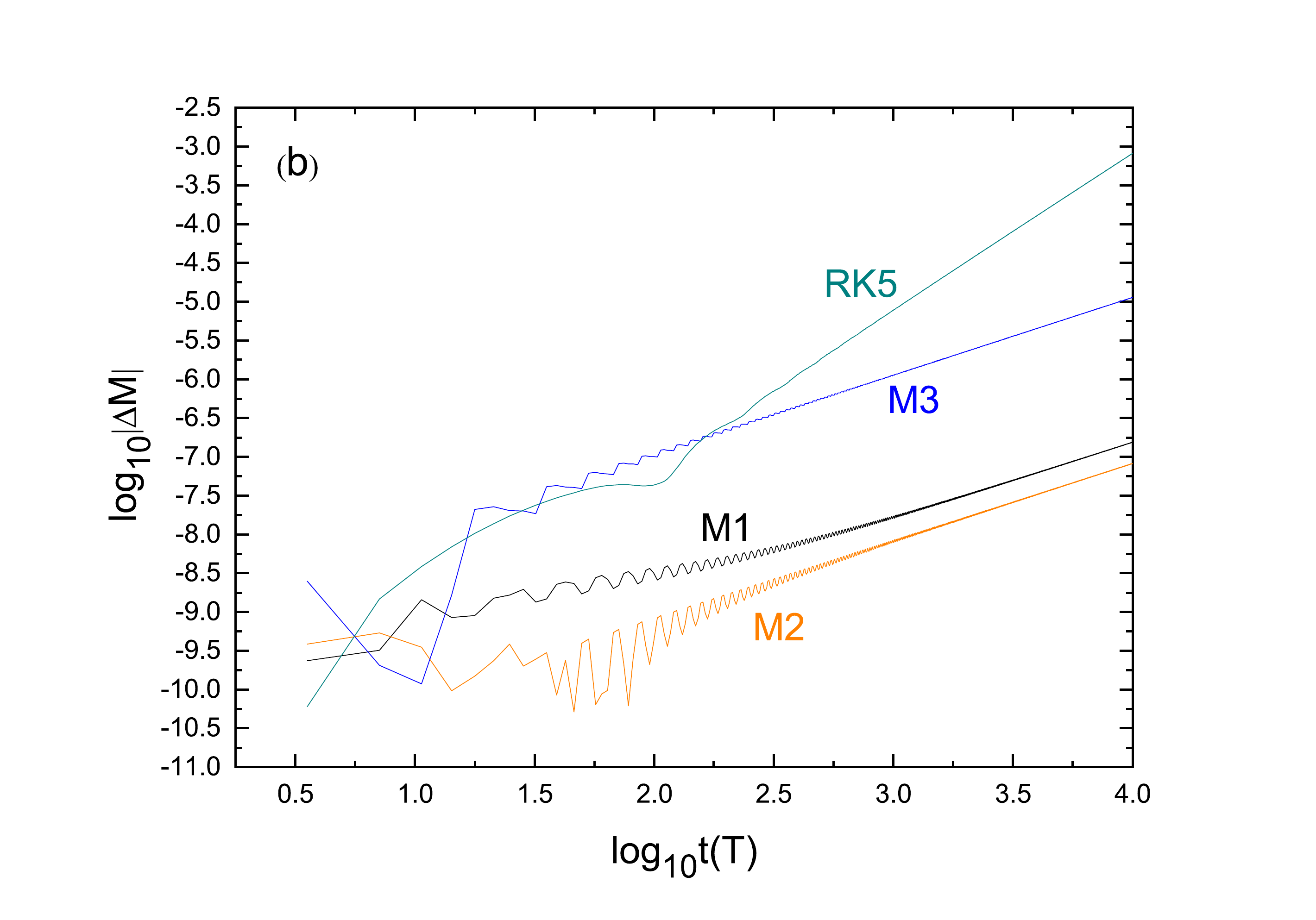} 
   \includegraphics[width=0.45\textwidth]{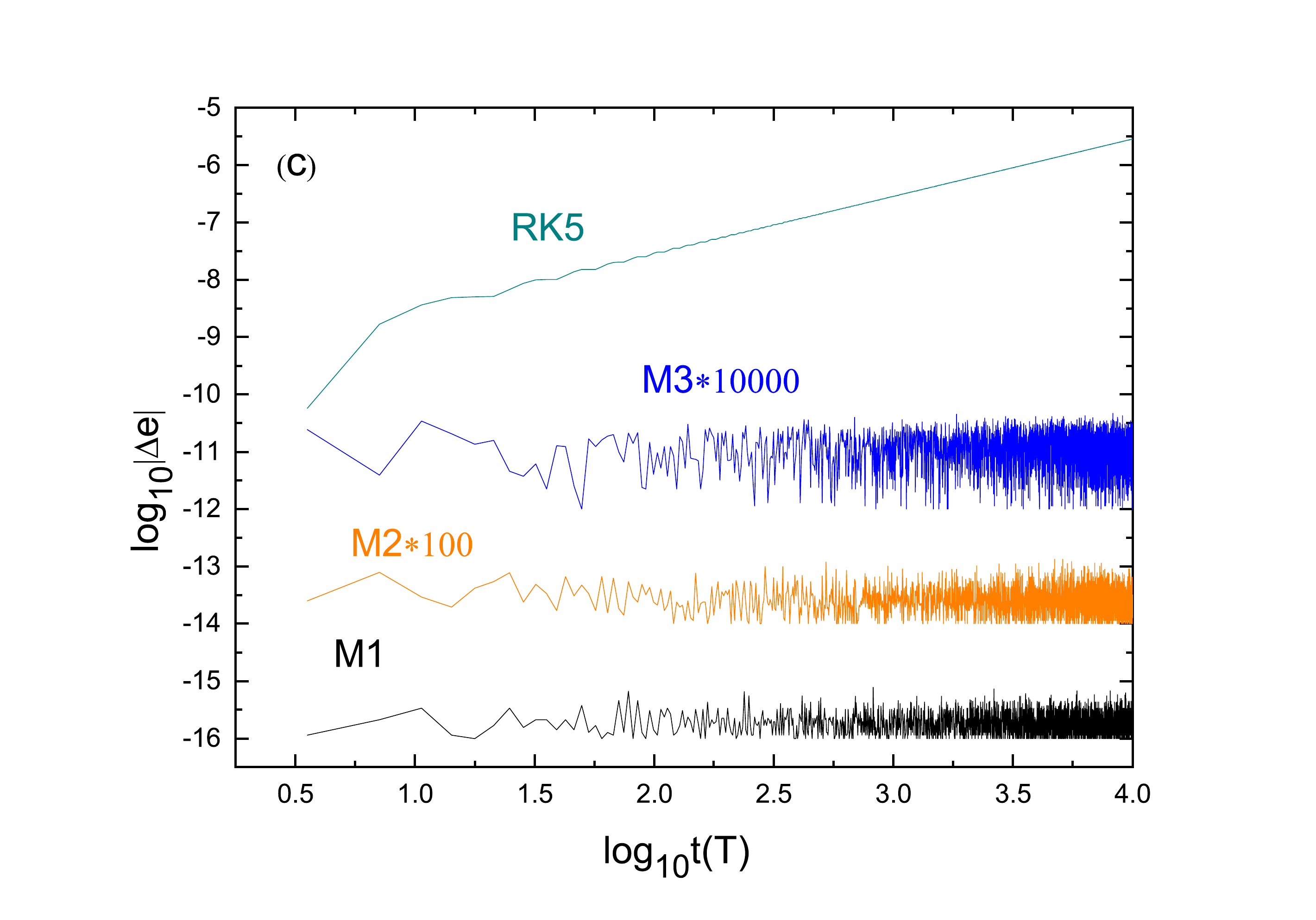}
   \includegraphics[width=0.45\textwidth]{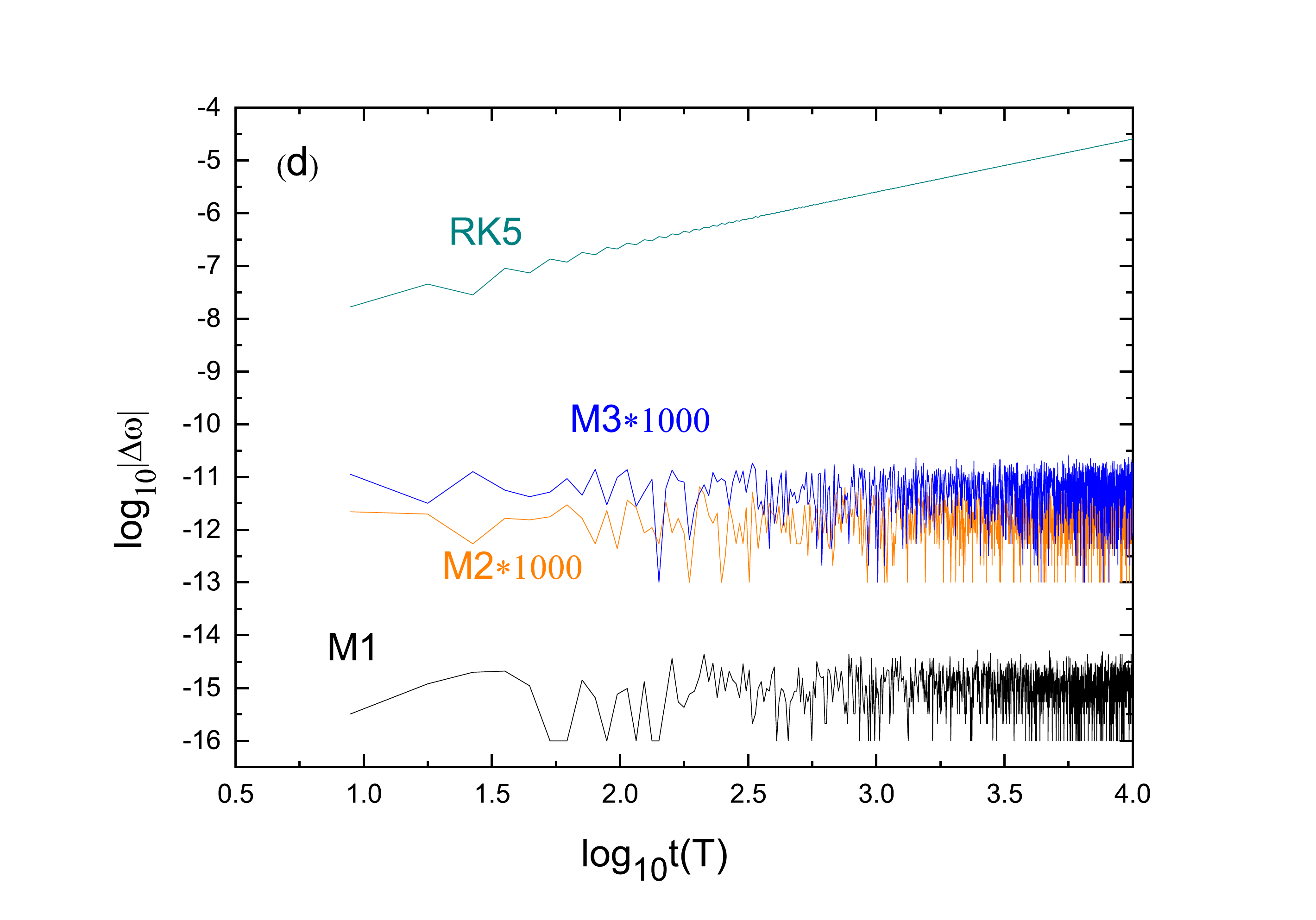}
   \includegraphics[width=0.45\textwidth]{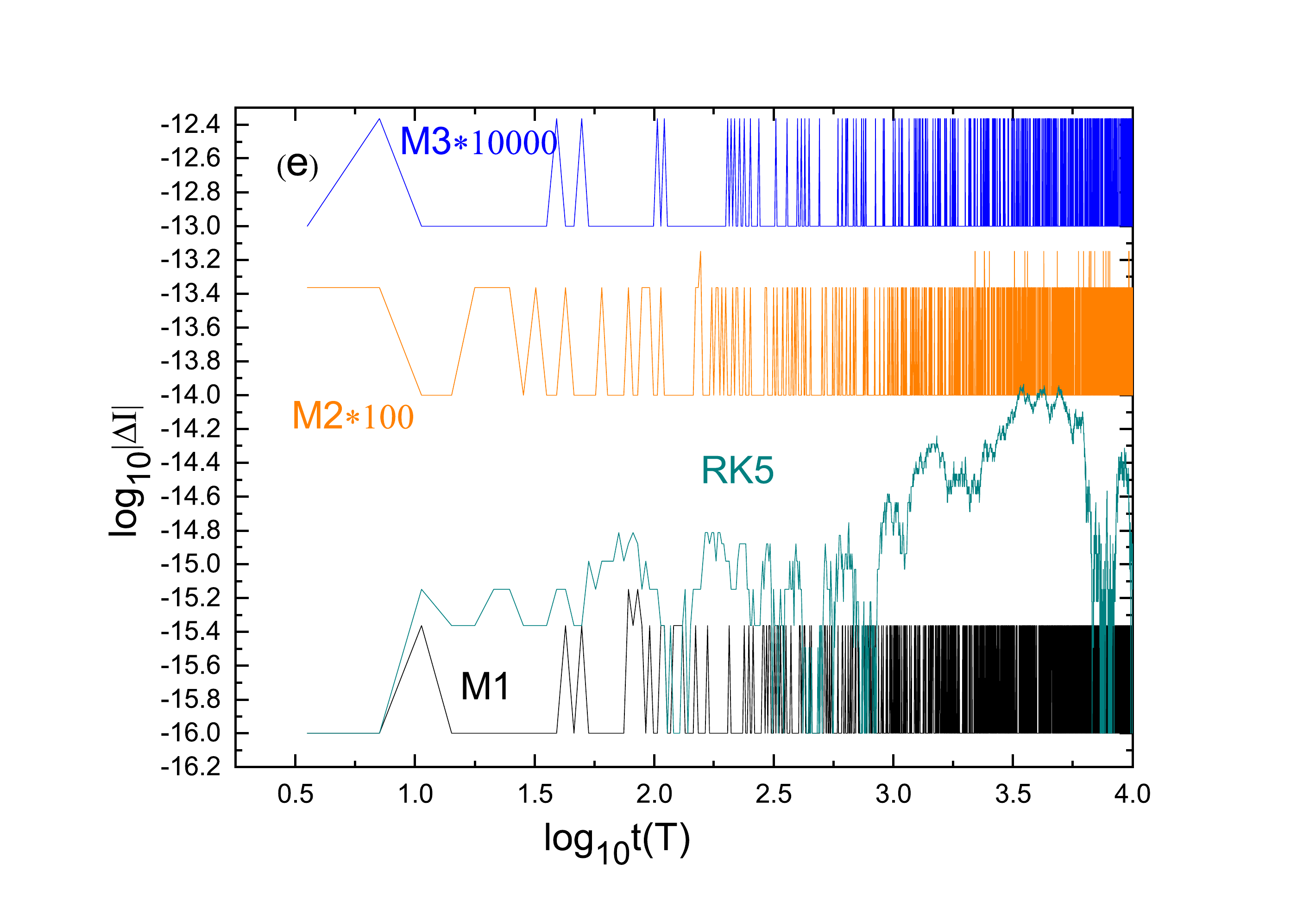}
   \includegraphics[width=0.45\textwidth]{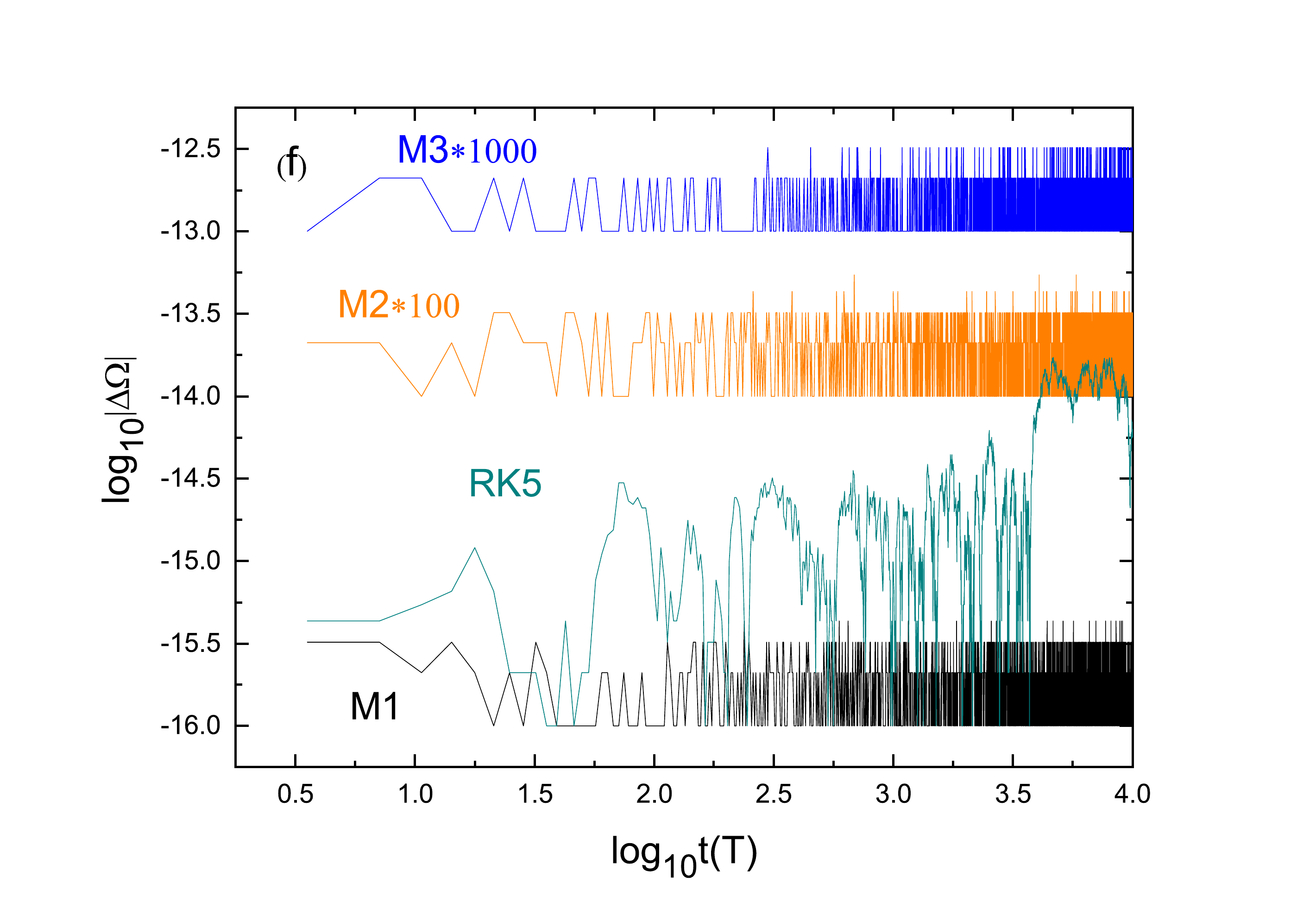}
\caption{Errors in the Keplerian elements of the pure Keplerian orbit integrated by RK5 and the manifold correction schemes M1, M2, and M3 with eccentricity $e$ = 0.3.
The units of all angle variables are radians, and the units of time are the orbital period $T$.
The step size of each method is $T/100$. The notations ``$\ast 100$, $\ast 1000$, $\ast 1000$'' mean that the errors are multiplied by a factor of 100, 1000, 10000 for M1, M2, or M3.}
\label{fig:1}
\end{figure}

 As shown in Fig.~\ref{fig:1}, the accuracies of all orbital elements are greatly improved for M1, M2, and M3, compared to those for RK5. However, M1, M2, and M3 have some differences in the corrections of individual orbital elements. In Fig.~\ref{fig:1}$a$ and~\ref{fig:1}$b$, M1, M2, and M3 have the same performance in suppressing the error of the semimajor axis $a$ to the order of the machine epsilon. However, at the end of integration, the accuracy of the mean anomaly $M$ for M3 is lower in magnitude of about two orders than that for M2, and the error in $M$ for M1 gradually approaches to that for M2. The reason is that the adjusted numerical solution accurately satisfies the Keplerian energy $K$ in M2, while approximately satisfies the equation of $K$ through the iterative method in M1. However, M3 linearly satisfies the equation of $K$, so the correction of $K$ by M3 is poorer than that by M1 and M2. In Fig.~\ref{fig:1}$c$ and~\ref{fig:1}$d$, M1 is slightly better than M2 and M3 in the correction of eccentricity $e$, and the readjustment of the argument of perihelion $\omega$ by M1 and M2 is slightly better than by M3. That is, M3 is the poorest in correcting the Laplace vector $P$. Here is an explanation for these different results. The three components of $\bm{P}$ are kept for M1, and the two related components are conserved for M2. However, only $P_z$ is preserved for M3. Finally, as shown in Figs.~\ref{fig:1}$e$--~\ref{fig:1}$f$, M1, M2, and M3 have the same effect on the errors of orbital inclination $I$ and longitude of ascending $\Omega$, and the errors almost reach the order of the machine epsilon. That means that M1, M2, and M3 have the same performances in the conservation of the angular momentum vector $L$.

\begin{figure}
  \centering
    \includegraphics[width=0.5\textwidth]{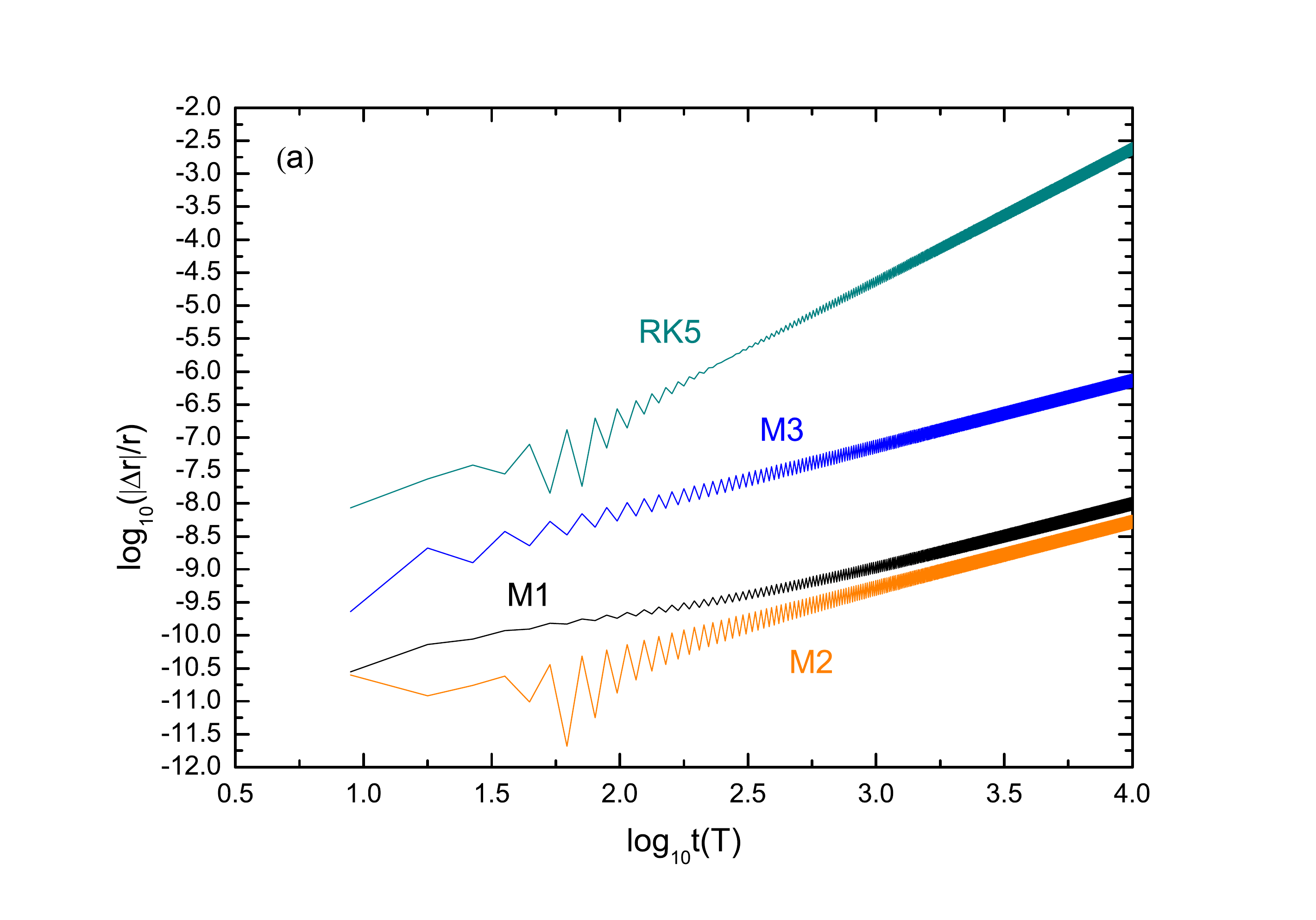}
  \caption{The errors of the relative positions for the pure Keplerian orbit.}
  \label{fig:3}
\end{figure}

Seen from the relative position errors in Fig.~\ref{fig:3}, the difference between M1 and M2 is not obvious when the integration time spans $10000T$. The methods M1 and M2 have higher accuracies in magnitude of about six orders than RK5, and in magnitude of about two orders than M3. Therefore, M1 and M2 are superior to M3 in the correction of the relative position.

\begin{figure}
  \centering
    \includegraphics[width=0.45\textwidth]{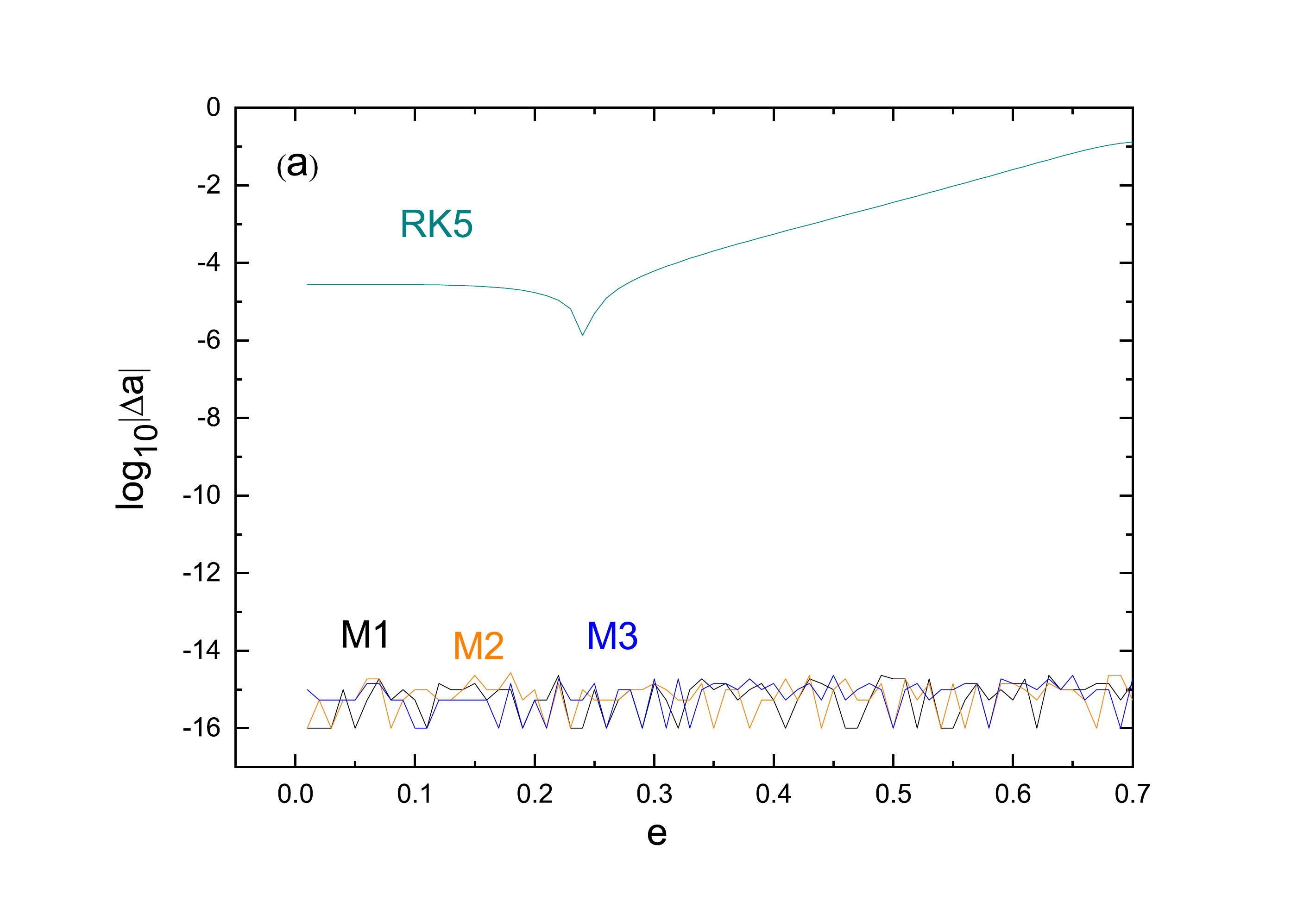}
    \includegraphics[width=0.45\textwidth]{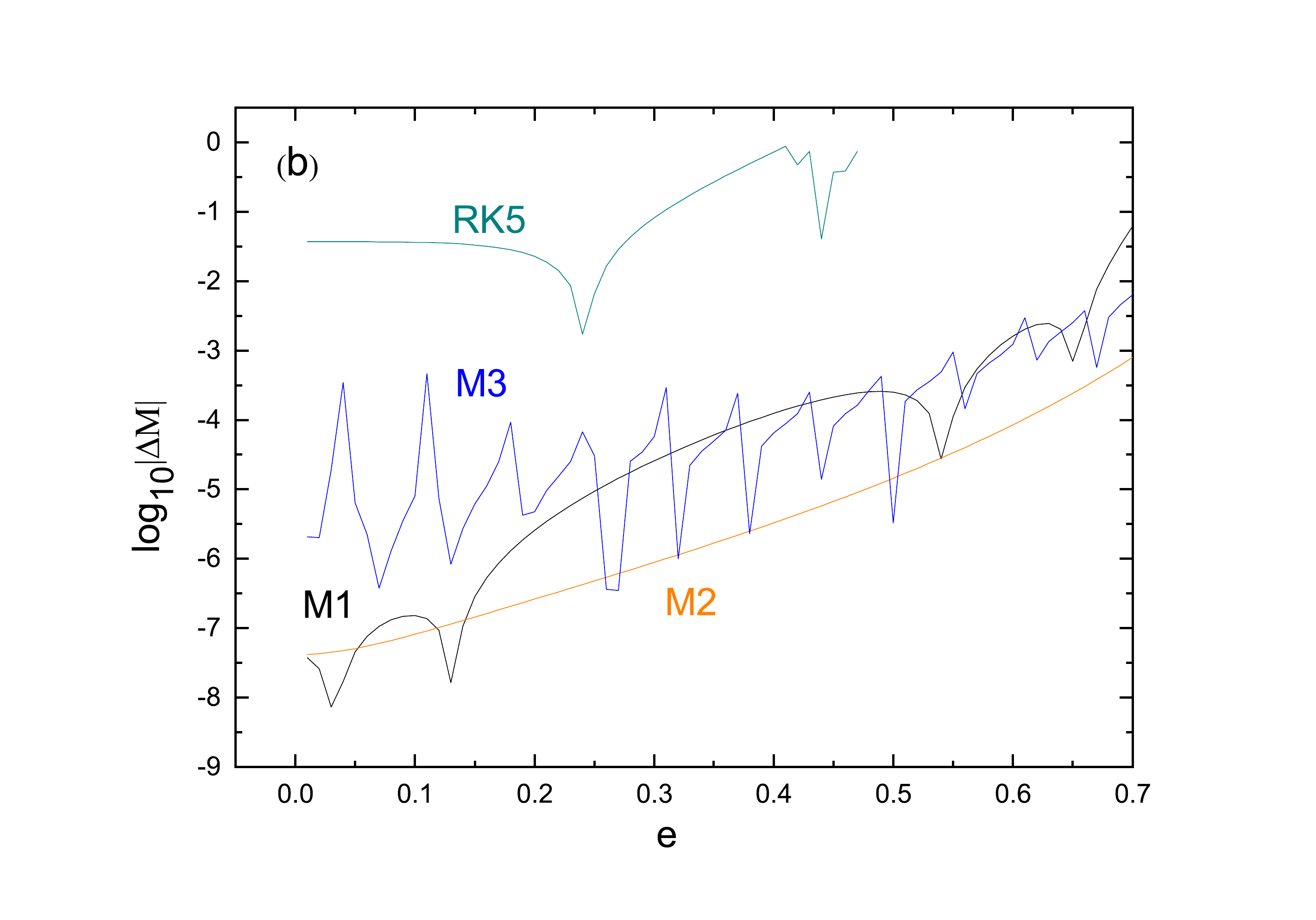}
   \includegraphics[width=0.45\textwidth]{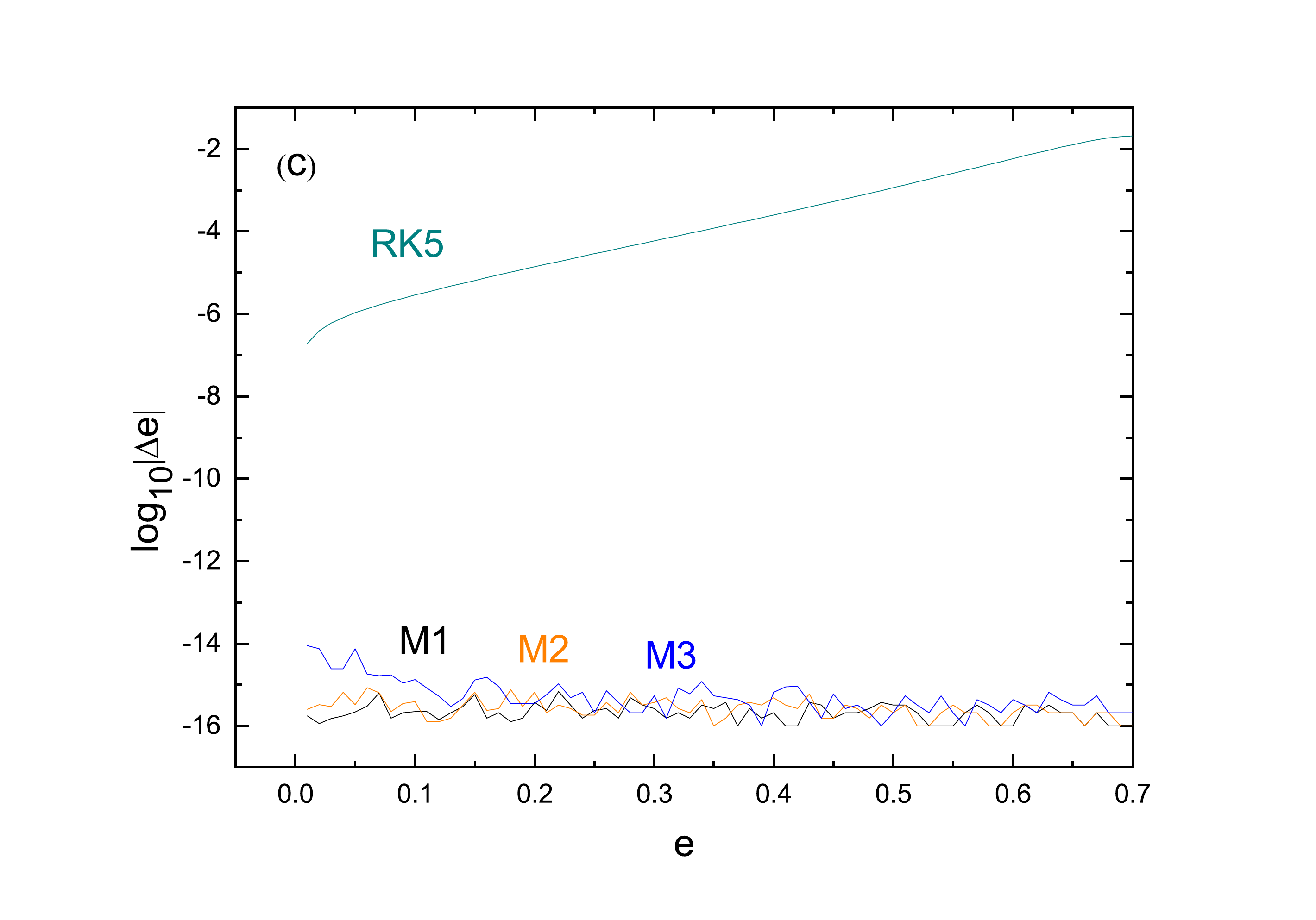}
   \includegraphics[width=0.45\textwidth]{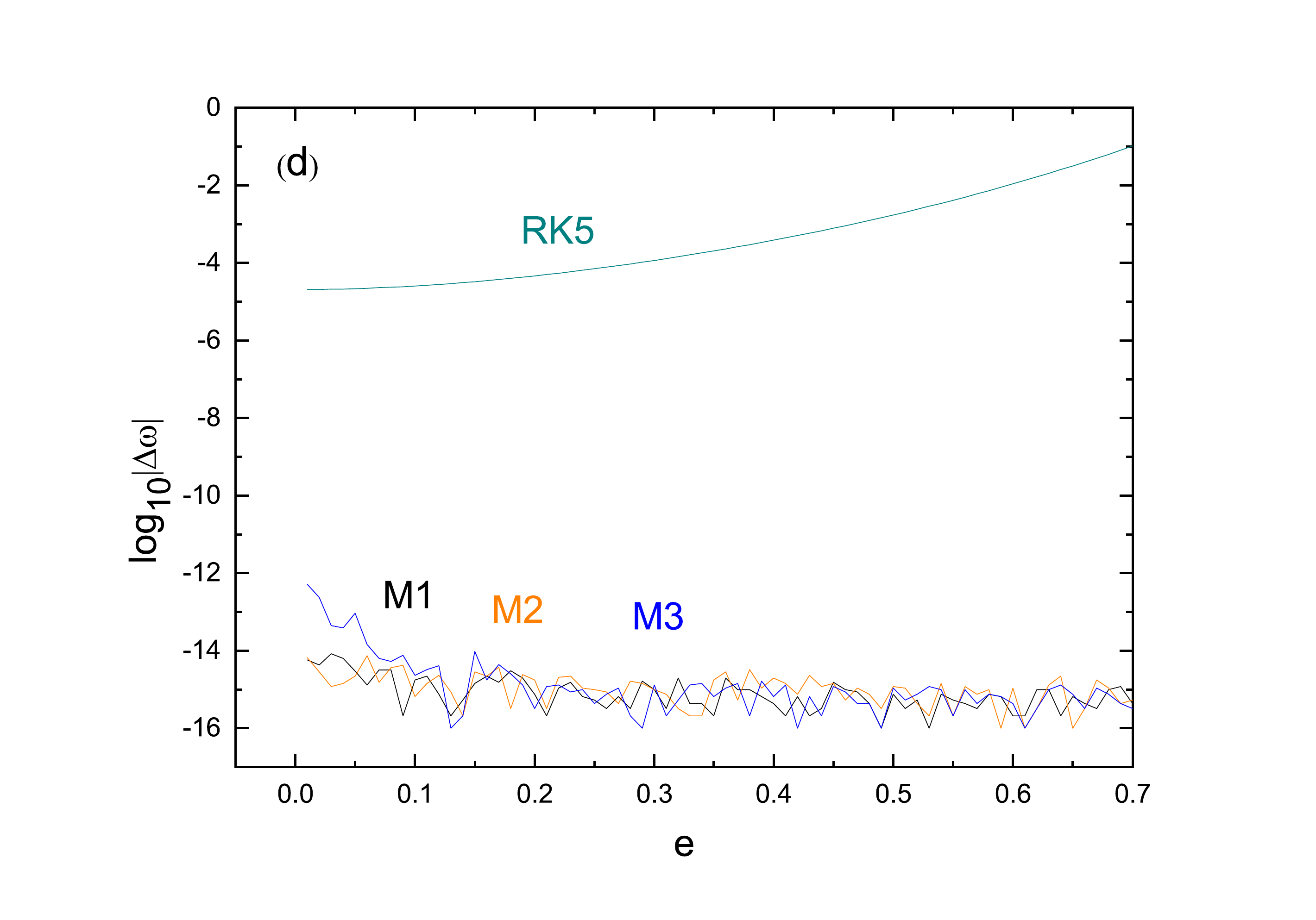}
   \includegraphics[width=0.45\textwidth]{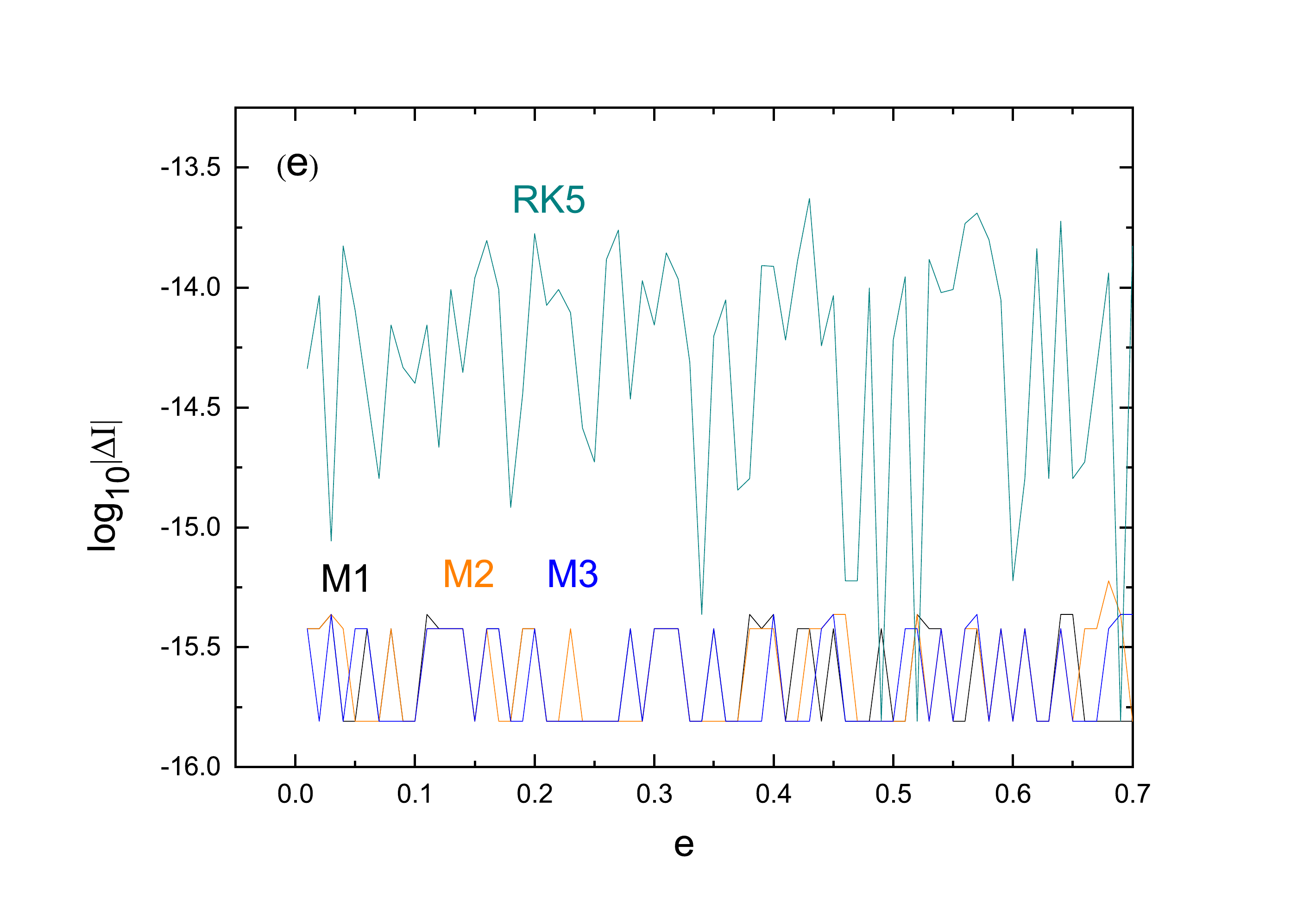}
   \includegraphics[width=0.45\textwidth]{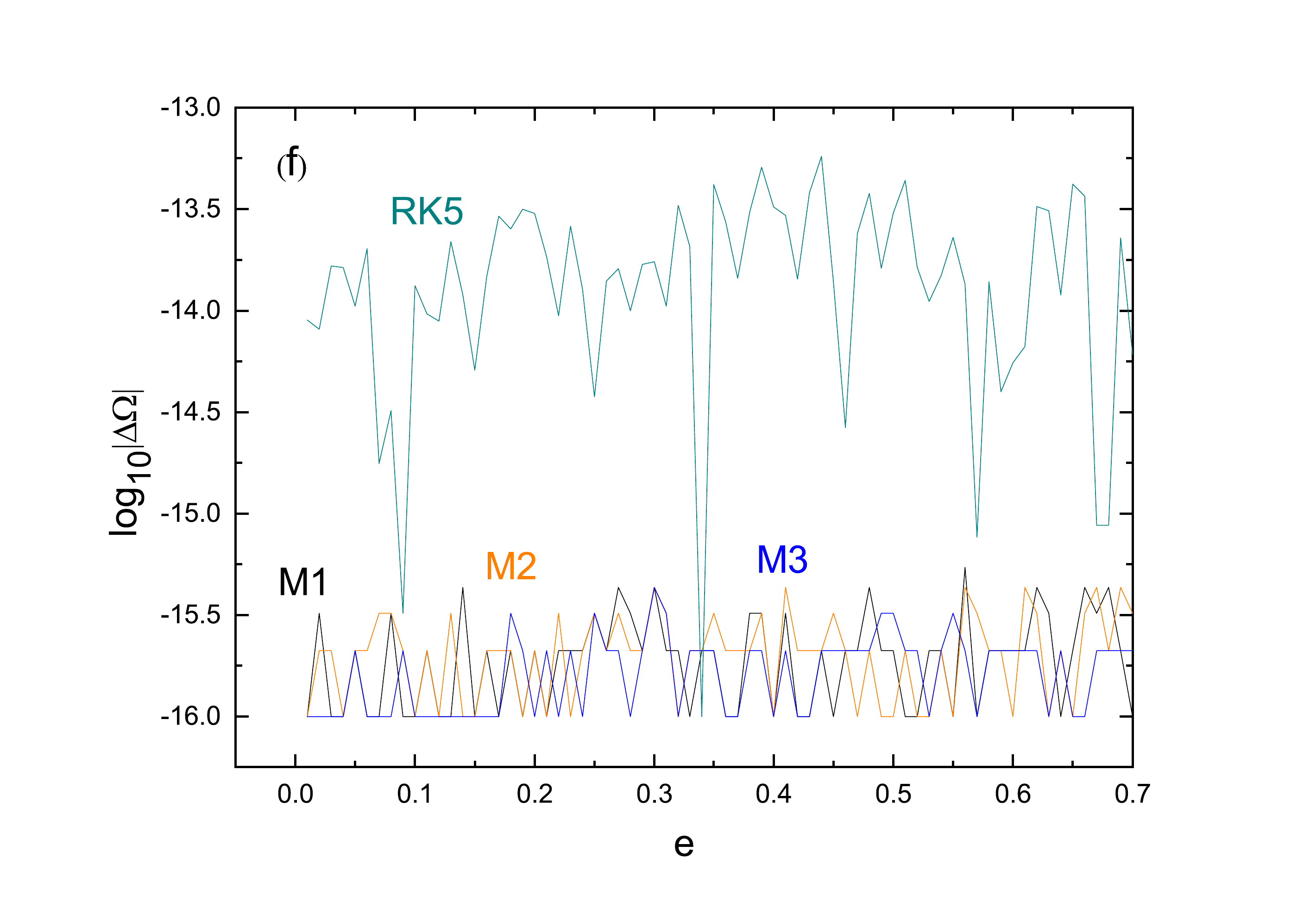}
  \caption{Dependence of the eccentricity on the integration errors in the Keplerian elements for several algorithms. The magnitude of the orbital eccentricity varies from 0.1 to 0.7 with an interval of 0.01, and each orbit is integrated till the time reaches $10000T$.}
  \label{fig:4}
\end{figure}

Next, let's consider the influence of the variation of eccentricities on the correction effectiveness. We fix initial orbital elements $a=2$, $I=23^{\circ}$, $\Omega=50^{\circ}$, $\omega=30^{\circ}$, and $M=40^{\circ}$, but let the orbital eccentricity be altered from 0.1 to 0.7 with an interval of 0.01. The integration time of each orbit is $10000T$. It can be seen from Fig.~\ref{fig:4} that for any one of the three correction methods, the errors of $a$, $I$ and $\Omega$ can remain stable with an increase of the eccentricity e, and the accuracies of $\omega$ and $e$ gradually increase, but that of M decreases.

\subsection{Discussions}
\label{sec:parameter}

\begin{table}[tp]
  \centering
  \begin{threeparttable}
  \caption{The forms of M1, M1$^{\prime}$, and M1$^{\prime\prime}$ are presented in this table. Here, $\bm{s}$ is a parameter vector. $\bm{\varepsilon}$ is the correction vector, and $\bm{\phi}$ is a set of conserved quantities.}
  \label{tab:3}
    \begin{tabular}{cccc}
    \toprule
    \textbf{Method}&$\bm{s}$&$\bm{\varepsilon}$&$\bm{\phi}$\cr
  
    \midrule
     \multirow{2}*{M1}&\multirow{2}*{$(s_1,s_2,s_3,s_4,s_5,s_6,s_7)^{\mathrm{T}}$}&$(s_1x_1,s_2y_1,s_3z_1,s_4\dot{x_1}+s_7x_1,$&$K$, $L_x$, $L_y$, $L_z$,\cr
     &&$s_5\dot{y_1}+s_7y_1,s_6\dot{z_1}+s_7z_1)^{\mathrm{T}}$& $P_x$, $P_y$, $P_z$\cr
   \midrule
        \multirow{2}*{M1$^{\prime}$}&\multirow{2}*{$(s_1^{\prime},s_2^{\prime},s_3^{\prime},s_4^{\prime},s_5^{\prime},s_6^{\prime})^{\mathrm{T}}$}&\multirow{2}*{$(s_1^{\prime}x_1,s_2^{\prime}y_1,s_3^{\prime}z_1,s_4^{\prime}\dot{x_1},s_5^{\prime}\dot{y_1},s_6^{\prime}\dot{z_1})^{\mathrm{T}}$}&$K$, $L_x$, $L_y$, $L_z$,\cr
        &&&$P_x$, $P_z$\cr
        \midrule
      \multirow{2}*{M1$^{\prime\prime}$}& \multirow{2}*{$(s_1^{\prime\prime},s_2^{\prime\prime},s_3^{\prime\prime},s_4^{\prime\prime},s_5^{\prime\prime})^{\mathrm{T}}$}&$(s_1^{\prime\prime}x_1,s_2^{\prime\prime}y_1,s_3^{\prime\prime}z_1,s_4^{\prime\prime}\dot{x_1}+s_5^{\prime\prime}x_1,$&$K$, $L_x$, $L_y$, $P_x$,\cr      
    &&$s_4^{\prime\prime}\dot{y_1}+s_5^{\prime\prime}y_1,s_4^{\prime\prime}\dot{z_1}+s_5^{\prime\prime}z_1)^{\mathrm{T}}$& $P_z$\cr
    \bottomrule
    \end{tabular}
    \end{threeparttable}
\end{table}

\begin{figure}
\centering
    \includegraphics[width=0.45\textwidth]{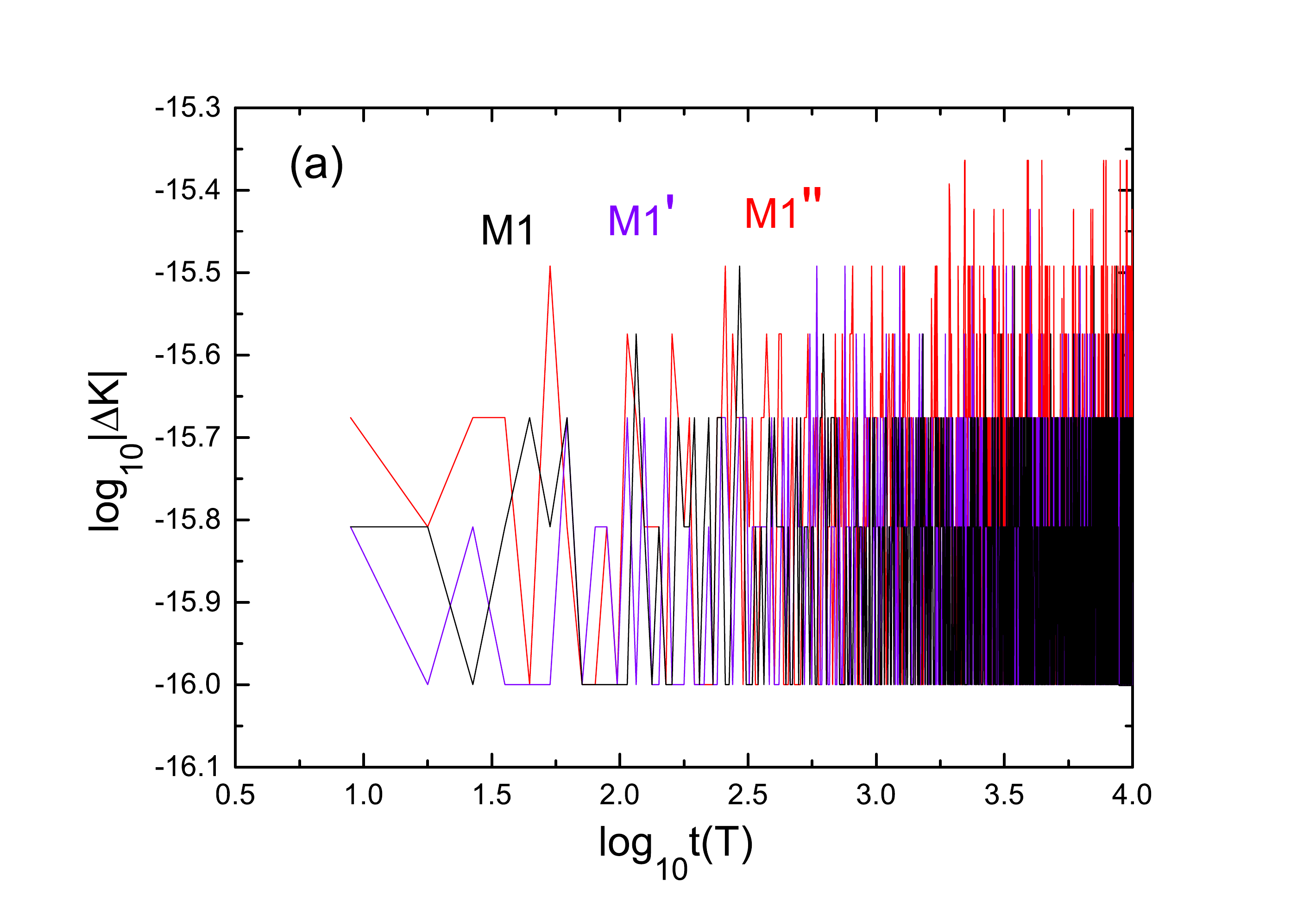}
    \includegraphics[width=0.45\textwidth]{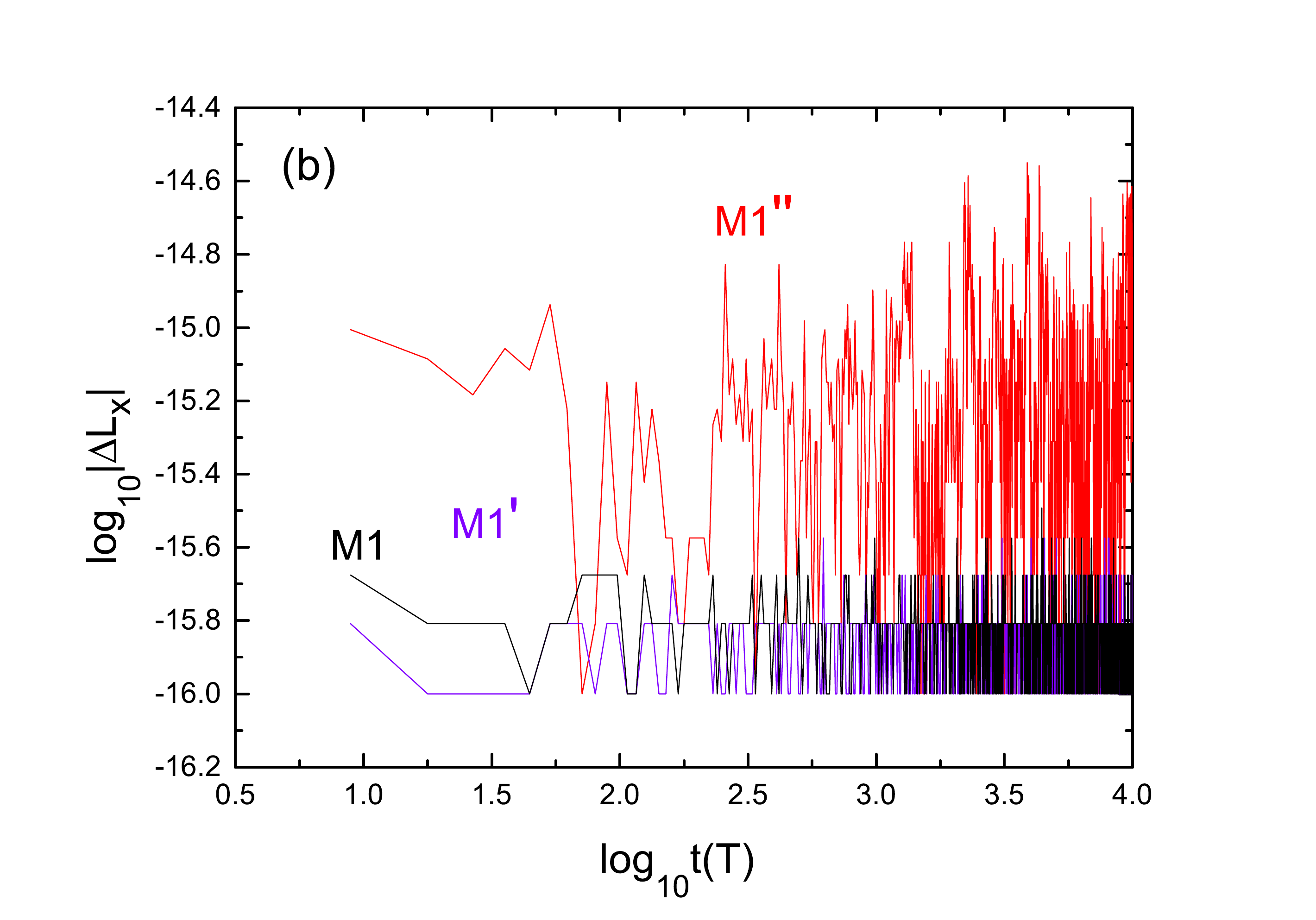}
   \includegraphics[width=0.45\textwidth]{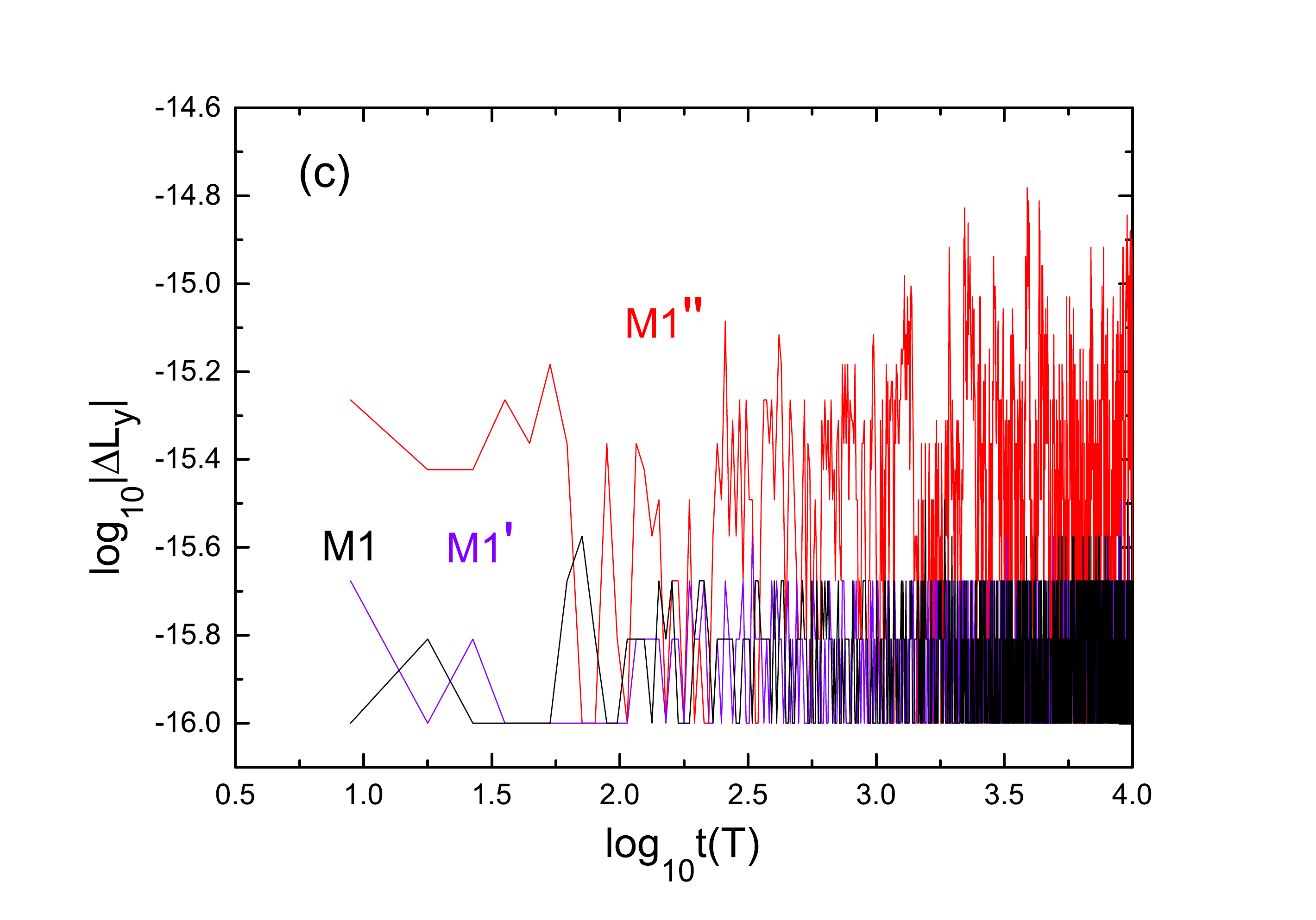}
   \includegraphics[width=0.45\textwidth]{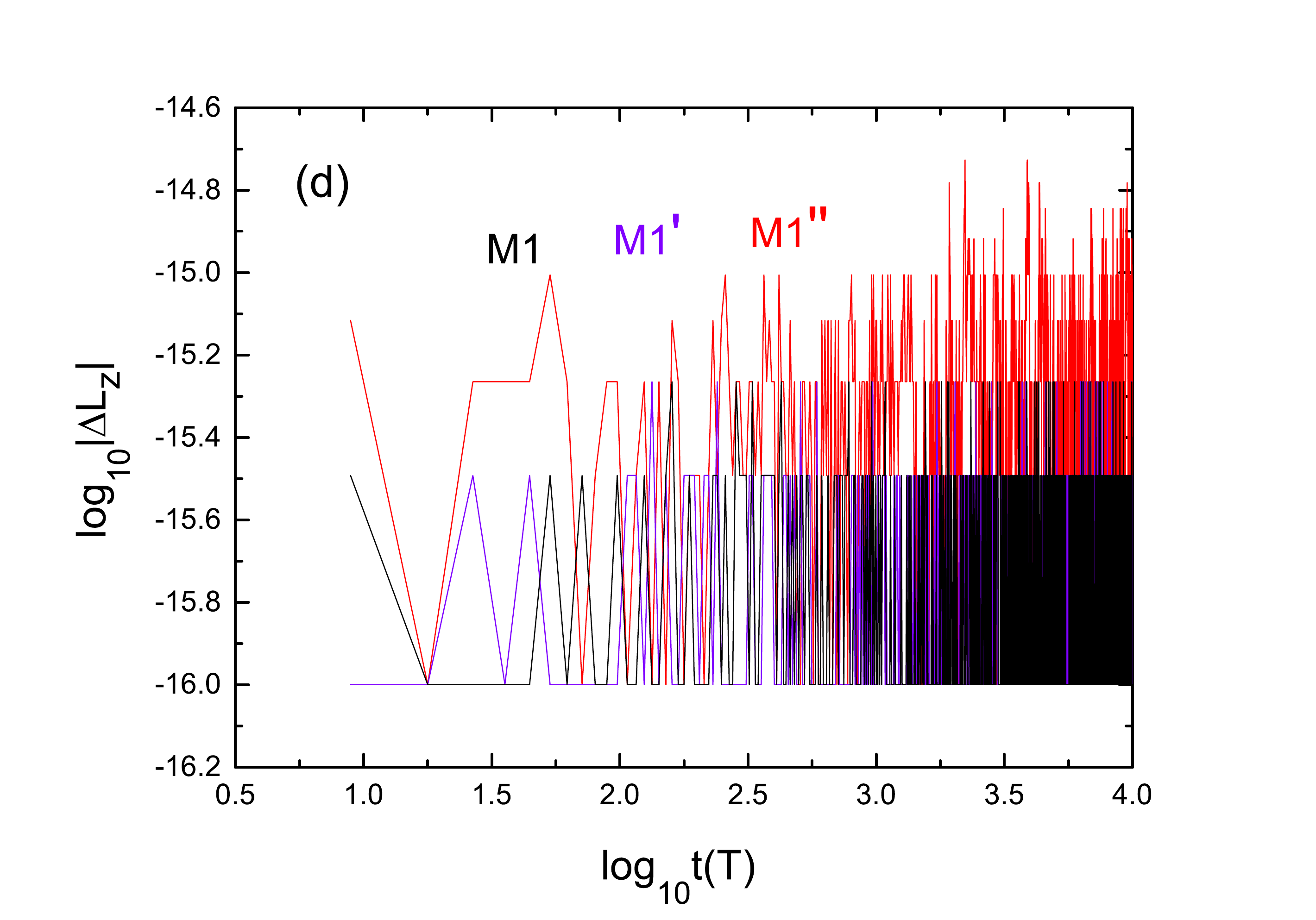}
   \includegraphics[width=0.45\textwidth]{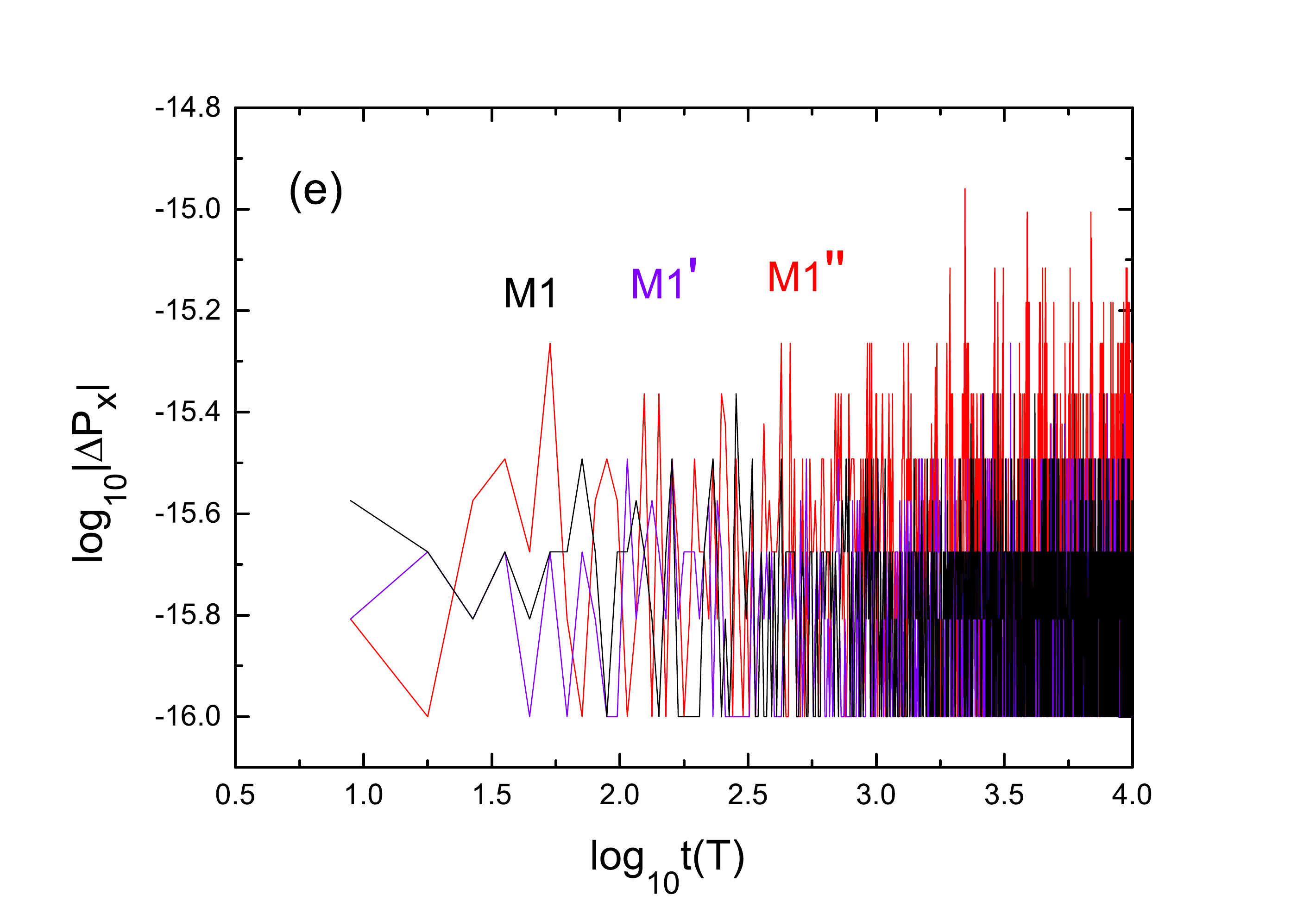}
   \includegraphics[width=0.45\textwidth]{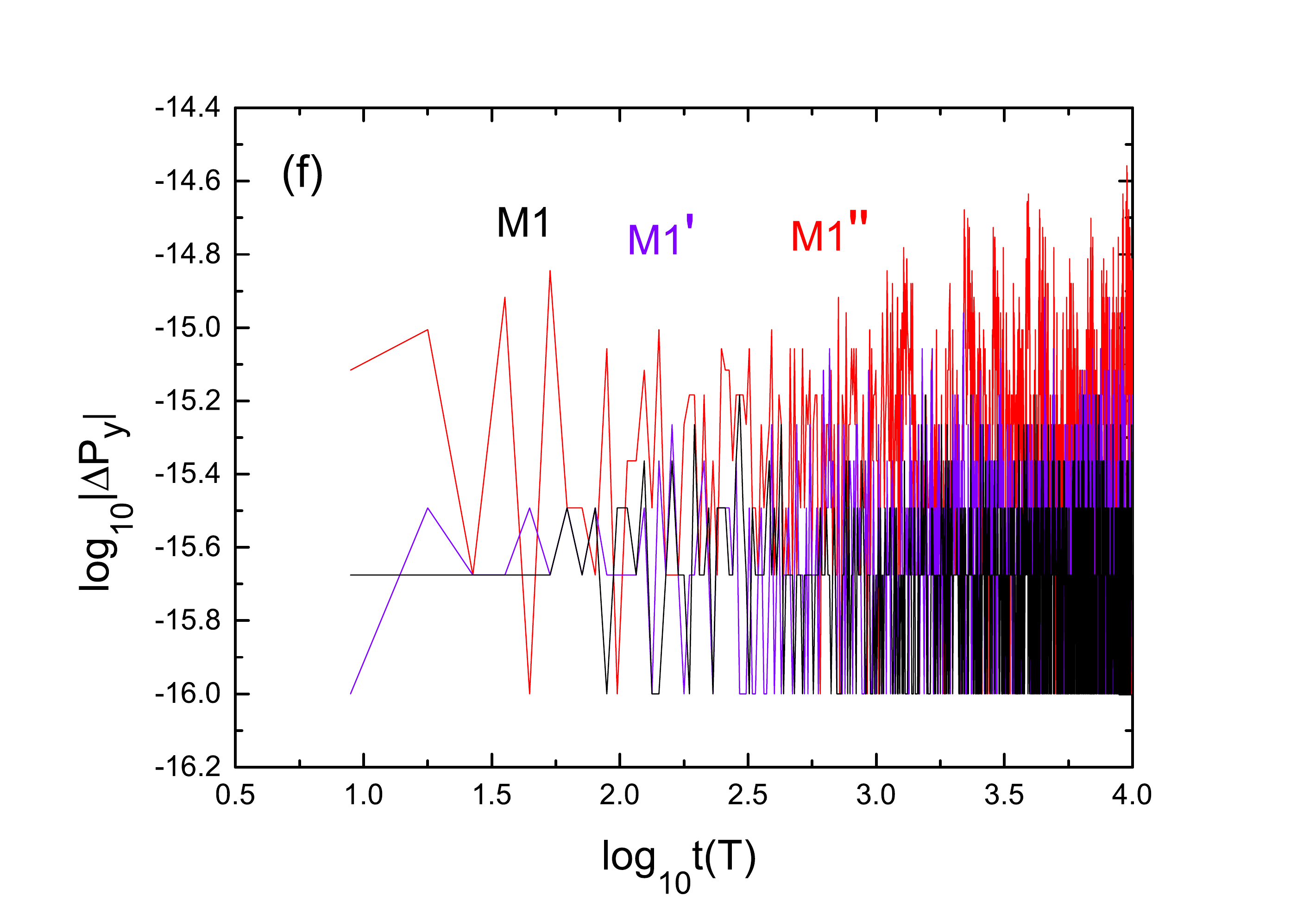}
   \includegraphics[width=0.45\textwidth]{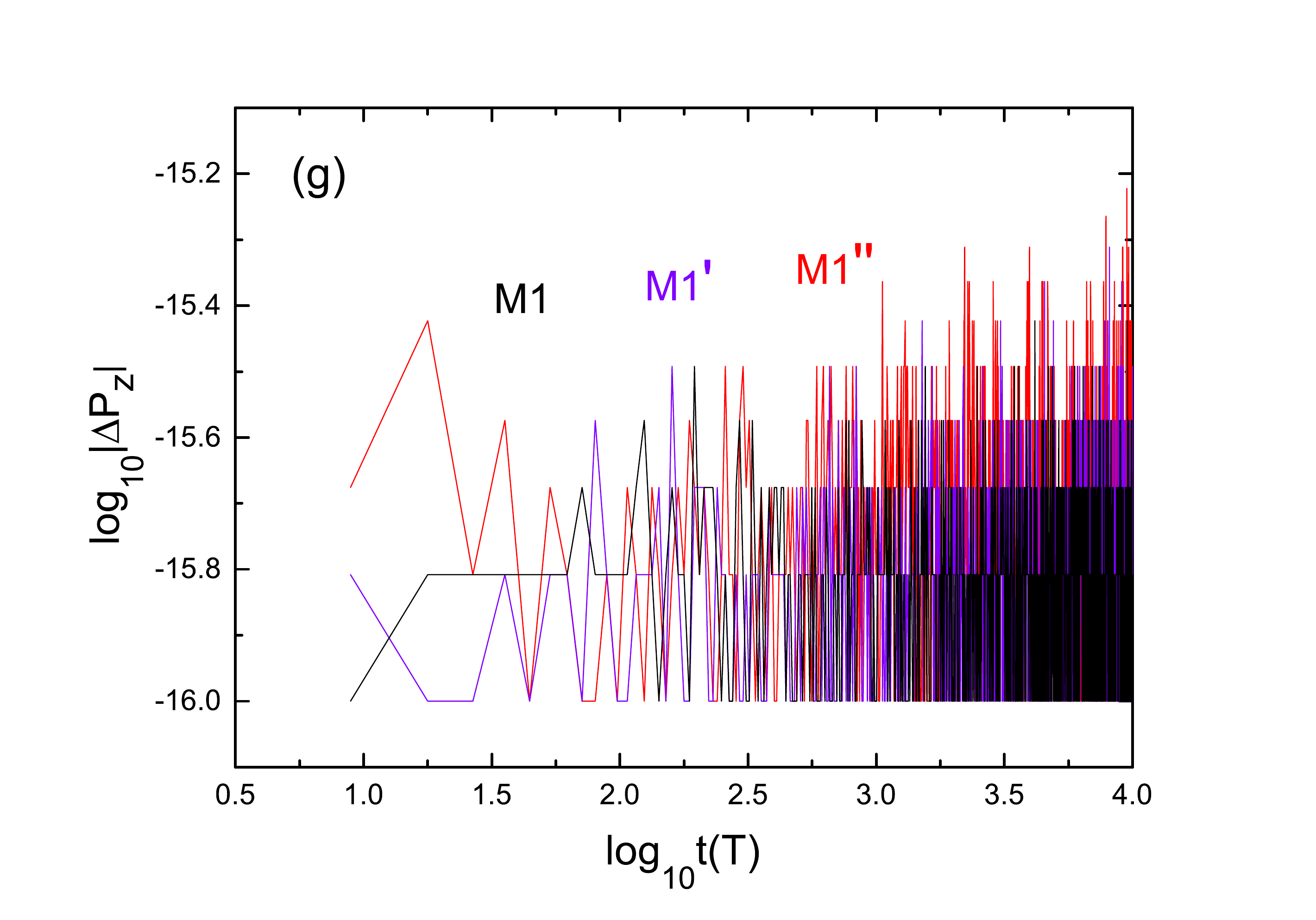}
\caption{The errors in the conserved quantities for a pure Keplerian orbit with eccentricity e=0.1, given by RK5 and its correction M1$^{\prime}$, M1$^{\prime\prime}$, and M1.}
\label{fig:2}
\end{figure}

\begin{table}[tp]

  \centering
  \begin{threeparttable}
  \caption{The singular values of the equations are listed from the largest to the smallest when decomposing the linear equations with SVD at some times.}
  \label{tab:1}
    \begin{tabular}{cccccc}
    \toprule
    \multicolumn{6}{c}{\textbf{The singular values}}\cr
    \midrule
    \multirow{2}{*}{\textbf{Method}}&
    \multicolumn{5}{c}{time (T)}\cr
    \cmidrule(lr){2-6}
    &9&3252&5927&8308&1000\cr
    \midrule
      \multirow{7}*{M1}&3.12201220558961&2.78181812405576&3.12201221280423&3.12201221566548&2.78181816346293\cr
        &1.94250895238320&1.86827315178044&1.94250885182877&1.94250881212345&1.86827303596444\cr
        &0.707305620343333&0.754983459415598&0.707305622708845&0.707305623643913&0.754983466482571\cr
        &0.481660111360378&0.464906262903175&0.481660102613845&0.481660099160655&0.464906230152241\cr
        &0.288114413481869&0.346893992951316&0.288114423914503&0.288114428031266&0.346894025161191\cr
        &0.000000000000000&0.000000000000000&0.000000000000000&0.000000000000000&0.000000000000000\cr
        &0.000000000000000&0.000000000000000&0.000000000000000&0.000000000000000&0.000000000000000\cr
        \midrule
        \multirow{6}*{M1$^{\prime}$}   &1.90585641255004&1.82513228900990&1.90586003470393&1.90586149214838&1.82513644377474\cr
        &0.623384135039900&0.760341114341583&0.623383418473375&0.623383130147222&0.760338527764295\cr
        &0.580340314386519&0.665518084870883&0.580337478485576&0.580336337386044&0.665518161683883\cr
        &0.300543512542193&0.304760732845583&0.300543973939492&0.300544159591971&0.304760891157126\cr
        &0.100451312700609&0.119914367532845&0.100451257030468&0.100451234632653&0.119914324278903\cr
        &0.000000000000000&0.000000000000000&0.000000000000000&0.000000000000000&0.000000000000000\cr
      \midrule
     \multirow{5}*{M1$^{\prime\prime}$}&2.85428942302834&2.19466753168822&2.85428935167753&2.85428932374321&2.19466741781362\cr
        &1.23576224456404&1.48642223104468&1.23576221513671&1.23576220357739&1.48642219737270\cr
        &0.594403383793914&0.644938698109366&0.594403400703252&0.594403407327637&0.644938708695499\cr
        &0.134194567655502&0.110528910893960&0.134194567996513&0.134194568130963&0.110528910340996\cr
        &0.000000000000000&0.000000000000000&0.000000000000000&0.000000000000000&0.000000000000000\cr
    \bottomrule
    \end{tabular}
    \end{threeparttable}
\end{table}

An appropriate choice of the conserved quantities and the modified vector is very important. As stated in Eq.~(\ref{eq:611}), there are seven conserved quantities, but only five of them are completely independent.
In the theory, if the five independent integrals are kept well, the two other dependent integrals are, too. However, they may not be from a numerical viewpoint. This is why the conservation of the seven dependent and independent integrals, called the method M1, is considered. Besides M1, the conservation of six dependent and independent integrals (called method M1$^{\prime}$) and that of five independent integrals (called method M1$^{\prime\prime}$) are also considered. The details of M1, M1$^{\prime}$, and M1$^{\prime\prime}$ are listed in table~\ref{tab:3}. For M1$^{\prime}$, we use six parameters $\bm{s}^{\prime} = (s_1^{\prime},s_2^{\prime}...,s_6^{\prime})^{\mathrm{T}}$ to construct new correction vector $\bm{\varepsilon}^{\prime}(\bm{s}^{\prime})=(s_1^{\prime}x_1,s_2^{\prime}y_1,s_3^{\prime}z_1,s_4^{\prime}\dot{x_1},s_5^{\prime}\dot{y_1},s_6^{\prime}\dot{z_1})^{\mathrm{T}}$, and obtain the corrected solution $\bm{x}^{\ast}=\bm{x}_1+\bm{\varepsilon}^{\prime}(\bm{s}^{\prime})$
to satisfy the six conserved quantities $K$, $L_x$, $L_y$, $L_z$, $P_x$, and $P_z$. Then, a problem is how to solve such a set of nonlinear equations about $s_i^{\prime} (i=1,2...,6)$. The iterative method described in Appendix~\ref{sec:iteration} is still used. In this way, the readjusted solution is obtained. For M1$^{\prime\prime}$, the modified solutions $\bm{x}^{\ast}=\bm{x}_1+\bm{\varepsilon}^{\prime\prime}(\bm{s}^{\prime\prime})$ satisfy the  five integrals $K$, $L_x$, $L_y$, $P_x$, and $P_z$. Here, $\bm{s}^{\prime\prime}= (s_1^{\prime\prime},s_2^{\prime\prime},...,s_5^{\prime\prime})^{\mathrm{T}}$ is the parameter vector and $\bm{\varepsilon}^{\prime\prime}(\bm{s}^{\prime\prime})=(s_1^{\prime\prime}x_1,s_2^{\prime\prime}y_1,s_3^{\prime\prime}z_1,s_4^{\prime\prime}\dot{x_1}+s_5^{\prime\prime}x_1,s_4^{\prime\prime}\dot{y_1}+s_5^{\prime\prime}y_1,s_4^{\prime\prime}\dot{z_1}+s_5^{\prime\prime}z_1)^{\mathrm{T}}$ is the corresponding new correction vector. Similarly, the corrected solution is obtained by the iterative method solving a five-dimensional nonlinear system. Now, let us estimate the correction effectiveness of the three methods.

The above Kepler problem is still used to check the numerical performance of M1$^{\prime}$, M1$^{\prime\prime}$, and M1. The initial conditions and the basic numerical integrator are the same as those in section~\ref{sec:experimental}.
 Fig.~\ref{fig:2} shows the growth of the errors in the conserved quantities for an orbit with eccentricity e=0.1.
First, M1$^{\prime}$ and M1 achieve almost the same good effectiveness in controlling the errors of all the conserved quantities in Fig.~\ref{fig:2}. The error $\triangle P_y$ is slightly larger in magnitude of 0.1$\sim$0.2 orders for M1$^{\prime}$ than for M1 in Fig.~\ref{fig:2}$f$. That is to say, although $P_y$ is not directly contained in M1$^{\prime}$, it can be auto-corrected to a large degree. However, the higher-precision results are obtained in this case can not be guaranteed in any other cases. However, they are always ensured for M1, and the cost of additional computation is negligible. Therefore, M1 is a prior choice. By comparing M1$^{\prime\prime}$ and M1, we find that the accuracies of M1$^{\prime\prime}$ are lower in magnitude of about one order than those of M1 in the correction of $L_x$ and $L_y$. In addition, M1$^{\prime\prime}$ is slightly poorer than M1 for the corrections of $L_z$, $P_x$, and $P_y$. In fact, five integrals are not well maintained by M1$^{\prime\prime}$. 
To clearly show this, we list the singular values of the three methods when the linear equations are decomposed by SVD at some times in Table~\ref{tab:1}. Five singular values of the  equations are nonzero in M1 and M1$^{\prime}$. Equivalently, the corresponding five integrals can be maintained well. 
However, one of the five singular values is zero in M1$^{\prime\prime}$. This implies that only four of the five integrals are validly preserved in the calculation. It is obvious that M1 is superior to M1$^{\prime\prime}$. These results are consistent with those in Fig.~\ref{fig:2}. As a consequence, the selection of the correction vector $\bm{\varepsilon}(\bm{s})$ in M1 is appropriate.
\section{Extension to quasi-Keplerian systems}

In the following, our new scheme is extended to quasi-Keplerian orbits. Here, the quasi-Keplerian orbits represent the Kepler orbits affected by small perturbations.

\subsection{The perturbed two-body system}

For a perturbed two-body problem, the relative motion is controlled by
\begin{equation}
\frac{d\bm{v}}{dt} = -\left(\frac{\mu}{r^3}\right)\bm{r}+\bm{a}.
\label{eq:511}
\end{equation}
Here $\bm{a}$ is a perturbing acceleration.

It should be noted that $K$, $\bm{P}$, and $\bm{L}$ are no longer integral constants and become slowly-varying quantities in this system. Like those in section~\ref{subsec:construction}, the seven slowly-varying quantities can be written as $\phi_i(t,\bm{x}) \equiv c_i(t)(i=1,2...7)$, where $c_i(t)$ is a set of slowly varying quantities with time. The integral-invariant relations of $K$, $\bm{P}$, and $\bm{L}$ were given in \citep{Fukushima2004}, by,
\begin{equation}\label{eq:8}
\begin{split}
\frac{dK}{dt}=&\bm{v}\cdot\bm{a},\quad \frac{d\bm{L}}{dt}=\bm{r}\times\bm{a},\\ \frac{d\bm{P}}{dt}=&2(\bm{a}\cdot\bm{v})\bm{r}-(\bm{r}\cdot\bm{a})\bm{v}-(\bm{r}\cdot\bm{v})\bm{a}.
\end{split}
\end{equation}
The right-hand sides of Eqs.~(\ref{eq:8}) are usually small quantities, so we use $\Delta K=K-K_0$, $\Delta\bm{L}=\bm{L}-\bm{L}_0$, $\Delta\bm{P}=\bm{P}-\bm{P}_0$ instead of $K$, $\bm{L}$, $\bm{P}$ at the left-hand sides of the equations so as to reduce round-off errors. $K_0$, $\bm{L}_0$, and $\bm{P}_0$ are the initial values of $K$, $\bm{L}$, and $\bm{P}$ which are respectively given by the initial positions and velocities. It has been reported that the values of $K$, $\bm{L}$, and $\bm{P}$ obtained by simultaneously integrating Eqs.~(\ref{eq:511}) and~(\ref{eq:8}) are more precise than those given by substituting the numerical solution $(\bm{r}_1, \bm{v}_1 )$ into Eq.~(\ref{eq:111}) and Eqs.~(\ref{eq:6}) \citep{huang1983,Mikkola2002}. Therefore, the reference values of the slowly-varying quantities are given by the integral-invariant relations~(\ref{eq:8}). For the perturbed two-body problem, the calculations are the same as those in the pure Keplerian problem, but the difference lies in that the conserved quantities for the latter are replaced by the slowly-varying quantities from the integral-invariant relations for the former.

Each body in a multi-body problem is a perturbed two-body problem. Similarly, it has the equations of motion like Eq.~(\ref{eq:511}) and the evolution equations of the slowly-varying quantities like Eq.~(\ref{eq:8}). Therefore, the correction method of the solution of the perturbed two-body problem is also suitable for that of each body of the multi-body problem.

\subsection{The inner solar system}
To compare the effects of M1, M2, and M3 in a multi-body system, we take the inner solar system composed of Sun, Mercury, Venus, Earth, and Mars (here, ``Earth'' refers to the Earth-Moon barycenter) as an example of the multi-body problems. In a heliocentric frame, each planet is viewed as a point mass $m_i$ with position $\bm{r_i}$. The Newtonian equation of motion of each planet writes 
\begin{equation}
\frac{d^2\bm{r_i}}{dt^2} = -\frac{G(M_{\odot}+m_i)}{r_{1}^3}\bm{r_i}+\bm{a_i}\quad(i=1,2,...,4).
\end{equation}
\begin{equation}
\bm{a_i} = \sum_{j=1,j\not=i}^{4}\frac{Gm_j}{|\bm{r_i}-\bm{r_j}|^3}(\bm{r_j}-\bm{r_i})-\sum_{j=1,j\not=i}^{4}\frac{Gm_j}{\bm{r_j}^3}\bm{r_j},
\end{equation}
where $\bm{a}_i$ is the perturbed acceleration of each planet.

\begin{figure}
  \centering
    \includegraphics[width=0.45\textwidth]{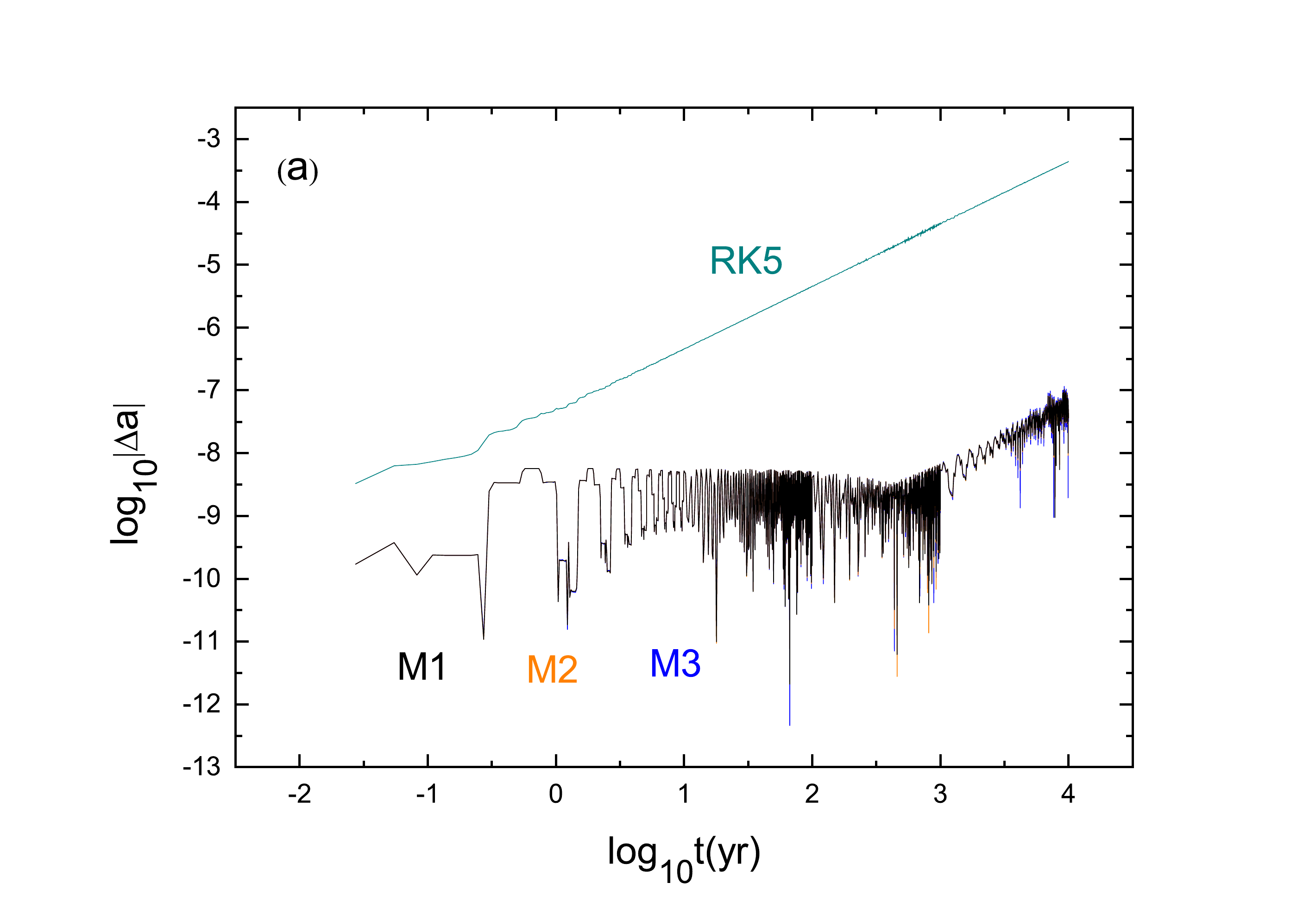}
    \includegraphics[width=0.45\textwidth]{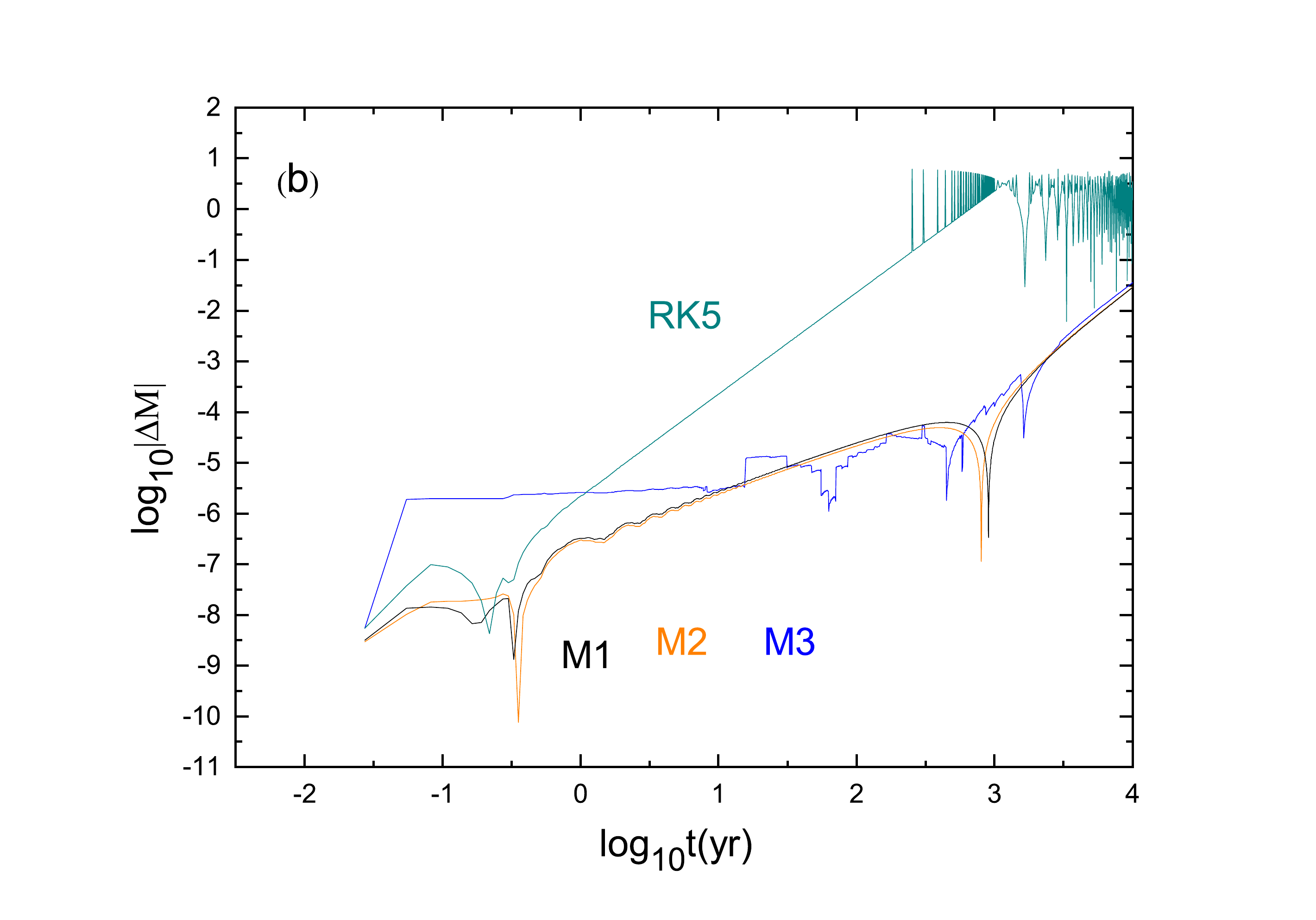}
   \includegraphics[width=0.45\textwidth]{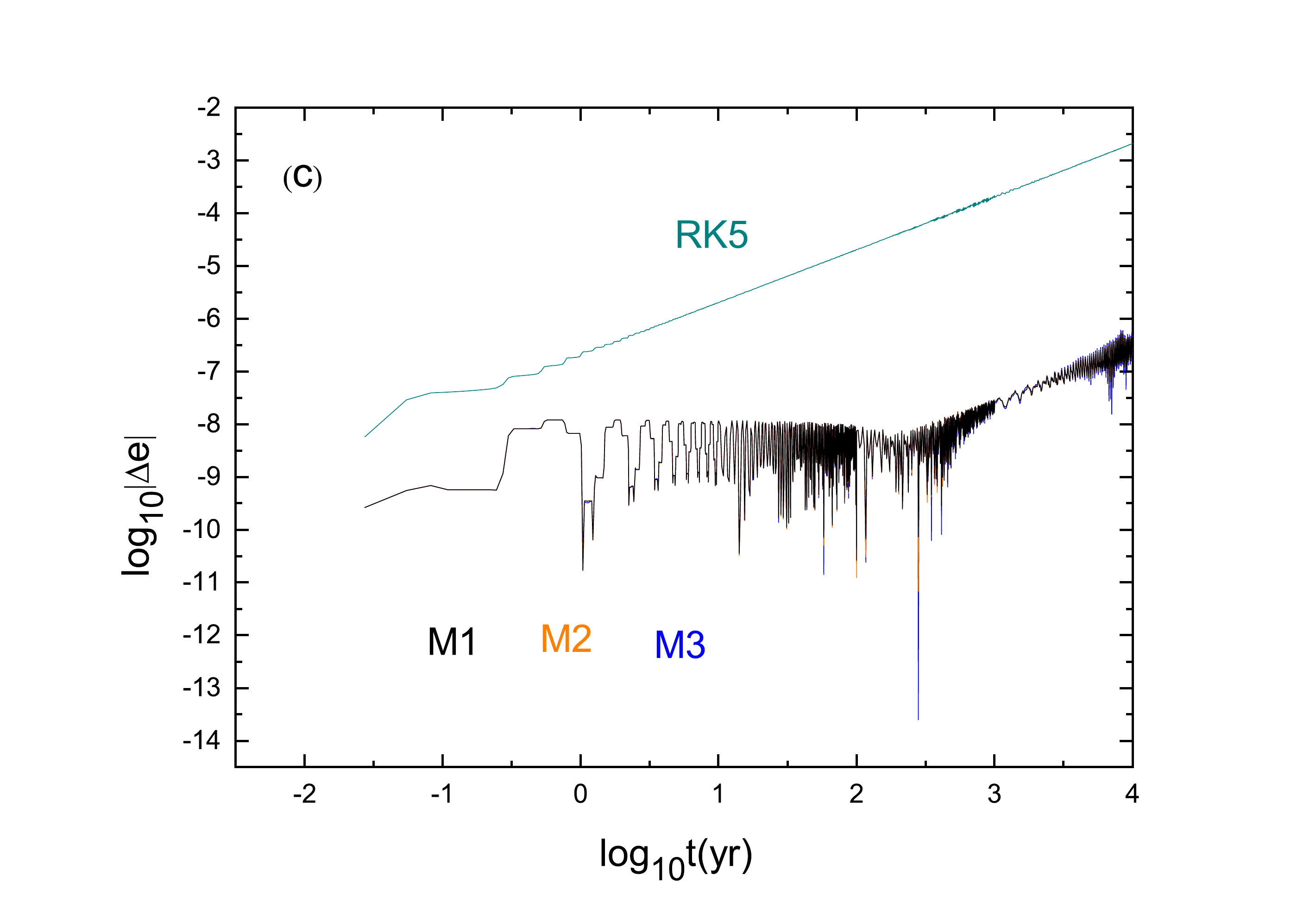}
   \includegraphics[width=0.45\textwidth]{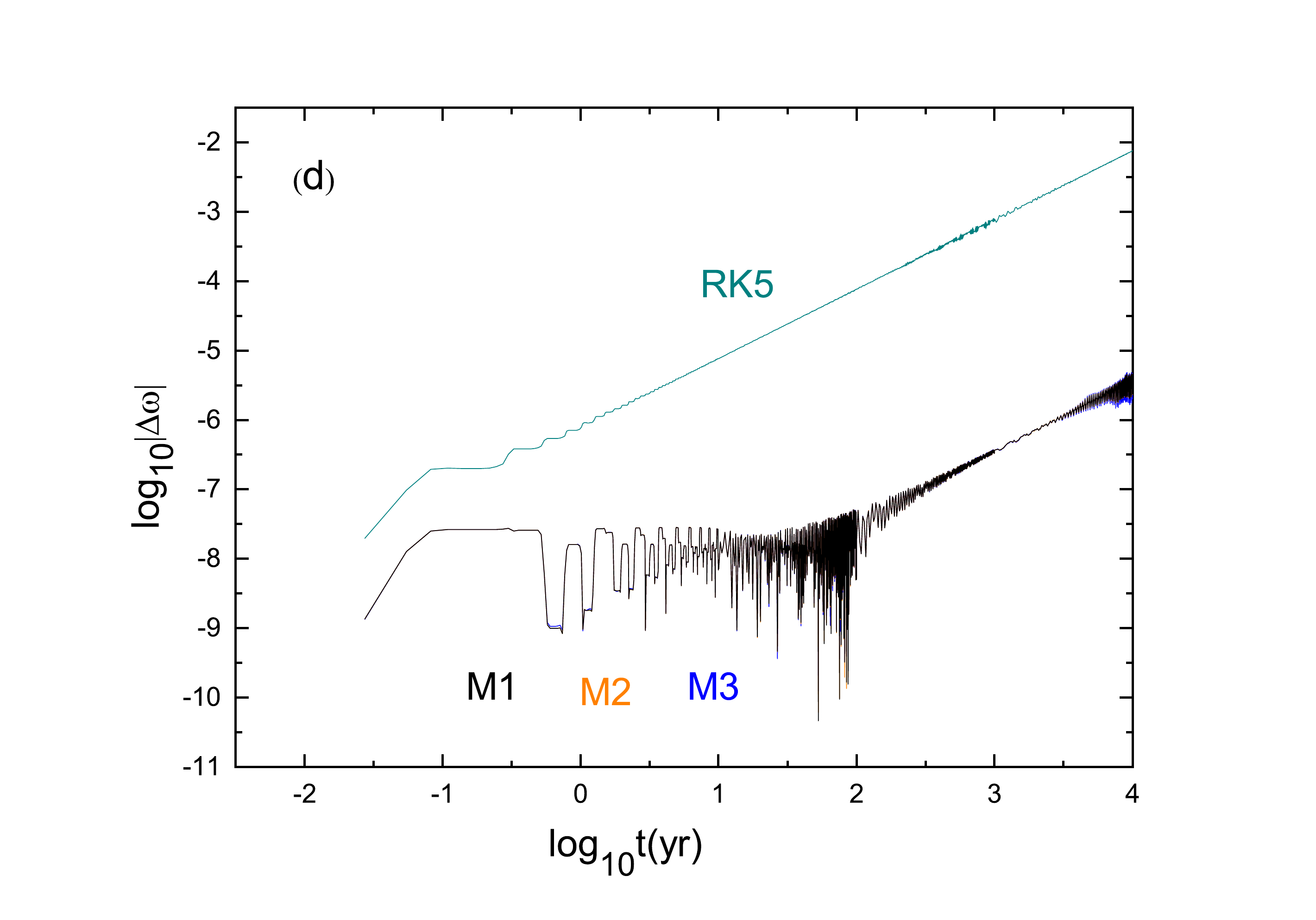}
   \includegraphics[width=0.45\textwidth]{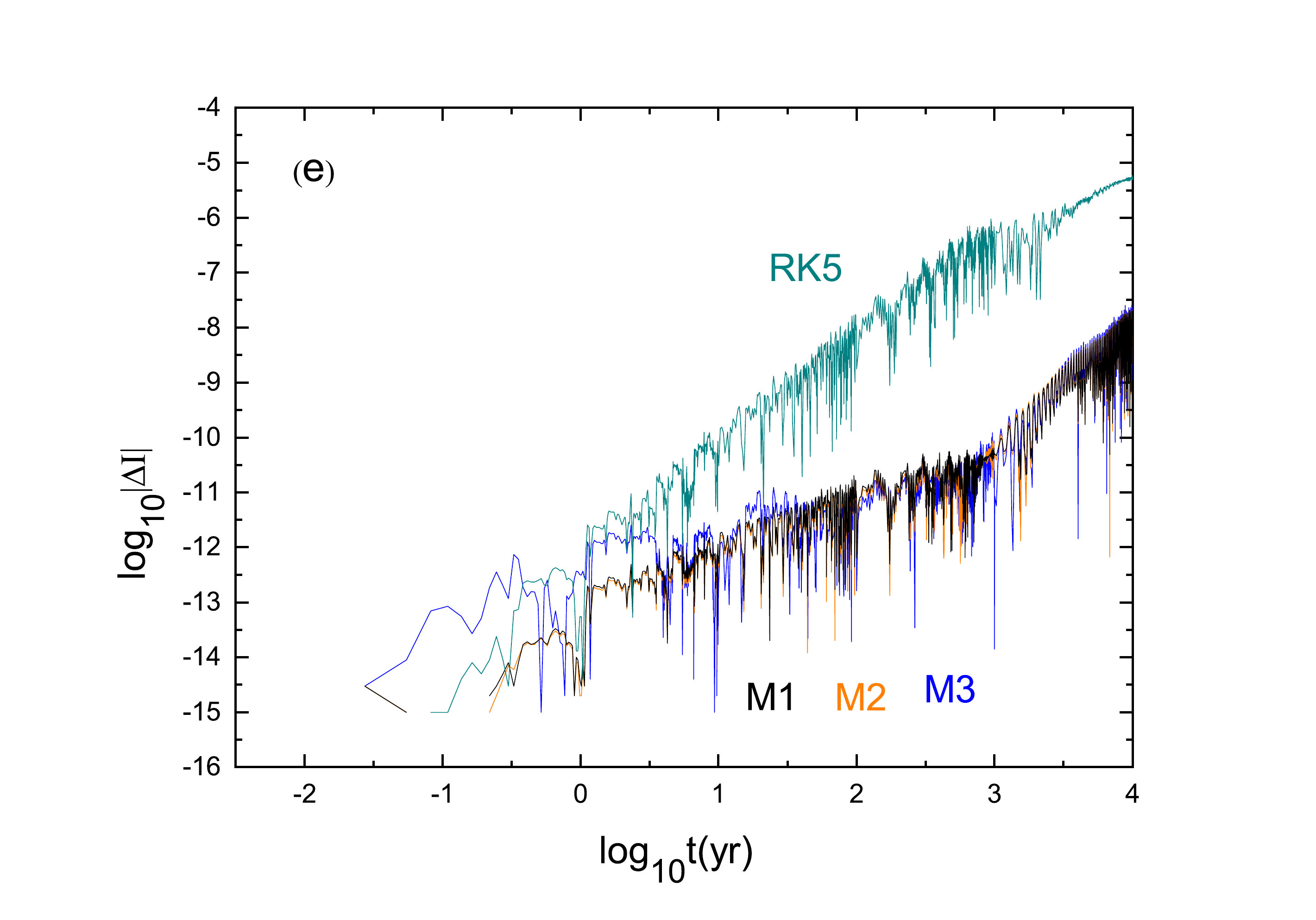}
   \includegraphics[width=0.45\textwidth]{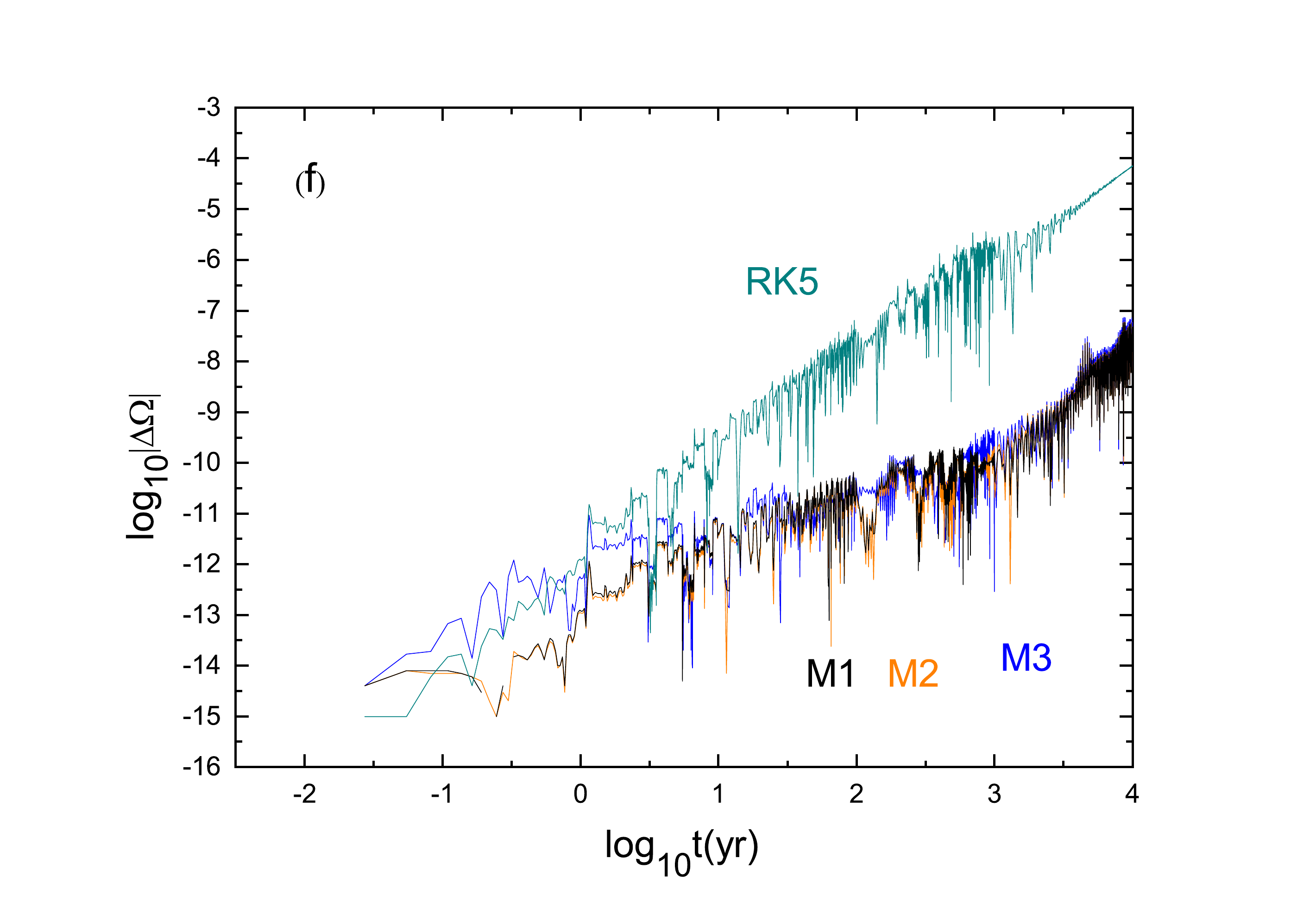}
  \caption{Errors of all orbital elements for Mercury in the inner solar system when several methods are used.}
  \label{fig:5}
\end{figure}

\begin{figure}
  \centering
    \includegraphics[width=0.45\textwidth]{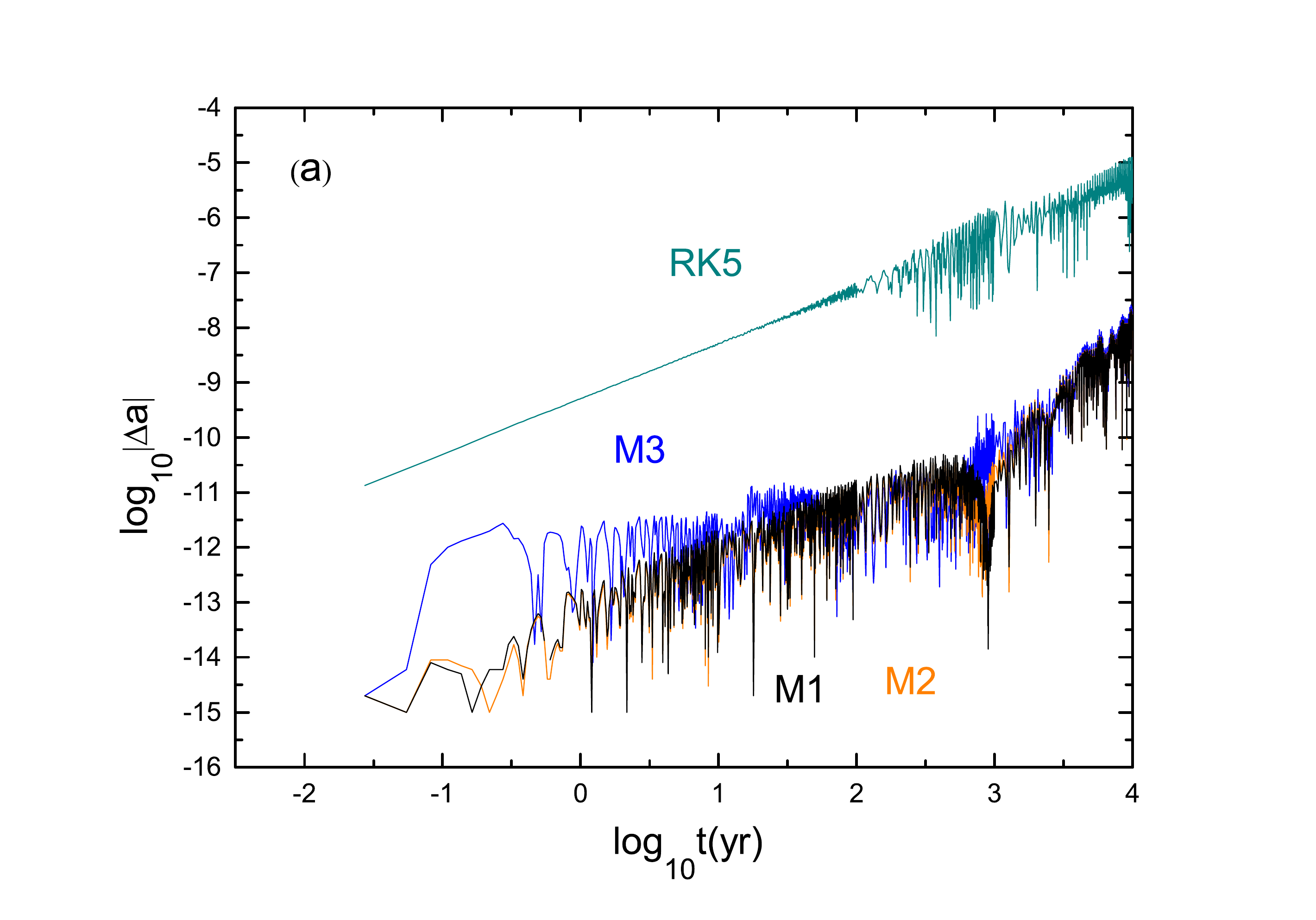}
    \includegraphics[width=0.45\textwidth]{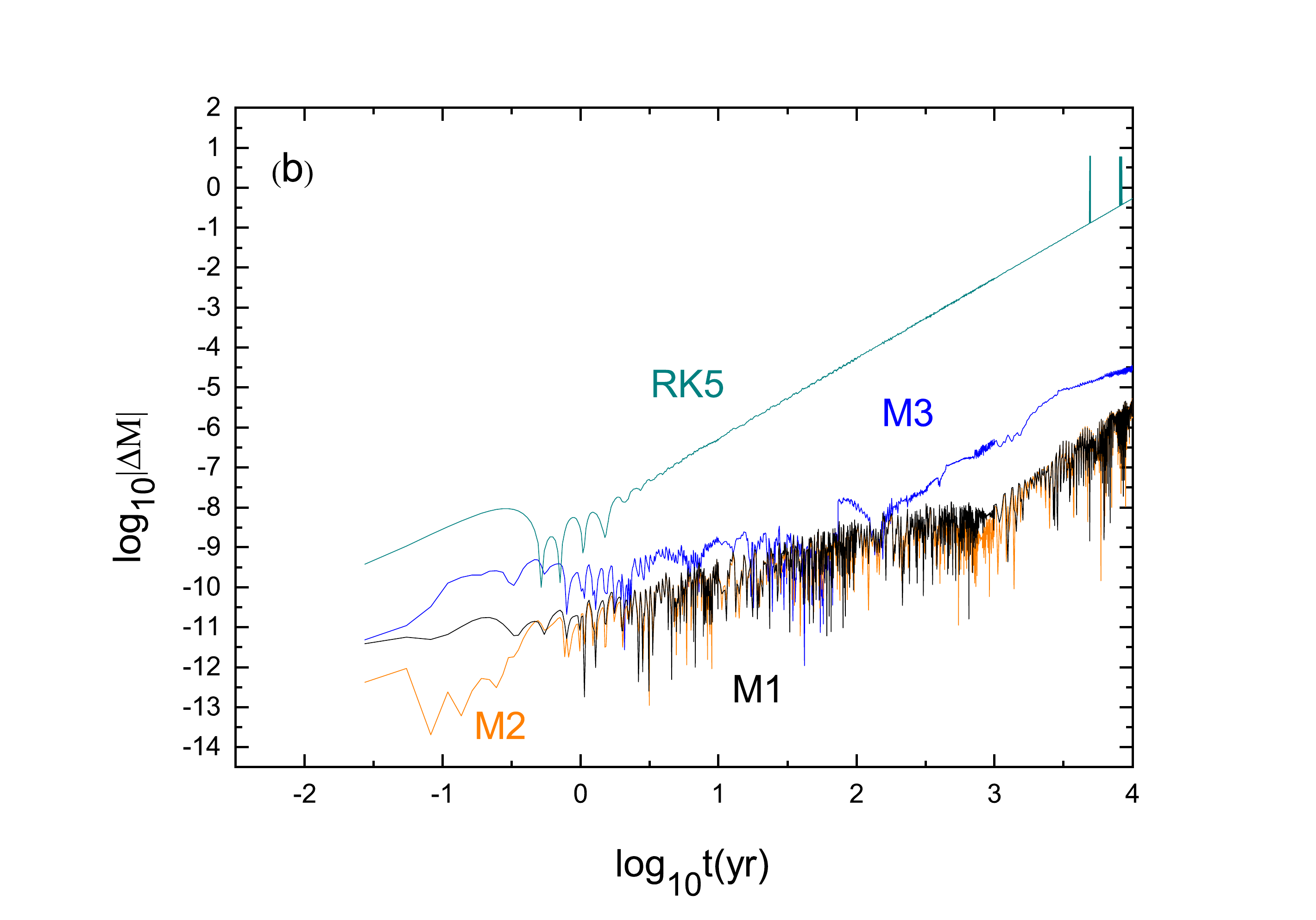}
   \includegraphics[width=0.45\textwidth]{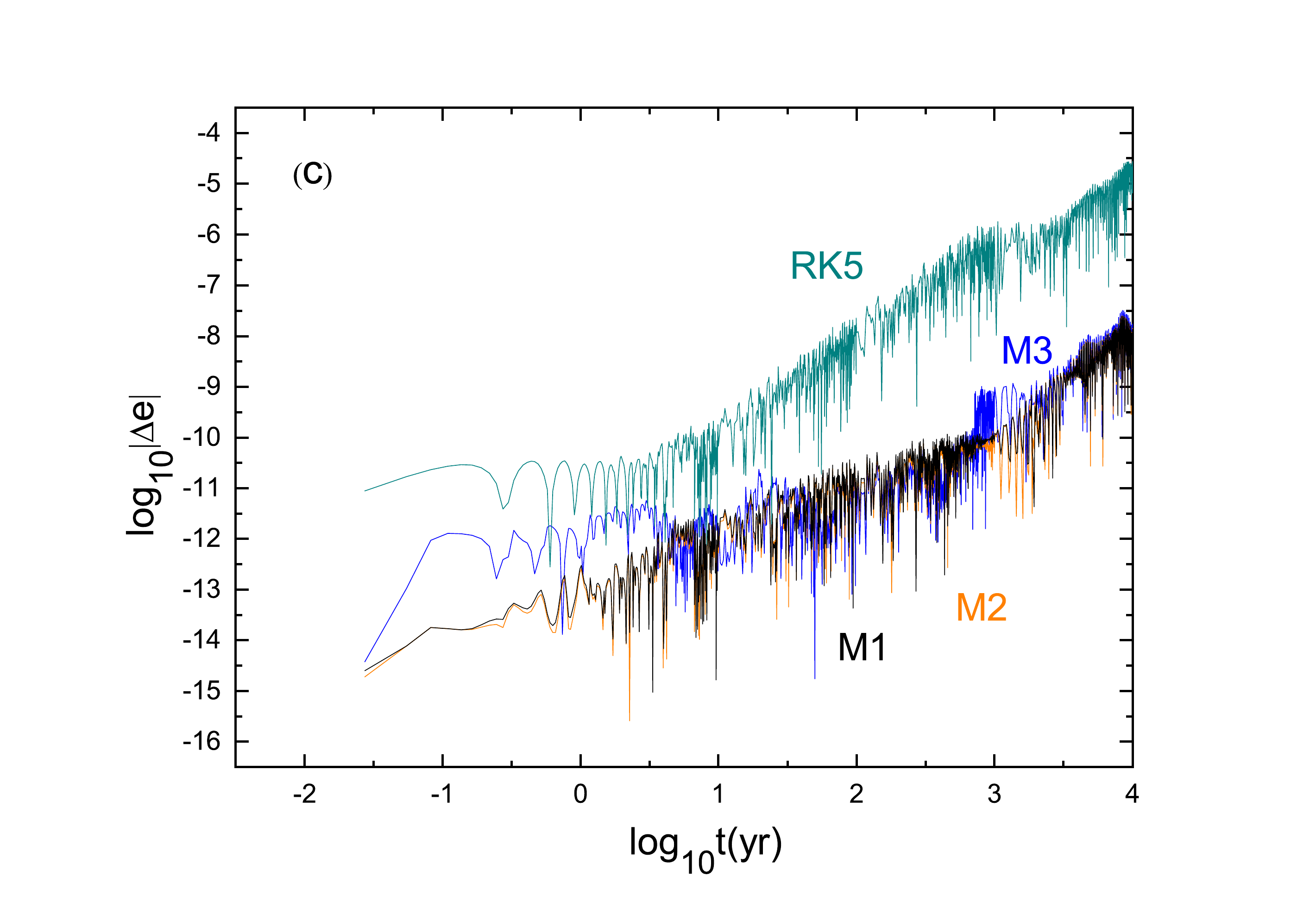}
   \includegraphics[width=0.45\textwidth]{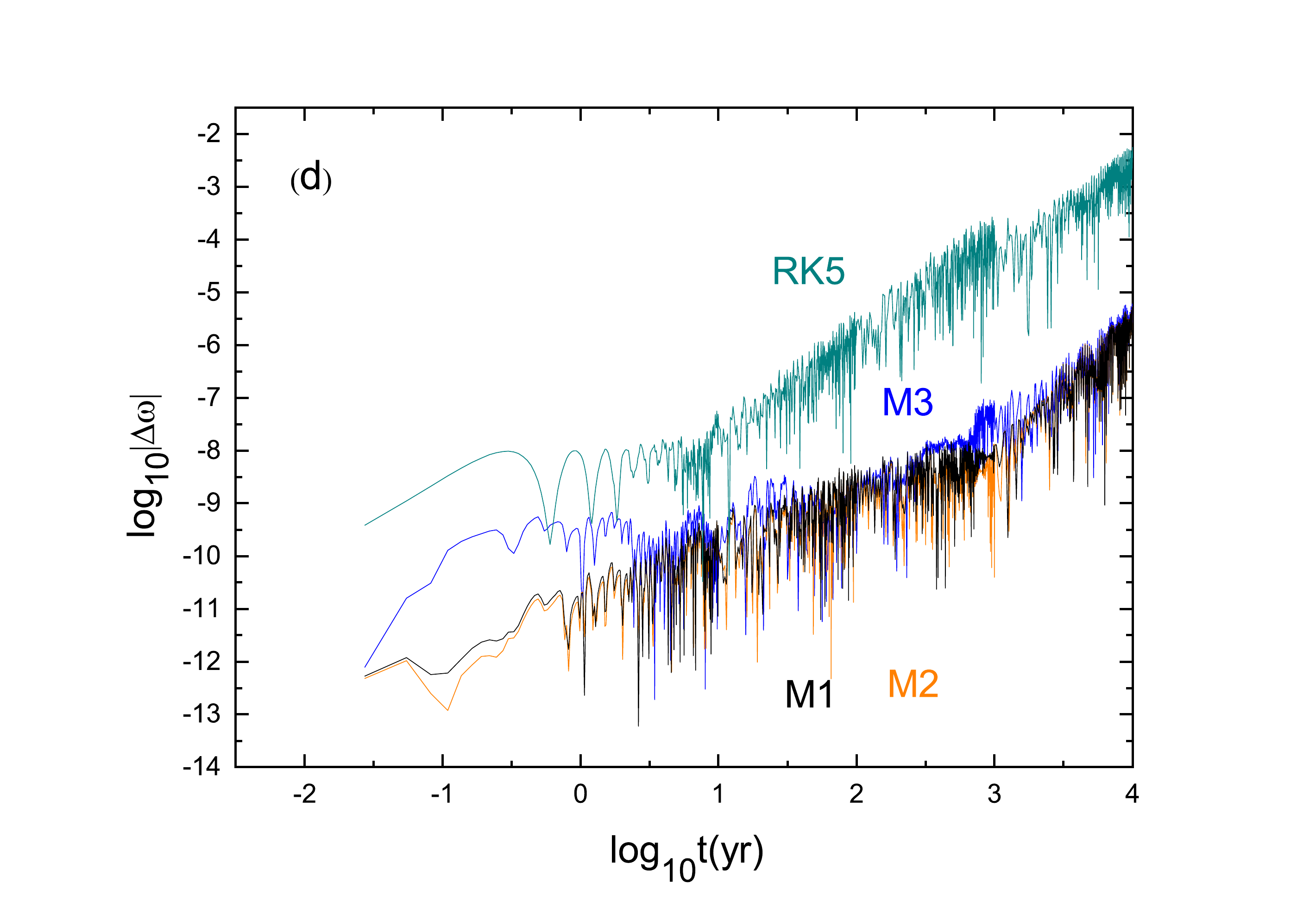}
   \includegraphics[width=0.45\textwidth]{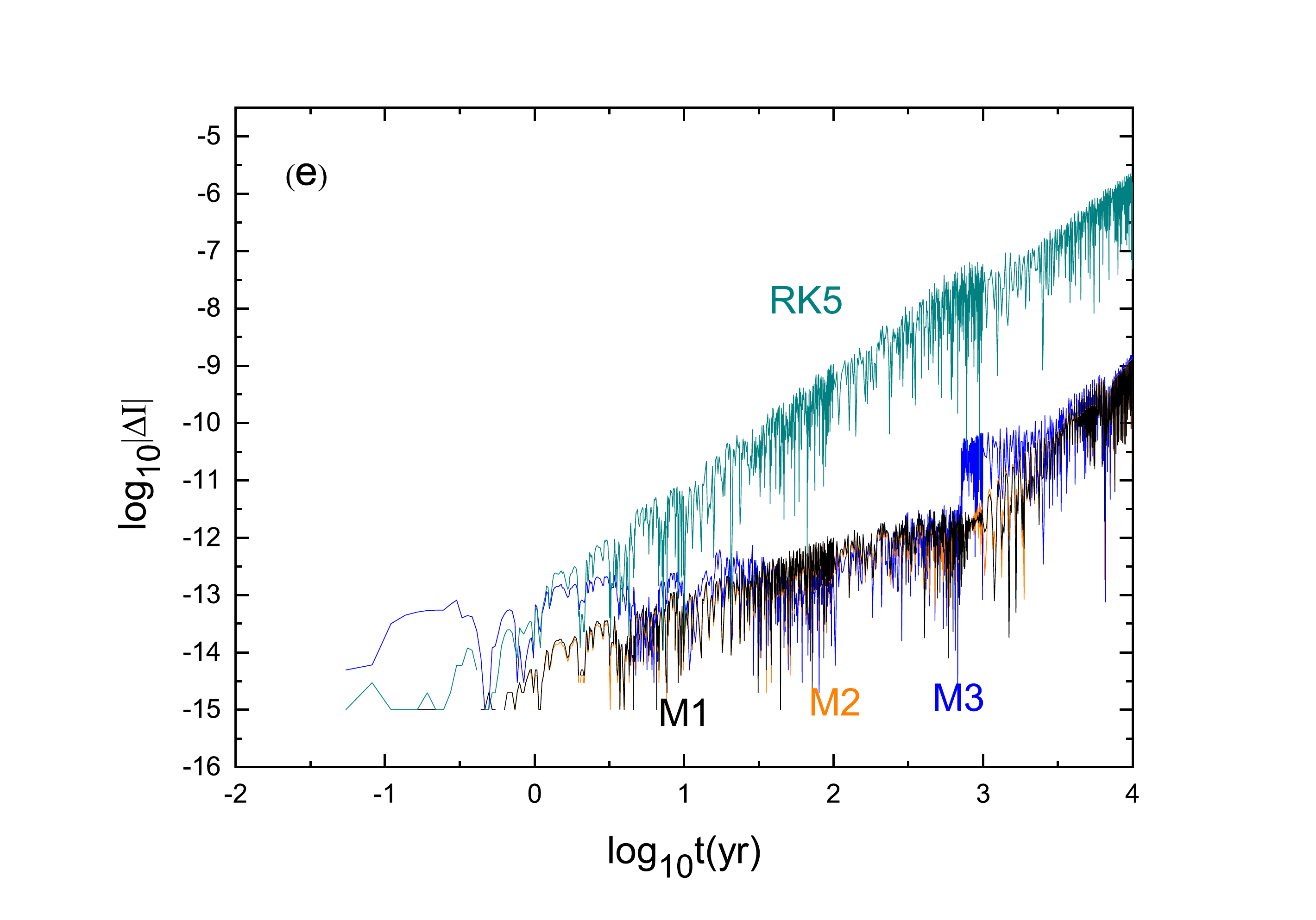}
   \includegraphics[width=0.45\textwidth]{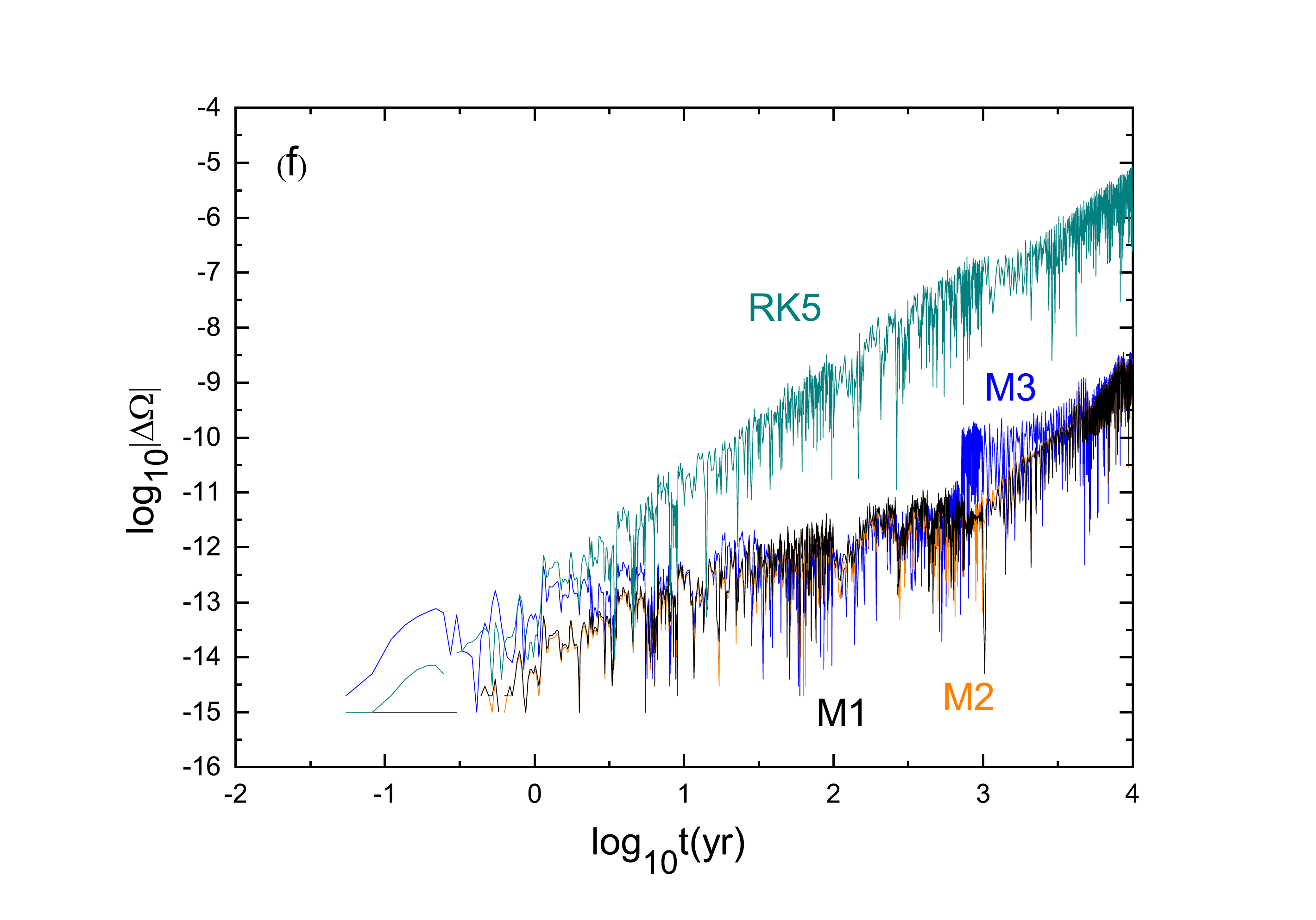}
  \caption{Same as Fig.~\ref{fig:5}, but for Venus.}
  \label{fig:6}
\end{figure}

\begin{figure}
  \centering
    \includegraphics[width=0.45\textwidth]{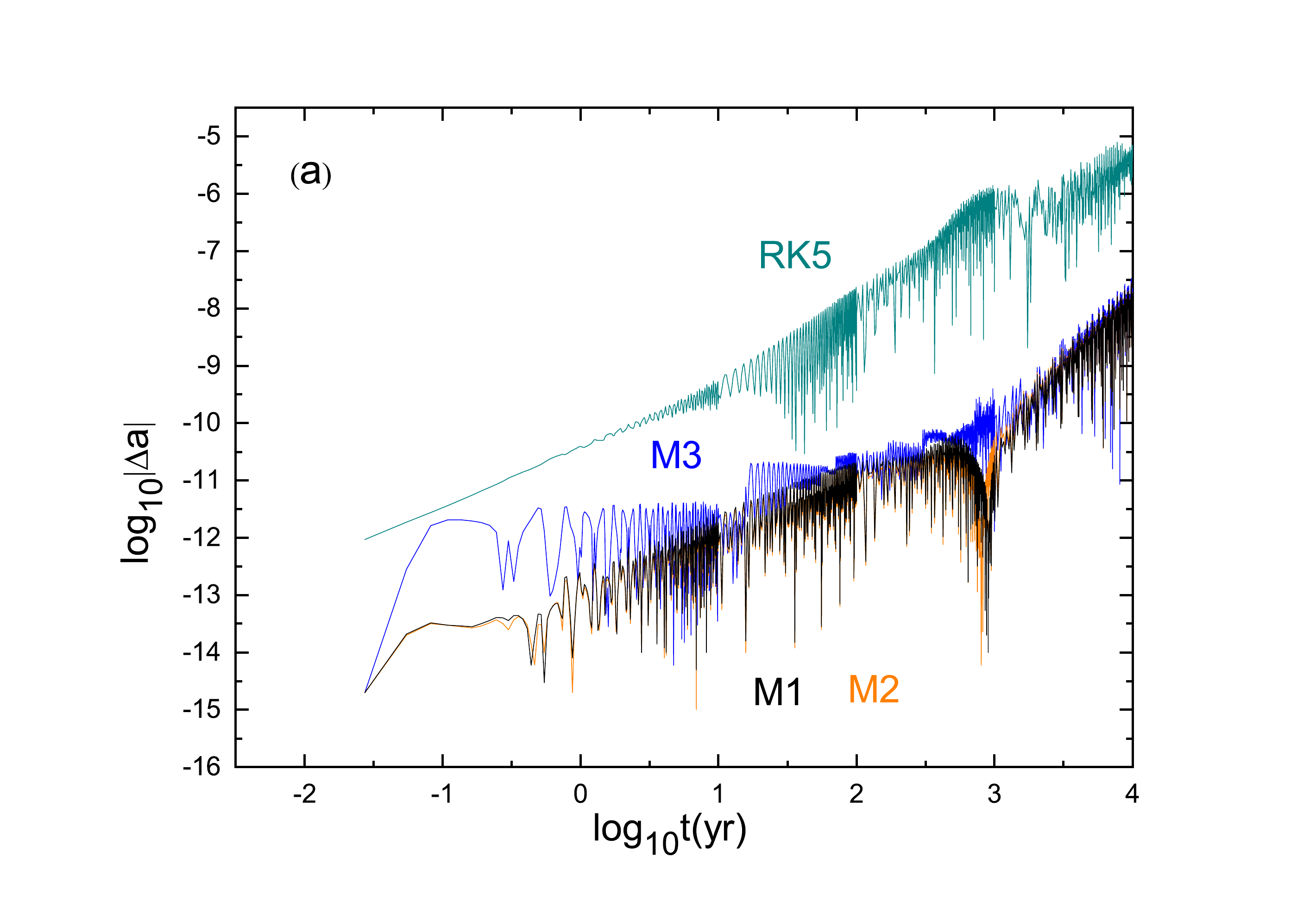}
    \includegraphics[width=0.45\textwidth]{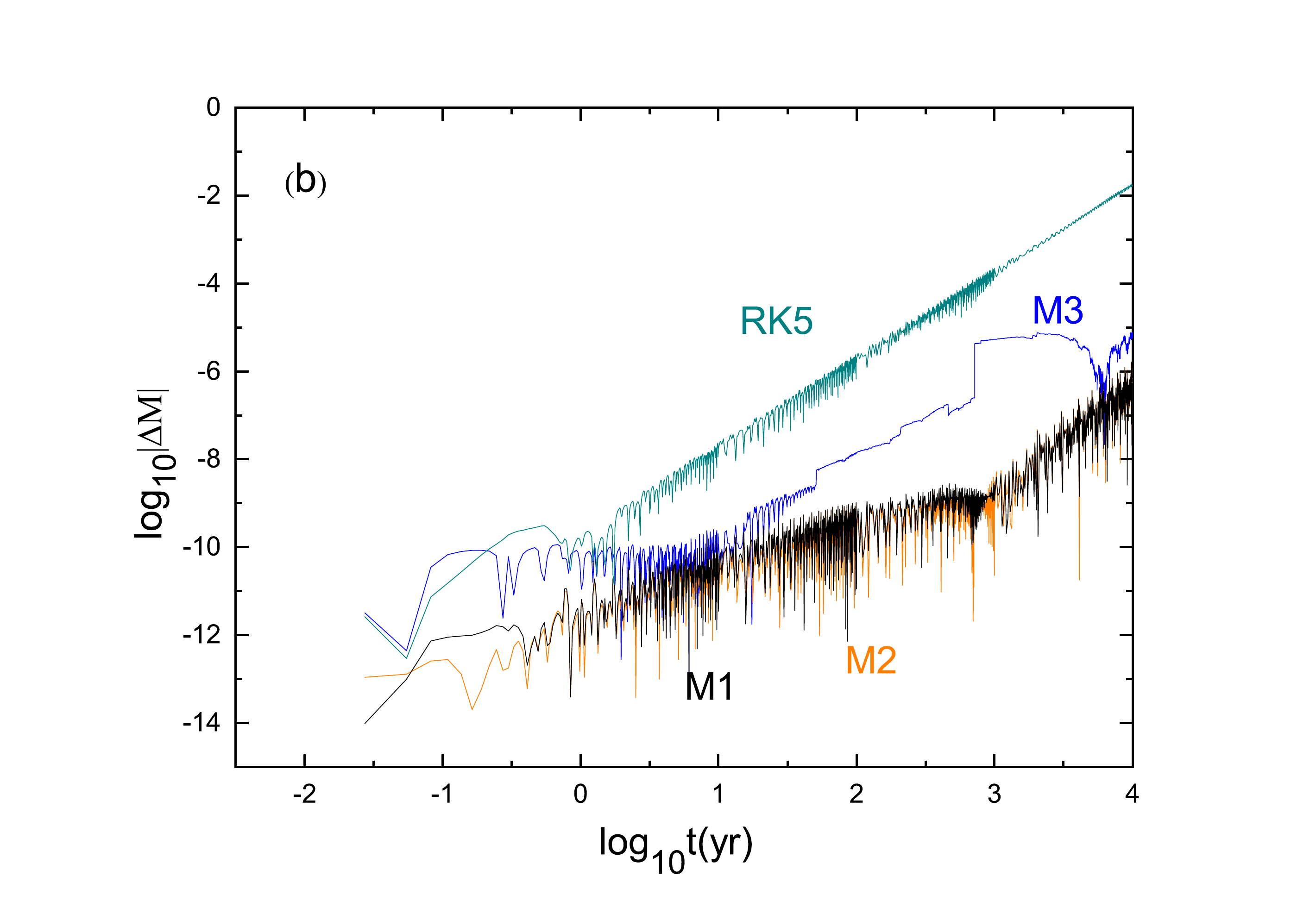}
   \includegraphics[width=0.45\textwidth]{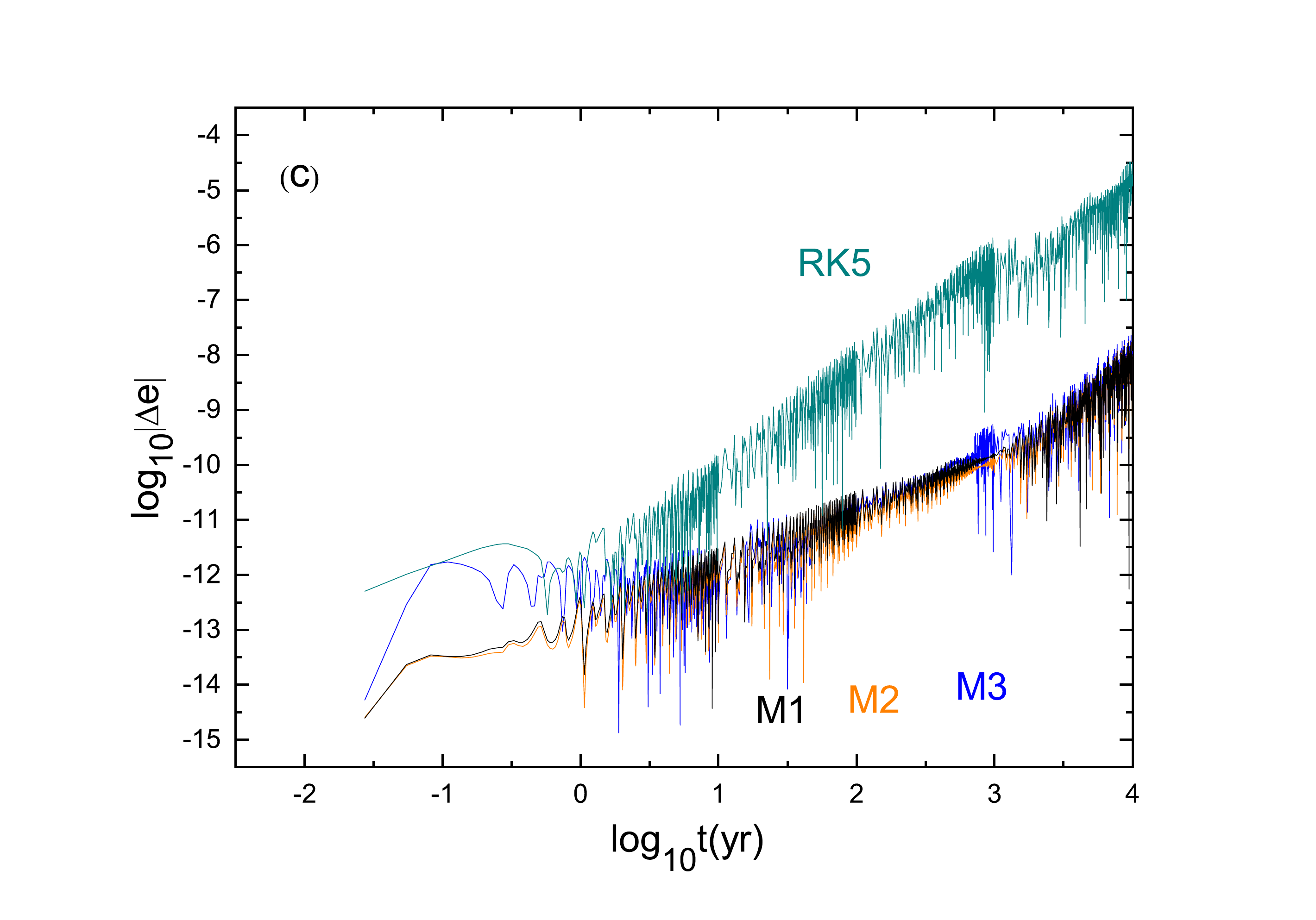}
   \includegraphics[width=0.45\textwidth]{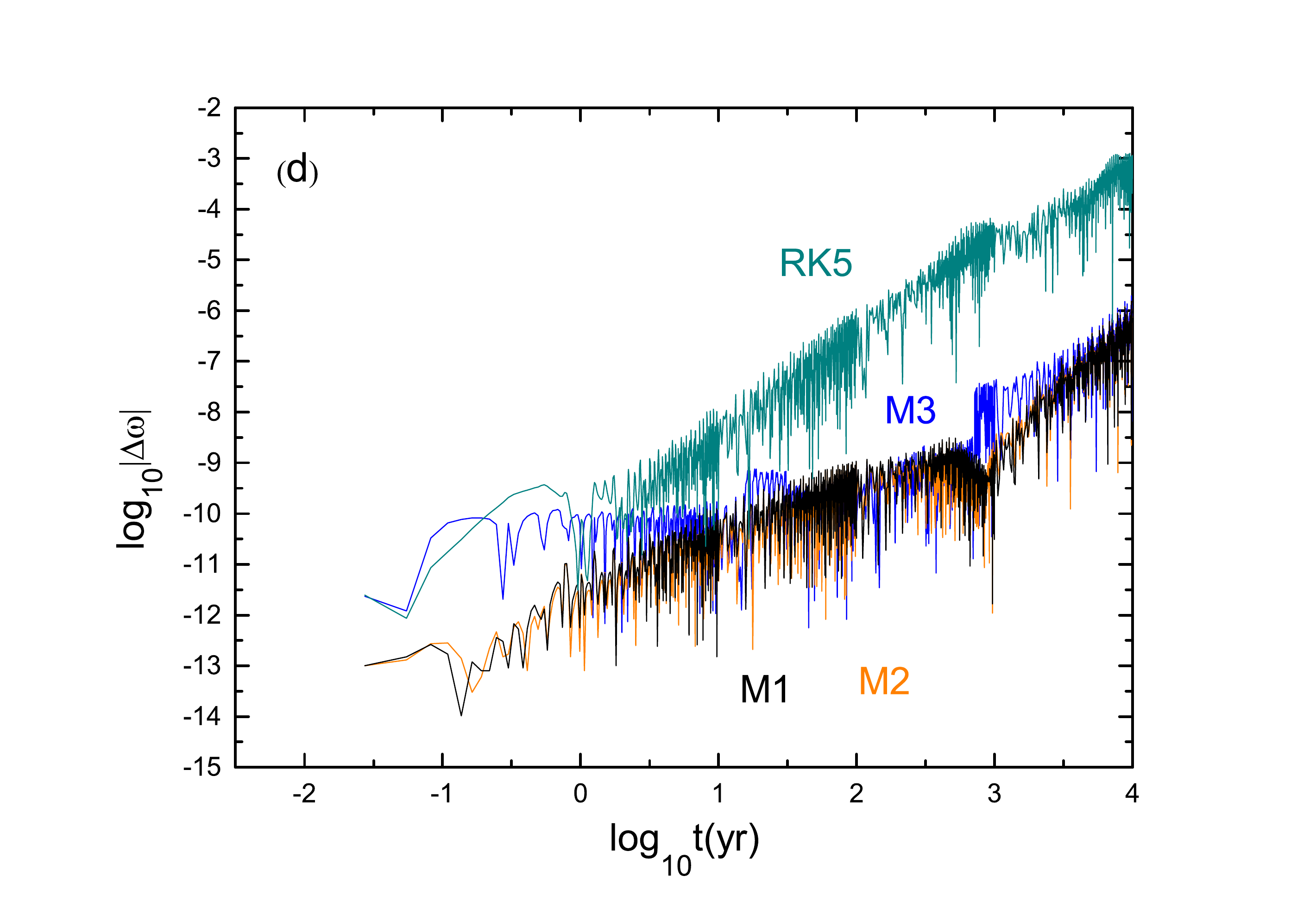}
   \includegraphics[width=0.45\textwidth]{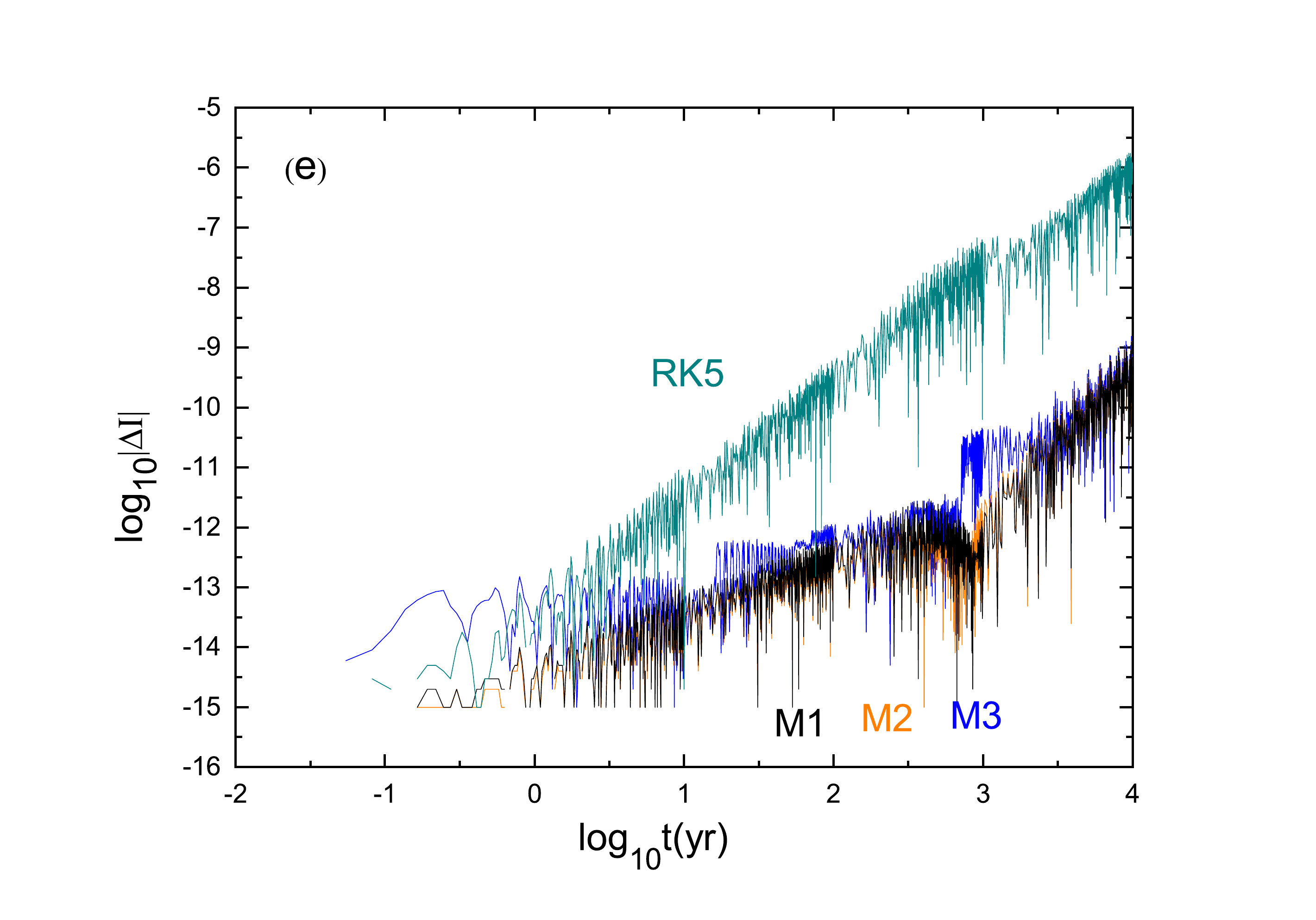}
   \includegraphics[width=0.45\textwidth]{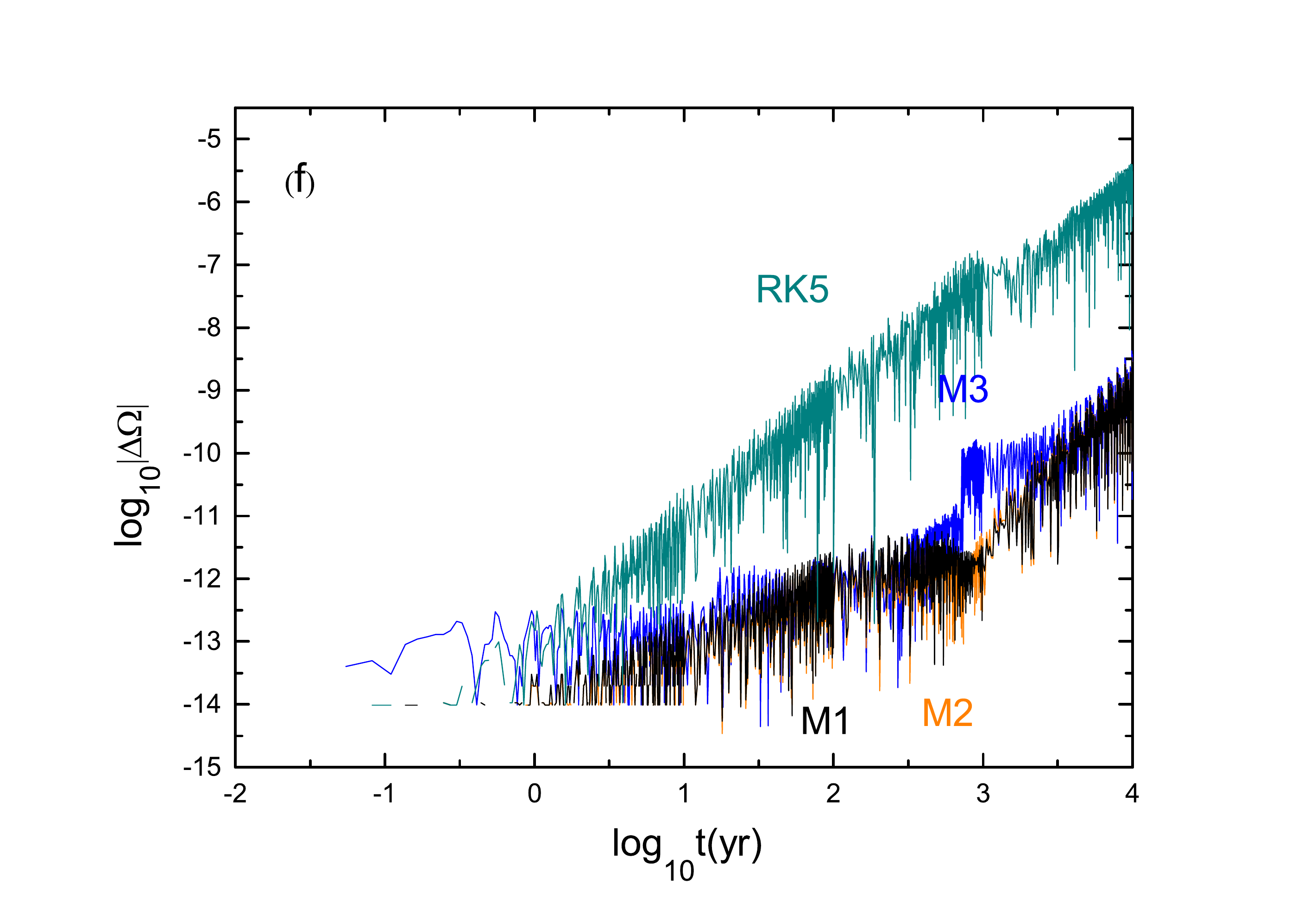}
  \caption{Same as Fig.~\ref{fig:5}, but for the Earth-Moon system. We simply take the Earth-Moon system as a point mass at the Earth-Moon barycenter. The Earth-Moon separation is small compared to the interplanetary separations.}
  \label{fig:7}
\end{figure}

\begin{figure}
  \centering
    \includegraphics[width=0.45\textwidth]{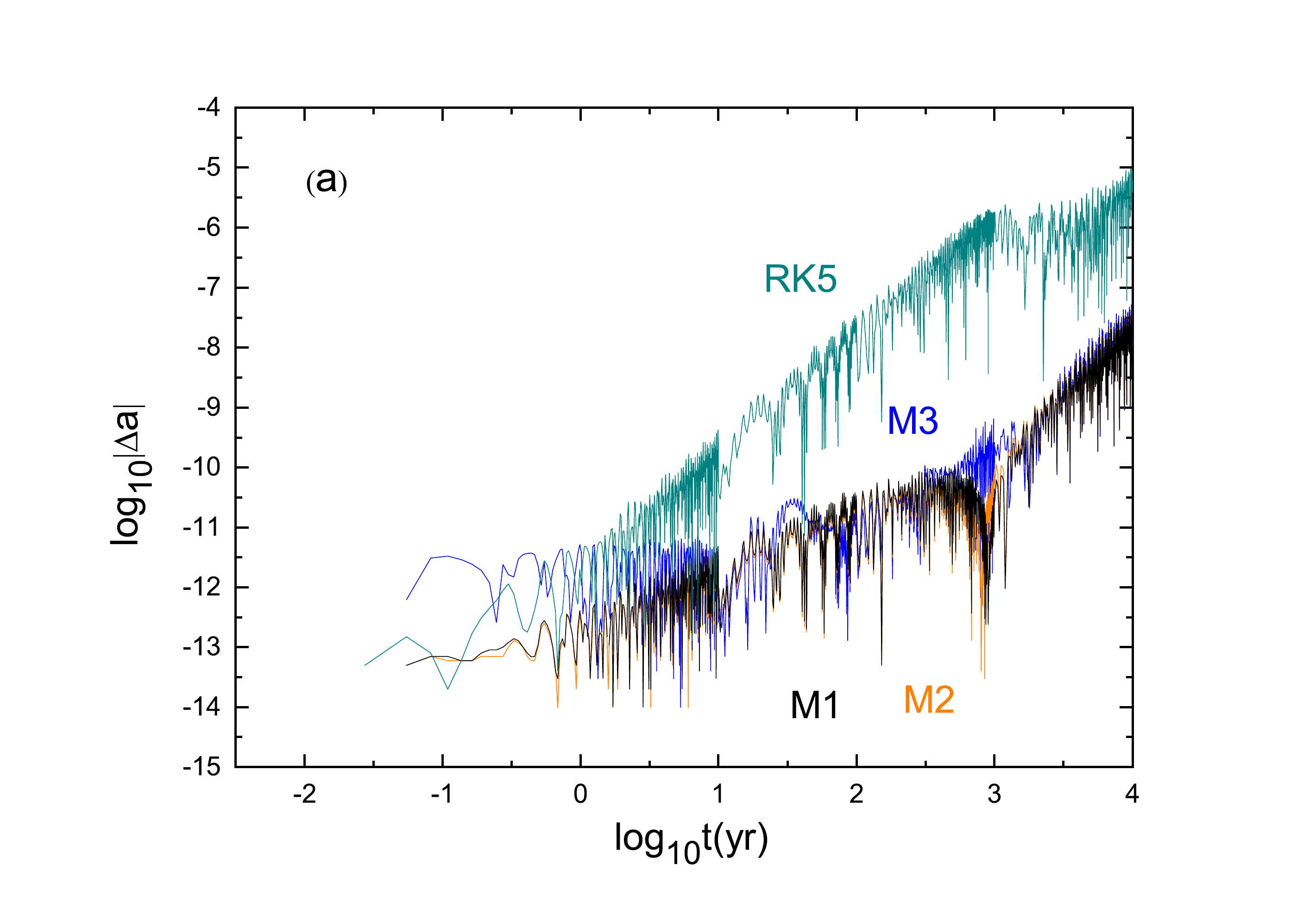}
    \includegraphics[width=0.45\textwidth]{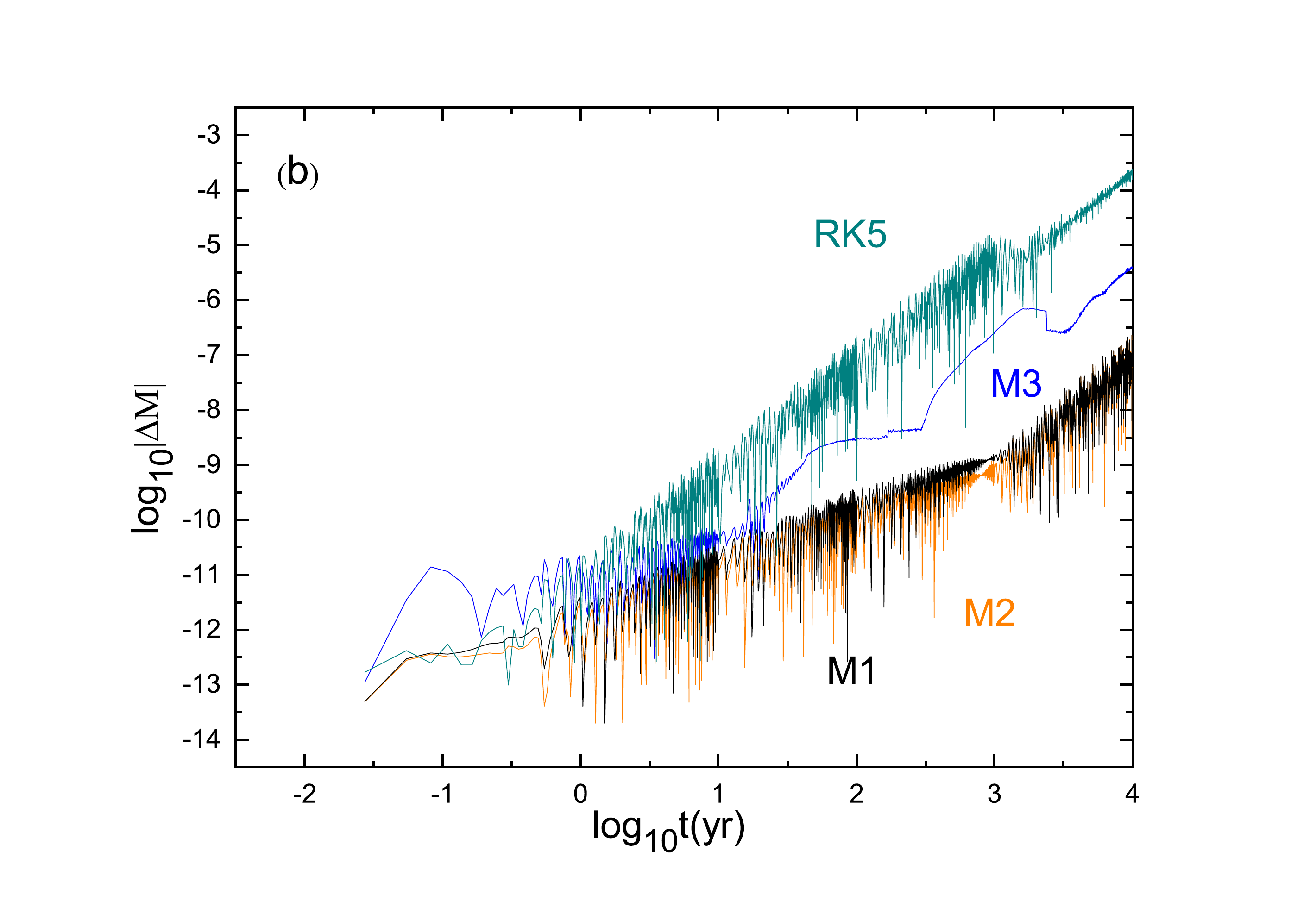}
   \includegraphics[width=0.45\textwidth]{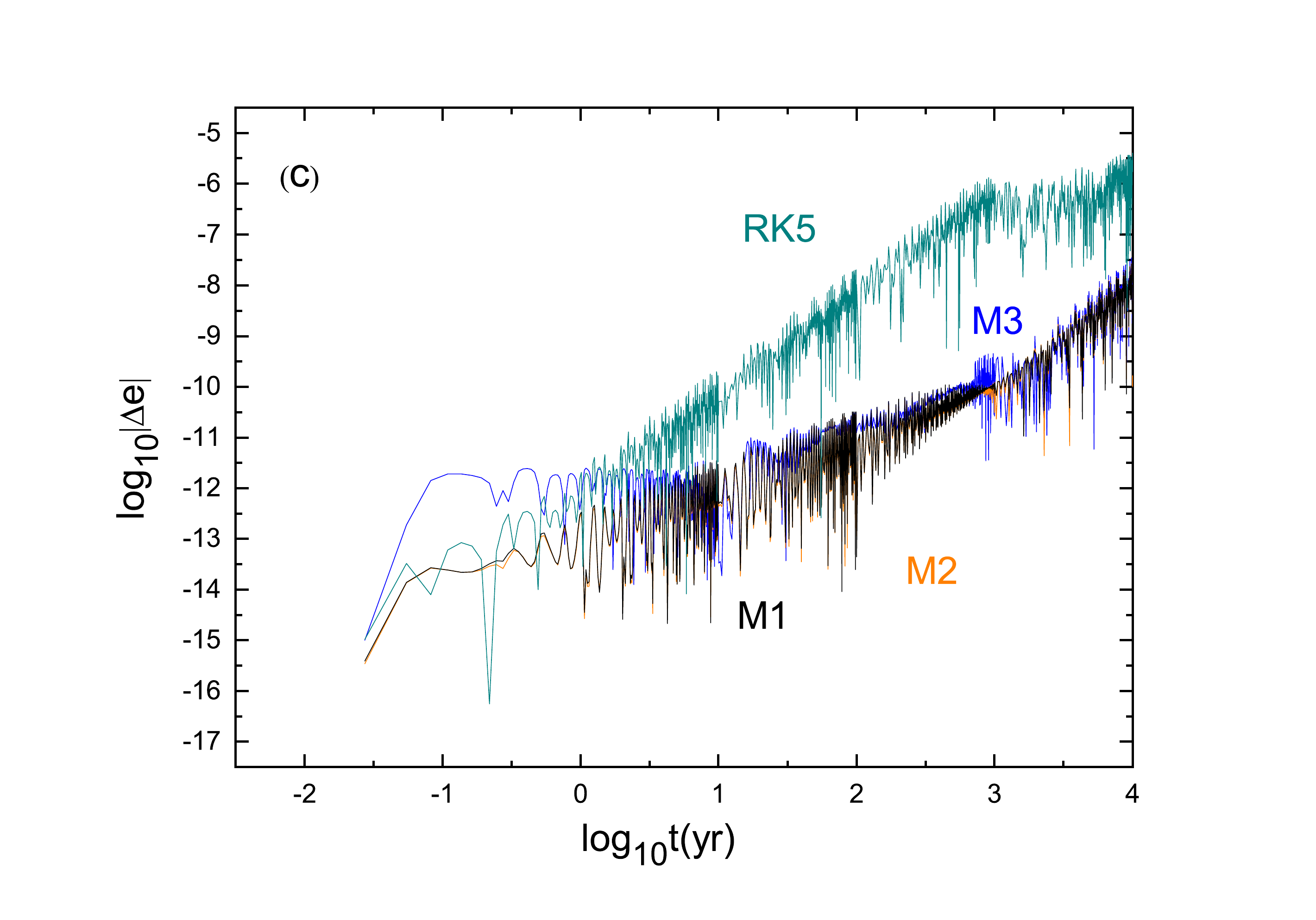}
   \includegraphics[width=0.45\textwidth]{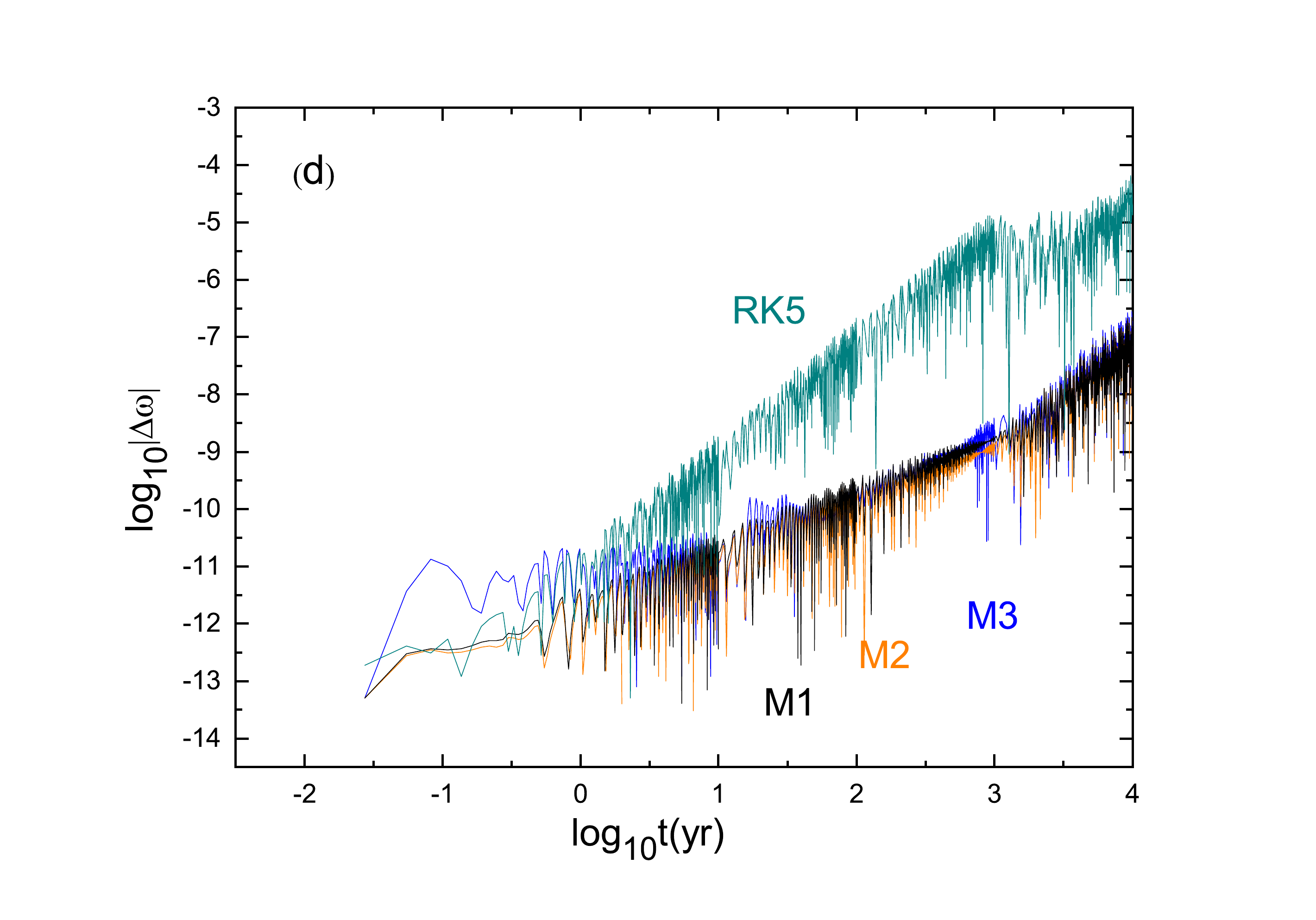}
   \includegraphics[width=0.45\textwidth]{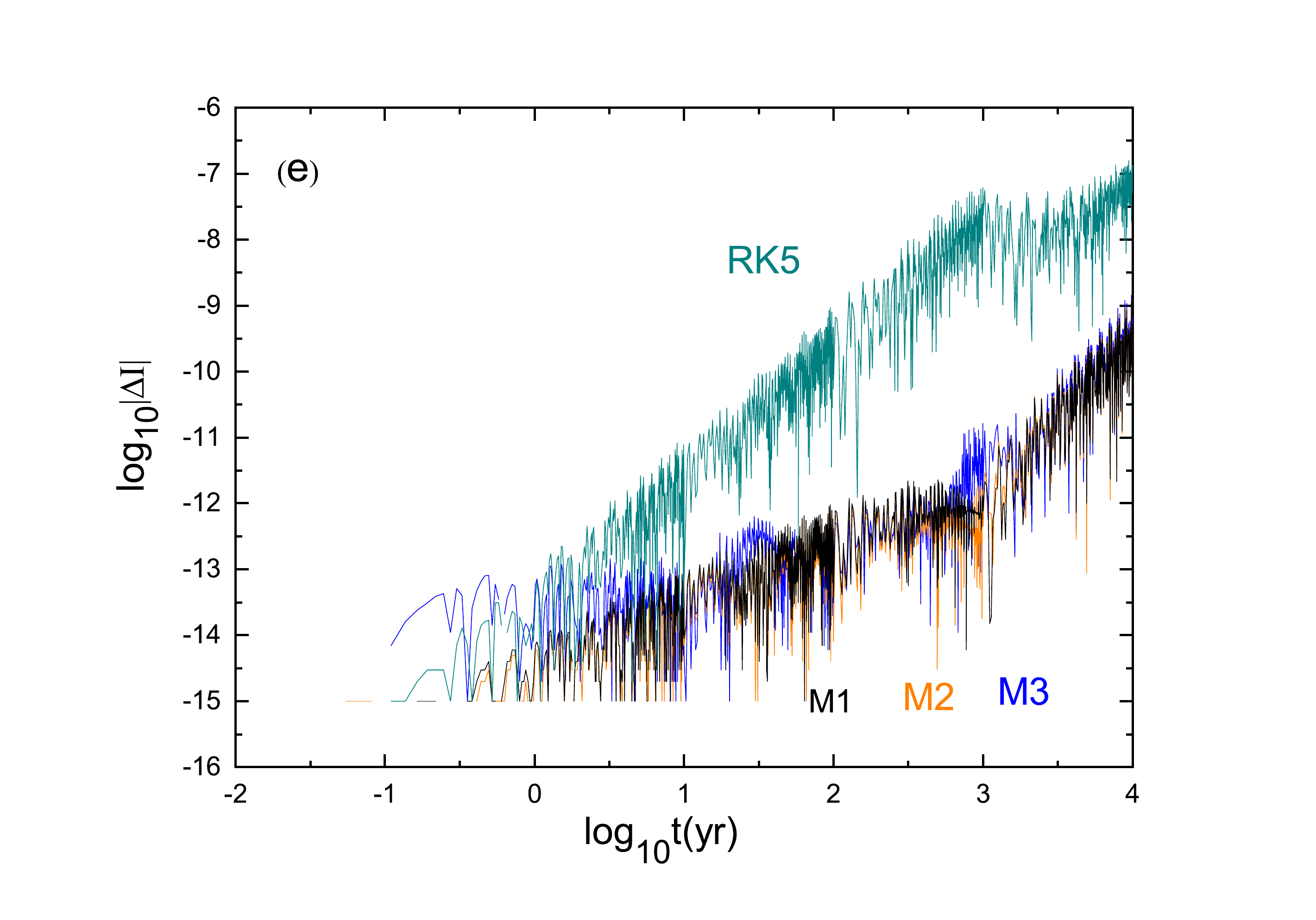}
   \includegraphics[width=0.45\textwidth]{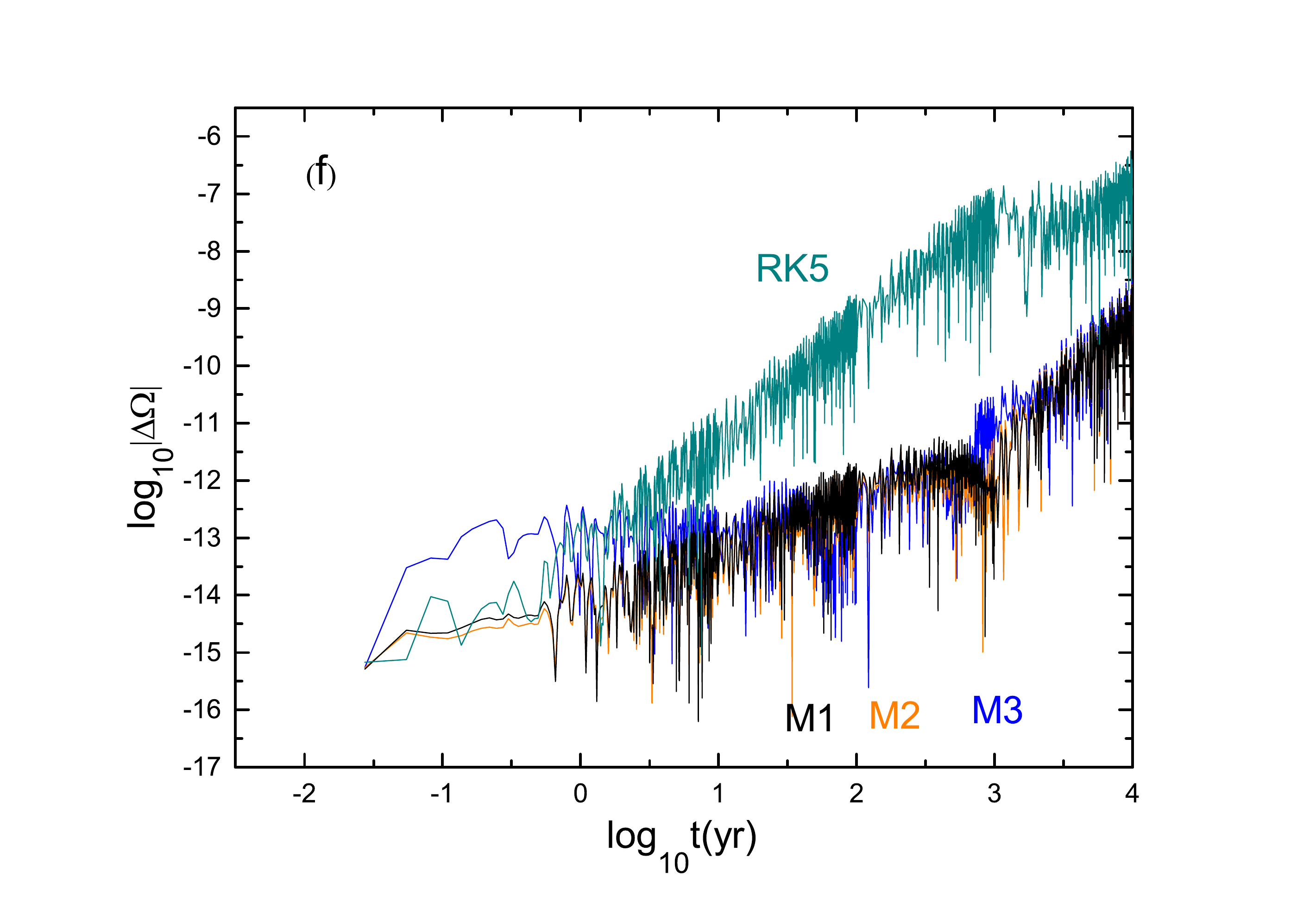}
  \caption{Same as Fig.~\ref{fig:5},but for Mars.}
  \label{fig:8}
\end{figure}

 The initial conditions of each planet and the related physical parameters are obtained from those in JPL planetary ephemeris ($t_0$=JD2440400.5), DE430. The basic integrator still uses RK5. The fixed time step is one day which is about 1/88 of Mercury’s orbital period, and the length of integration time is $10^4$ yr. The higher precision reference solutions are provided by a $12th$-order $Adams-Cowell$ method. The errors in the orbital elements of each planet are shown in Figs.~\ref{fig:5}--\ref{fig:8}. The related results are presented in the following.

The accuracies of all the orbital elements for M1 are consistent with those for M2 and are higher in magnitude of about three orders than those for RK5. M3 is not as good as M1 and M2 in the accuracies of some orbital elements. This is because the seven slowly-varying quantities of each body are satisfied simultaneously in M1 and M2, but not in M3.

As stated in \cite{Ma2008c}, the effects for improving all elements of every planet in the perturbed problem are less than that those in the pure Keplerian problem. The effects are also influenced by the semimajor axes. Mercury, Venus, Earth, and Mars have different semimajor axes corresponding to different periods. For the same step size, a smaller period means that the uncorrected integrator shows poorer performance, but the corrected method has better effects. 

\begin{figure}
  \centering
    \includegraphics[width=0.45\textwidth]{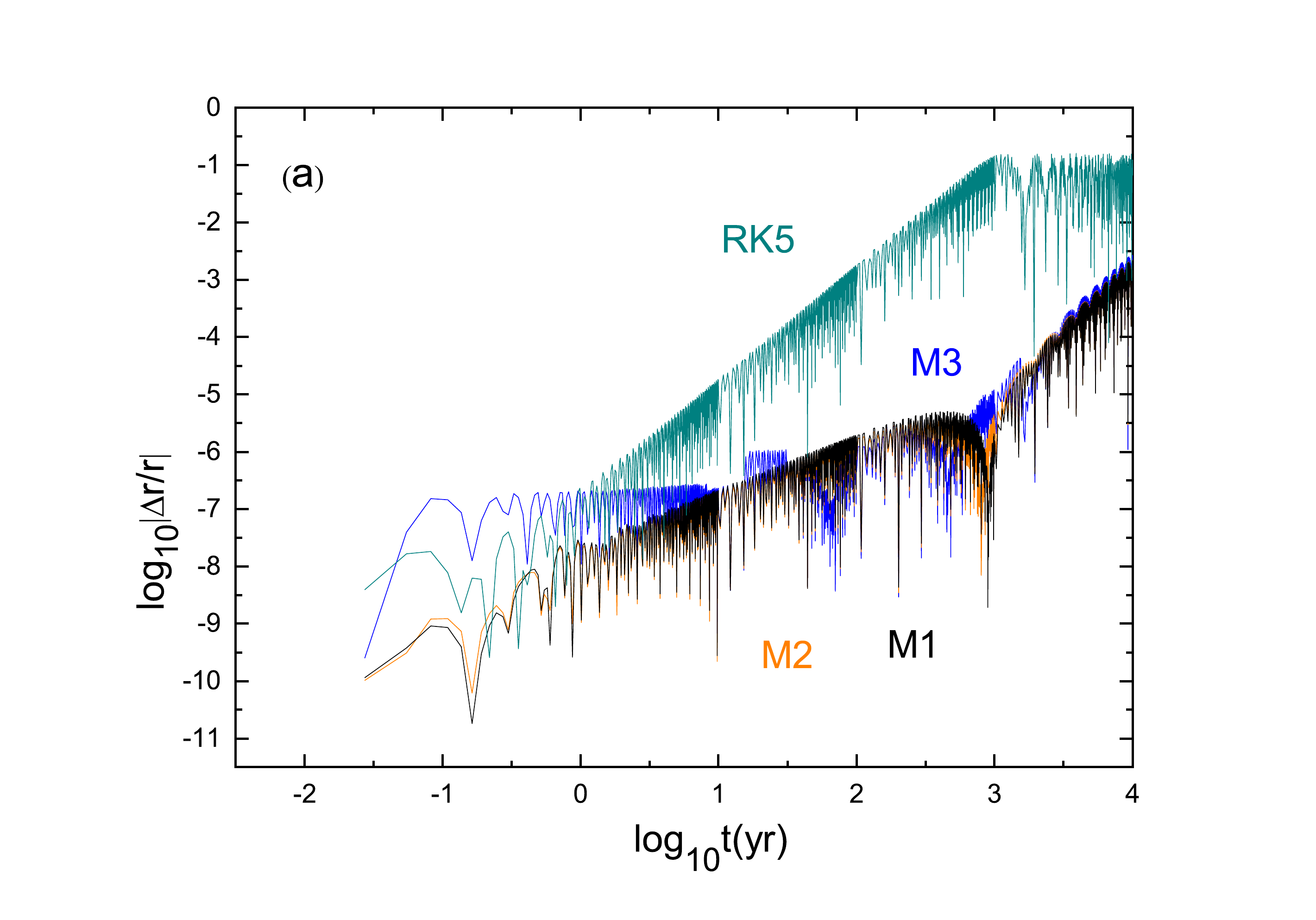}
    \includegraphics[width=0.45\textwidth]{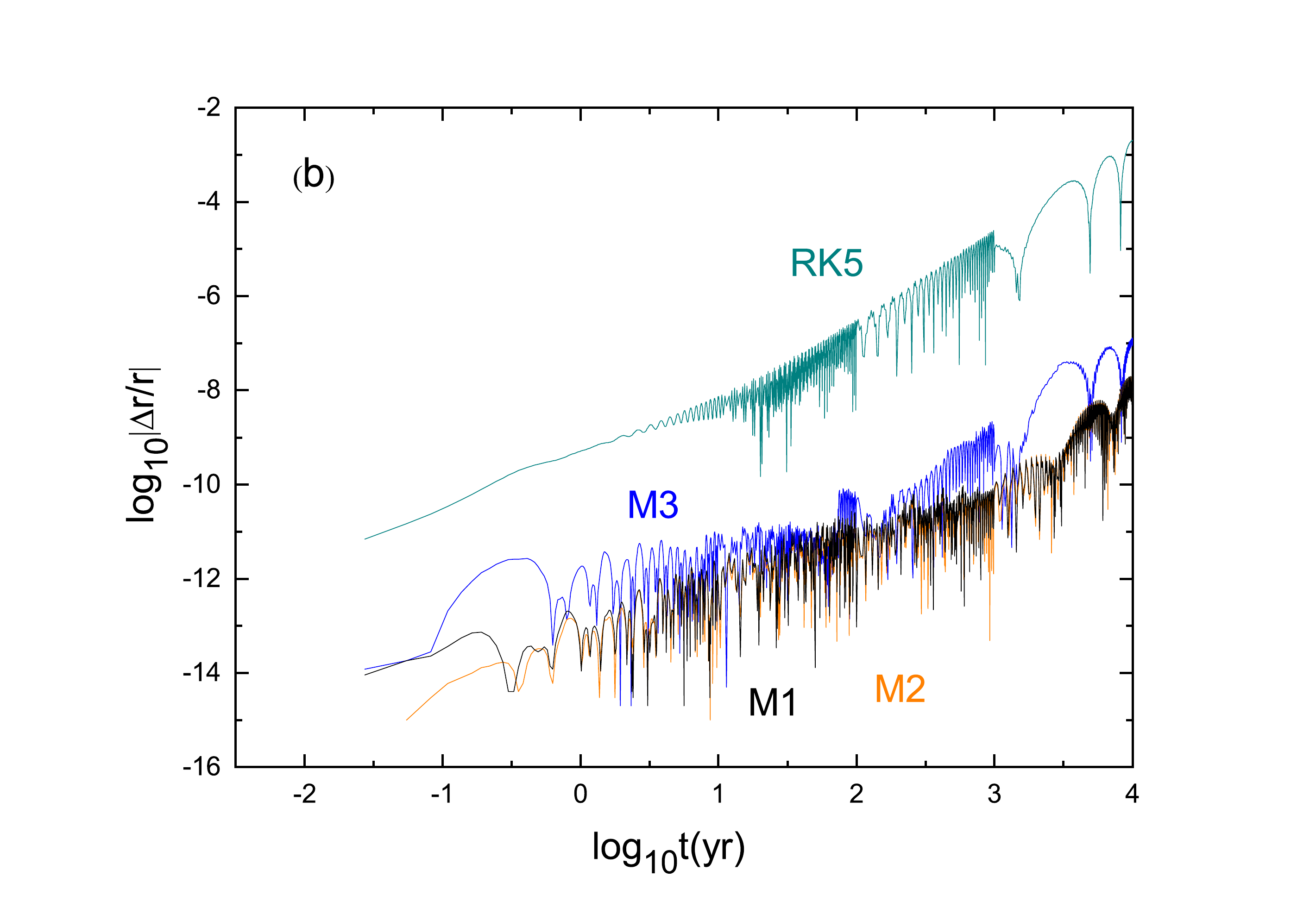}
   \includegraphics[width=0.45\textwidth]{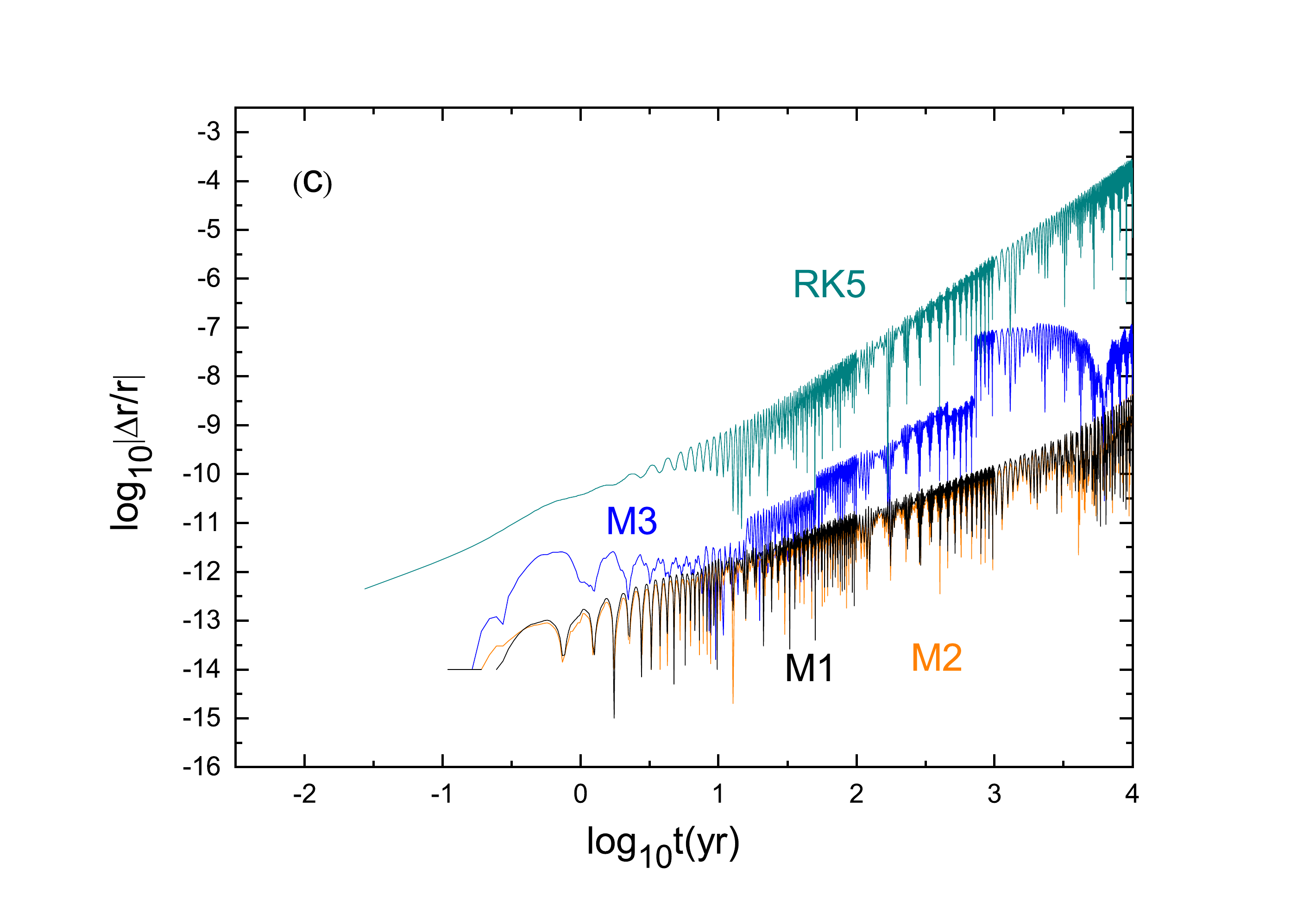}
   \includegraphics[width=0.45\textwidth]{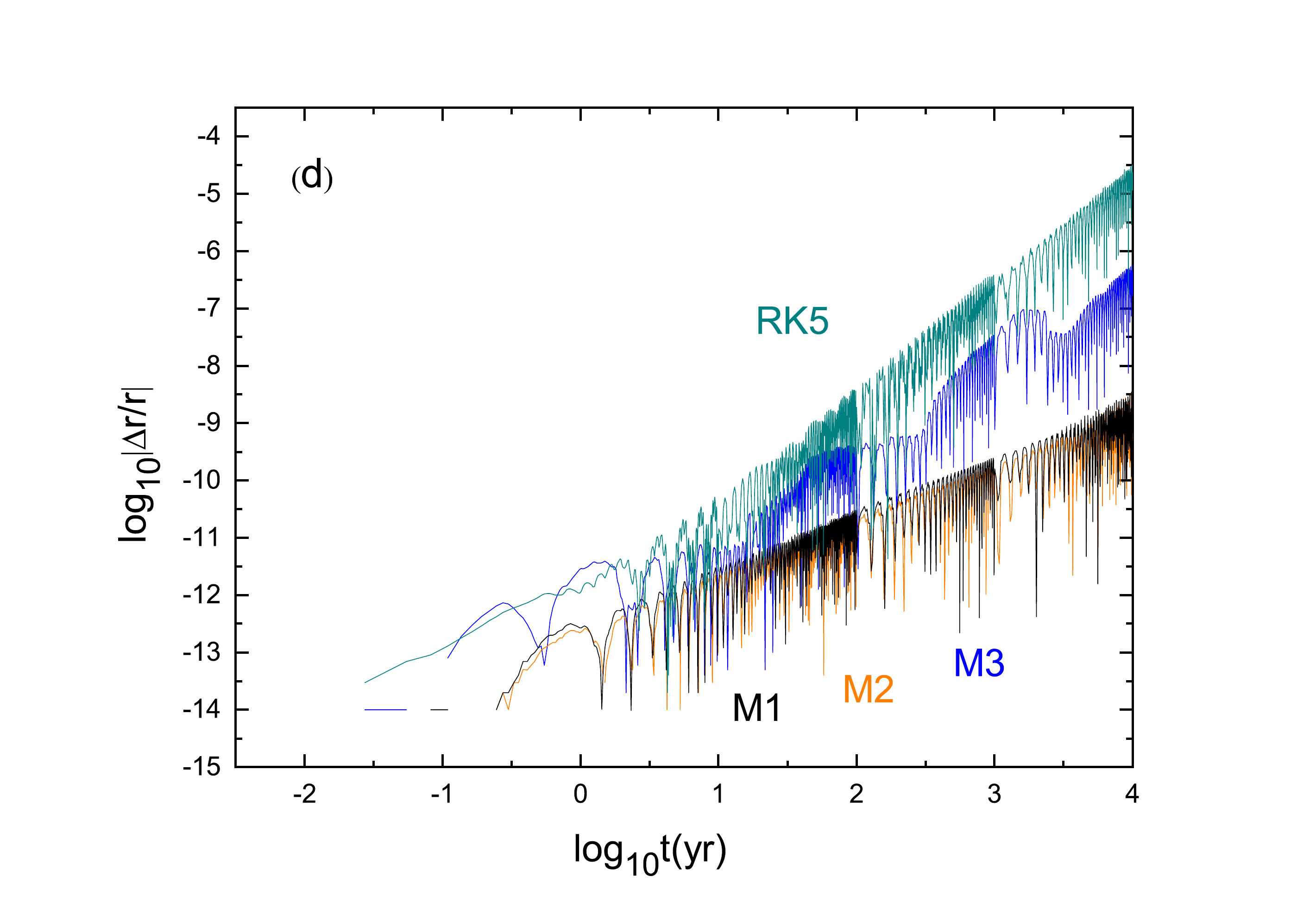}
  \caption{Errors of the relative positions for the four inner planets. Subfigures (a), (b), (c), and (d) respectively show the errors of the relative positions of Mercury, Venus, Earth, and Mars.}
  \label{fig:9}
\end{figure}

The errors in the relative positions of Mercury, Venus, Earth, and Mars are shown in Fig.~\ref{fig:9}.
Here, the results in Fig.~\ref{fig:9} are almost the same as those in Figs.~\ref{fig:5}--\ref{fig:8}. To more clearly show the effectiveness of the correction schemes, we list the errors of the position and velocity in Table~\ref{tab:2}. As expected, M1 and M2 exhibit typically better performance than M3. However, M1 and M2 have no obvious differences.

\renewcommand{\arraystretch}{0.8} 
\begin{table}[tp]

  \centering
  \fontsize{6.5}{8}\selectfont
  \begin{threeparttable}
  \caption{The errors of the positions and velocities for each inner planet, produced by RK5 and its manifold methods M1, M2, and M3 at some times.}
  \label{tab:2}
    \begin{tabular}{cccccccccccc}
    \toprule
   \multicolumn{3}{c}{\multirow{2}{*}{Time(yr)}}&\multirow{2}{*}{Method}&
    \multicolumn{4}{c}{$\vert\triangle\bm{r}\vert$}&\multicolumn{4}{c}{$\vert\triangle\bm{v}\vert$}\cr
    \cmidrule(lr){5-8} \cmidrule(lr){9-12}
    &&&&Mercury&Venus&Earth&Mars&Mercury&Venus&Earth&Mars\cr

    \midrule
    \multicolumn{3}{c}{\multirow{4}{*}{1}}&RK5&1.15E-07&5.17E-10&3.77E-11&1.17E-12&9.03E-09&7.31E-12&2.93E-13&1.54E-14\cr 
    &&&M1&1.15E-09&1.11-14&1.50E-13&2.70E-13&1.70E-10&2.08E-15&3.55E-15&3.70E-15\cr
    &&&M2&1.05E-09&1.39E-14&1.09E-13&2.30E-13&1.60E-10&2.51E-15&2.92E-15&3.08E-15\cr
    &&&M3&1.09E-08&1.86E-12&6.09E-13&2.91E-12&1.07E-09&1.85E-14&6.34E-15&6.26E-15\cr
    \midrule
     \multicolumn{3}{c}{\multirow{4}{*}{2370}}&RK5&1.54E-02&9.31E-05&1.72E-05&1.41E-06&1.20E-02&2.57E-06&2.96E-07&1.31E-08\cr
    &&&M1&8.31E-05&1.75E-10&5.96E-11&5.20E-10&6.62E-06&4.34E-12&6.49E-12&9.63E-12\cr
    &&&M2&8.89E-05&2.15E-10&4.44E-11&3.54E-10&7.09E-06&5.44E-12&6.62E-12&8.33E-12\cr
    &&&M3&8.27E-05&1.30E-08&1.17E-07&7.13E-08&6.59E-06&3.60E-10&2.02E-09&6.99E-10\cr
    \midrule
     \multicolumn{3}{c}{\multirow{4}{*}{7572}}&RK5&1.12E-01&6.37E-04&1.59E-04&1.77E-05&7.77E-03&1.77E-05&2.75E-06&1.67E-07\cr
    &&&M1&4.66E-04&1.93E-09&2.58E-10&1.22E-09&2.85E-05&5.67E-11&6.25E-11&3.88E-12\cr
    &&&M2&4.73E-04&1.56E-09&5.02E-10&8.03E-10&2.90E-05&4.67E-11&5.94E-11&8.12E-12\cr
    &&&M3&5.83E-04&5.72E-08&5.41E-08&3.29E-07&3.57E-05&1.59E-09&1.02E-09&3.09E-09\cr
    \midrule
     \multicolumn{3}{c}{\multirow{4}{*}{9954}}&RK5&6.63E-02&1.99E-03&1.91E-04&2.26E-05&4.34E-03&5.56E-05&3.29E-06&2.30E-07\cr
    &&&M1&9.538E-04&1.35E-08&1.62E-10&5.74E-11&5.86E-05&4.60E-10&1.17E-10&7.41E-11\cr
    &&&M2&9.65E-04&1.44E-08&5.70E-10&5.01E-10&5.94E-05&4.88E-10&1.12E-10&7.13E-11\cr
    &&&M3&1.19E-02&1.30E-07&8.03E-08&4.48E-07&7.33E-05&3.71E-09&1.54E-09&4.00E-09\cr
    \bottomrule
    \end{tabular}
    \end{threeparttable}
\end{table}

\section{Conclusions}
Unlike the rotation and linear transformation method of \cite{Fukushima2004} (M2) and the correction approach of \cite{Ma2008c} (M3), a new extension scheme has been established here.
For a pure Keplerian system, we introduce seven parameters $\bm{s}=(s_1,s_2,...,s_7)^{\mathrm{T}}$ to the modified vector $\bm{\varepsilon}$,
and make the readjusted solution satisfy the seven independent and dependent quantities including the Kepler energy,
three components of the angular momentum vector, and three components of the Laplace vector. Then, the problem is how to solve such a set of nonlinear equations about $\bm{s}$. The Newton iterative method combined with SVD is used to solve these underdetermined equations, and the corrected numerical solutions are obtained. The new method can be extended to a perturbed two-body or multi-body system. In the perturbed case, the reference solutions of $K$, $\bm{P}$, and $\bm{L}$ are calculated by the integral-invariant relations of $K$, $\bm{P}$, and $\bm{L}$.

To evaluate the performance of the new method, we take the pure two-body problem and the inner solar system as tested models.
For the new scheme, the errors of all orbital elements can achieve the order of the machine epsilon in the pure Keplerian problem. In addition, the accuracies of all the Keplerian elements for each planet in the inner solar system can be improved typically by the new correction method, compared with the uncorrected integrator. The numerical performance in the correction of the seven slowly-varying quantities is more effective than in that of the five integrals. Especially, the variation of eccentricity does not affect the effectiveness of M1.
Compared with M2, M1 almost has the same performance in suppressing the errors of all the orbital elements for each body in the inner solar system. 
It means that the new scheme is feasible and effective.

\normalem
\begin{acknowledgements}
The authors are very grateful to Prof. Xin Wu and Prof. Yan-Ning Fu for valuable suggestions and discussions. This research was supported by the National Natural Science Foundation of China under Nos. 11703005, 11533004, 11178006, 11673071, 11263003, and 11273066.
\end{acknowledgements}

\appendix

\section{An iterative method}
\label{sec:iteration}
Eq.~(\ref{eq:3}) is underdetermined. It means that the number of independent equations is less than that of unknown variables. The Newton iteration method cannot solve this kind of system of equations. Fortunately, the SVD method is helpful to solve the underdetermined linear equations \citep{1992nrfa.book.....P}. Thus, the Newton iterative combined with the SVD method is used to solve Eq.~(\ref{eq:3}). The specific operation process is as follows. Assume that $\Delta\bm{\phi}(\bm{s}) = \bm{0}$ has an approximate root $\bm{s_k}$, and the set of nonlinear equations are expanded at this root. Then, we have
\begin{equation}\label{eq:51}
\Delta\bm{\phi}(\bm{s})\approx\Delta\bm{\phi}(\bm{s_k})-\Delta\bm{\phi}^{\prime}(\bm{s_k})(\bm{s}-\bm{s_k}).
\end{equation}
In fact, $\Delta\bm{\phi}(\bm{s}) =\bm{0}$ can be approximated as
\begin{equation}\label{eq:61}
\Delta\bm{\phi}(\bm{s_k})-\Delta\bm{\phi}^{\prime}(\bm{s_k})(\bm{s}-\bm{s_k}) =\bm{0}.
\end{equation}
Eq.~\ref{eq:61} is an underdetermined system of linear equations. That is to say, $\Delta\bm{\phi}^{\prime}(\bm{s_k})$ is a singular matrix whose inverse does not exist. In this case, the SVD method is used to solve its pseudo inverse $\Delta\bm{\phi}^{\prime+}(\bm{s_k})$. That is
\begin{equation}\label{eq:71}
\Delta\bm{\phi}^{\prime+}(\bm{s_k})=\bm{V}(\bm{s_k})\bm{\Sigma}^{+}(\bm{s_k})\bm{U}^{T}(\bm{s_k}).
\end{equation}
Here, $\bm{V}$ is a $7\times 7$ unitary matrix, $\bm{U}$ is a $7\times 7$ unitary matrix and $\bm{\Sigma}$ is a $7\times 7$  diagonal matrix with positive or zero elements (the singular values). Setting $\bm{s_{k+1}}=\bm{s}$, we have the following iterative formula
\begin{equation}\label{eq:81}
\bm{s_{k+1}}=\bm{s_k}-\Delta\bm{\phi}^{\prime+}(\bm{s_k})\Delta\bm{\phi}(\bm{s_k}), \\k=0,1....
\end{equation}
In this way, the roots $\bm{s}^{\ast}$ of Eq.~(\ref{eq:3}) can be obtained. Finally, $\bm{x}^{\ast} = \bm{x}_1+\bm{\varepsilon}(\bm{s}^{\ast})$ is given.

\bibliographystyle{raa}
\bibliography{ref}{}

\end{document}